\documentclass[a4paper,11pt]{article}
\usepackage{amssymb,amsmath}
\usepackage{graphicx}\graphicspath{{Figures/}} 
\usepackage{color}
\usepackage{enumitem}
\usepackage{array}
\usepackage{braket}

\usepackage{wrapfig}
\usepackage{cancel}
\usepackage{float}
\usepackage{slashed}
\usepackage{subcaption}
\usepackage{tikz}
\usepackage{bigints}

\usepackage{cases}

\usepackage[table]{xcolor}
\usepackage[normalem]{ulem}
\usepackage{soul}
\usepackage{mathrsfs}
\usepackage{slashed}

\mathchardef\ordinarycolon\mathcode`\:
\mathcode`\:=\string"8000
\begingroup \catcode`\:=\active
\gdef:{\mathrel{\mathop\ordinarycolon}}
\endgroup

\pdfoutput=1

\usepackage{jheppub}\makeatletter\gdef\@fpheader{}\makeatother
\usepackage[T1]{fontenc}

\title{ The effects of polarization on the observables in the decay $\Xi_{cc}^{++} \rightarrow \Xi_{c}^{+} \bar{\ell}\nu_{\ell}$}

\author[a,b]{Qazi Maaz Us Salam,}
\author[a]{Anamta Asif,\footnote{Corresponding author}}
\author[b]{Ishtiaq Ahmed,}
\author[a]{Rizwan Khalid}

 \affiliation[a]{School of Science and Engineering, Lahore University of Management Sciences (LUMS), Opposite Sector U, D.H.A, Lahore 54792, Pakistan}
 \affiliation[b]{National Center for Physics,  Shahdra Valley Road, Islamabad 44000, Pakistan}

\emailAdd{qazimaaz92@gmail.com}
\emailAdd{25100098@lums.edu.pk}
\emailAdd{ishtiaq.ahmed@ncp.edu.pk}
\emailAdd{rizwan\_khalid@lums.edu.pk}

\abstract{We investigate the effects of polarization on several physical observables in the semileptonic decay $\Xi_{cc}^{++} \rightarrow \Xi_c^{+} \bar{\ell}\nu_{\ell}$. We analyze the polarization effects of the particles involved in the decay, namely $\Xi_{cc}^{++}$, $\Xi_c^{+}$, and the charged muon $\ell$. Using the form factors obtained from QCD sum rules, we compute the $q^{2}$-dependent observables including the differential branching ratio, forward–backward asymmetry, and polarization asymmetries for both longitudinal and transverse polarization states. We also define and examine several polarization ratios and discuss correlations among different observables. In addition, we evaluate the lepton flavor universality ratio defined as $\mathcal{R}_{\Xi_c^+}(\mu/e) \equiv \mathcal{D}(\Xi_{cc}^{++}\to \Xi_c^+\mu^+\nu_\mu)/\mathcal{D}(\Xi_{cc}^{++}\to \Xi_c^+e^+\nu_e)$ and analyze its behavior over the available dynamical range. {Our results show that these observables are quite sensitive to polarization effects, and can provide suitable probes for testing Standard Model predictions.} 
}
\makeatletter
\def\ps@plain{
  \let\@oddhead\@empty
  \let\@evenhead\@empty
  \def\@oddfoot{\hfil\thepage\hfil}
  \def\@evenfoot{\hfil\thepage\hfil}
}
\def\ps@titlepage{
  \let\@oddhead\@empty
  \let\@evenhead\@empty
  \def\@oddfoot{\hfil\thepage\hfil}
  \def\@evenfoot{\hfil\thepage\hfil}
}
\makeatother

\begin{document} 

\pagenumbering{arabic}
\setcounter{page}{1}

\maketitle
\thispagestyle{titlepage}
\pagestyle{plain}
\setcounter{page}{2}

\section{Introduction}

The semileptonic decays of heavy baryons offer valuable insight into the dynamics of heavy flavor and weak interactions. Among these processes, the study of flavor changing charged current (FCCC) interactions which occur at the quark level through decay of a heavy quark to a light one in conjunction with a lepton-antineutrino/antilepton-neutrino pair is interesting. Such studies provide information about the Cabibbo-Kobayashi-Maskawa (CKM) matrix elements, as well as the structure of the weak effective Hamiltonian. In this manuscript, we analyze in detail the decay $\Xi_{cc}^{++} \to \Xi_c^+ \ell^+ \nu_\ell$, which at the quark level corresponds to the transition $c \to s$, making it sensitive to the CKM matrix element $|V_{cs}|$. This decay also serves as a testing ground for heavy quark effective theory (HQET) symmetries in the baryon sector.

The quark model~\cite{Gell-Mann:1964ewy} has been crucial in describing various features of hadron spectroscopy. {While many theorized particles have yet to be experimentally confirmed, a large number of doubly heavy baryons have been identified.}
One such baryon is $\Xi_{cc}^+ (3520)$, which was first observed by the SELEX collaboration in $2002$ via its decay to $pD^+K^-$~\cite{SELEX:2002wqn}. Similarly, the doubly charmed baryon $\Xi_{cc}^{++} (3621)$ was confirmed in the LHCb experiment in 2017, making use of the $\Xi_{cc}^{++} \to \Lambda_c^+K^-\pi^+\pi^+$ channel~\cite{LHCb:2017iph}. Subsequent experiments had been carried out in search for the $\Xi_{cc}^+$ through the decay channels $\Xi_{cc}^{+} \to \Lambda_c^+K^-\pi^+$ in $2019$~\cite{LHCb:2019gqy} and $\Xi_{cc}^{+} \to \Xi_c^+\pi^+\pi^-$ in $2021$ by the LHCb collaboration~\cite{LHCb:2021eaf}. 
The combined result of these three experiments concluded with a local significance of $4.0$ standard deviations for the mass of $\Xi_{cc}^+$~\cite{LHCb:2021eaf} to be $3620$~MeV, showing a minimal gap between the masses of $\Xi_{cc}^+$ and $\Xi_{cc}^{++}$. The search for more doubly heavy baryons is ongoing with little success~\cite{LHCb:2021xba}.

{Motivated by these experimental developments, numerous theoretical studies have investigated the properties of doubly heavy baryons using a variety of approaches, including analyses of their masses and residues~\cite{ShekariTousi:2024mso,Aliev:2012ru,Aliev:2012iv,Aliev:2019lvd,Aliyev:2022rrf,Lu:2017meb,Wang:2018lhz,Ebert:2002ig,Zhang:2008rt,Wang:2010hs,Rahmani:2020pol,Yao:2018ifh,Padmanath:2019ybu,Brown:2014ena,Shah:2017liu,Giannuzzi:2009gh,Shah:2016vmd,Valcarce:2008dr,Wang:2010it,Yoshida:2015tia,Ortiz-Pacheco:2023kjn}, chiral effective Lagrangians~\cite{Qiu:2020omj}, mixing angles~\cite{Aliev:2012nn}, strong coupling constants~\cite{Aliev:2020aon,Olamaei:2020bvw,Aliev:2020lly,Aliev:2021hqq,Olamaei:2021hjd,Alrebdi:2020rev}, strong interactions and decays~\cite{Azizi:2020zin,Qin:2021zqx,Xiao:2017udy,Xiao:2017dly}, radiative decays~\cite{Rahmani:2020pol,Li:2017pxa,Ortiz-Pacheco:2023kjn,Aliev:2021hqq,Xiao:2017udy}, weak decays~\cite{Zhao:2018mrg,Xing:2018lre,Jiang:2018oak,Tousi:2024usi,Gerasimov:2019jwp,Wang:2017mqp,Sharma:2017txj,Patel:2024mfn,Gutsche:2019wgu,Gutsche:2019iac,Ke:2019lcf,Cheng:2020wmk,Hu:2020mxk,Li:2020qrh,Han:2021gkl,Wang:2017azm,Shi:2017dto,Ivanov:2020xmw,Shi:2020qde,Hu:2017dzi,Li:2018epz,Shi:2019hbf,Shi:2019fph,Gutsche:2017hux,Gutsche:2018msz,ShekariTousi:2025fjf}, magnetic moments~\cite{Ozdem:2018uue,Ozdem:2019zis}, and lifetimes~\cite{Berezhnoy:2018bde}, \emph{etc}.} The computation of these properties requires the use of non-perturbative approaches.

{One such approach that has been successful and effective in calculating different hadronic parameters is that of QCD sum rules, formulated by Shifman, Vainshtein, and Zakharov~\cite{Shifman:1978bx,Shifman:1978by}, where the correlation functions of decays are computed using relevant interpolating currents~\cite{Aliev:2010uy,Aliev:2009jt,Agaev:2016dev,Azizi:2016dhy,Wang:2007ys,ShekariTousi:2026efp,Aliev:2025dnz,Neishabouri:2025abl,Lu:2025gol,Ahmadi:2025oal,Yu:2026tbk}. This correlation function, in turn, helps to calculate the form factors of the decays that we use in our analysis~\cite{Tousi:2024usi}.}

Investigating polarization effects in heavy baryon decays provides a powerful mechanism for exploring the dynamics of weak interactions, revealing information that cannot be obtained from decay rates alone. We focus on the polarization structure of the decay $\Xi_{cc}^{++} \rightarrow \Xi_{c}^{+} \bar{\ell}\nu_{\ell}$, which serves as a sensitive probe of the Lorentz covariant decomposition of the hadronic weak current~\cite{Tousi:2024usi}. 
Furthermore, these polarization measurements are particularly valuable for testing Standard Model (SM) predictions and potentially revealing signatures of new physics (NP) effects beyond the SM.

A lot of research over the past decades has focused on indirect signatures of new physics (NP) effects beyond the SM in the B physics sector~\cite{Crivellin:2012ye,Sakaki:2012ft,Sakaki:2014sea,Calibbi:2015kma,Dumont:2016xpj,Feruglio:2016gvd,Bordone:2016gaq,Bordone:2017anc,Choudhury:2017qyt,Li:2018rax,Aarfi:2025qcp,Salam:2024nfv,Wang:2021xib,Wang:2019trs,Gomez:2019xfw,Yasmeen:2024cki,Bolognani:2024cmr,Fajfer:2015ixa,Leng:2020fei,Zhang:2025tki,BESIII:2023gbn,Zafar:2025mhe,Colangelo:2021dnv,Belfatto:2019swo,Boora:2024nsx,Huang:2025kof}. In particular, one popular test of the SM has been measurement of the lepton flavor universality (LFU) ratios $R_D$ and $R_{D^*}$~\cite{Li:2018lxi,Bifani:2018zmi,Bernlochner:2017jka,Bigi:2017jbd,Belle:2019rba,Belle-II:2025yjp}. {These ratios have attracted considerable attention because earlier measurements suggested deviations from the Standard Model at the level of $\sim 3\sigma$~\cite{BaBar2013RD}. More recent analyses, however, have reduced this discrepancy to $\sim 1.7\sigma$~\cite{Belle-II:2025yjp}.} Nonetheless, such LFU ratios continue to play an important role as a test for the SM and in the hunt for NP signatures. We, therefore, also compute the LFU ratio for our process 
$\Xi_{cc}^{++} \rightarrow \Xi_{c}^{+} \bar{\ell}\nu_{\ell}$. 

We now comment on the structure of the rest of this manuscript. In Section~\ref{TF}, we describe the theoretical framework used to study the various physical observables of the decay $\Xi_{cc}^{++} \rightarrow \Xi_c^{+} \bar{\ell} \nu_{\ell}$. We present the parameterization of the $\Xi_{cc}^{++} \rightarrow \Xi_c^{+}$ hadronic matrix elements in terms of form factors and use them to derive the branching ratio, forward-backward asymmetry, and polarization asymmetries. We also define and consider the corresponding polarization ratios. Section~\ref{PS} begins with a listing of the numerical input parameters and the form factors. We then go on to examine the phenomenological implications through various observables of the decay $\Xi_{cc}^{++} \rightarrow \Xi_c^{+} \bar{\ell} \nu_{\ell}$ in Section~\ref{PS}. Finally, we summarize our findings and present the conclusions in Section~\ref{conclude}.

\section{Theoretical Framework}\label{TF}

\begin{figure}[H]
    
    \centering
    
    \tikzset{every picture/.style={line width=0.75pt}}       
    
    \begin{tikzpicture}[x=0.75pt,y=0.75pt,yscale=-1,xscale=1]
        \draw  [fill={rgb, 255:red, 155; green, 155; blue, 155 }  ,fill opacity=0.46 ] (347.46,119.97) .. controls (358.63,119.83) and (367.92,140.37) .. (368.21,165.85) .. controls (368.5,191.33) and (359.68,212.1) .. (348.51,212.24) .. controls (337.34,212.38) and (328.05,191.84) .. (327.76,166.36) .. controls (327.47,140.88) and (336.29,120.11) .. (347.46,119.97) -- cycle ;
        
        \draw  [fill={rgb, 255:red, 155; green, 155; blue, 155 }  ,fill opacity=0.46 ] (154.42,120.46) .. controls (165.59,120.32) and (174.88,140.86) .. (175.17,166.34) .. controls (175.45,191.82) and (166.63,212.59) .. (155.46,212.73) .. controls (144.29,212.87) and (135.01,192.33) .. (134.72,166.85) .. controls (134.43,141.38) and (143.25,120.61) .. (154.42,120.46) -- cycle ;
        
        \draw    (164.88,142.02) -- (338.04,141.52) ;
        
        \draw    (164.88,166.6) -- (338.04,166.11) ;
        
        \draw    (164.88,191.18) -- (338.04,190.69) ;
        \draw   (215.68,137.1) -- (222.6,142.39) -- (215.68,147.67) ;
        \draw   (245.69,186.26) -- (252.62,191.55) -- (245.69,196.83) ;
        \draw   (245.69,160.45) -- (252.62,165.74) -- (245.69,171.02) ;
        \draw   (273.4,137.1) -- (280.32,142.39) -- (273.4,147.67) ;
        
        \draw   (285.48,88.19) .. controls (283.14,87.13) and (280.92,86.13) .. (280.47,86.74) .. controls (280.02,87.35) and (281.47,89.41) .. (282.99,91.58) .. controls (284.52,93.74) and (285.97,95.8) .. (285.52,96.41) .. controls (285.07,97.03) and (282.84,96.02) .. (280.5,94.96) .. controls (278.17,93.91) and (275.94,92.9) .. (275.49,93.52) .. controls (275.04,94.13) and (276.49,96.19) .. (278.01,98.35) .. controls (279.54,100.51) and (280.99,102.58) .. (280.54,103.19) .. controls (280.09,103.8) and (277.86,102.8) .. (275.52,101.74) .. controls (273.19,100.68) and (270.96,99.68) .. (270.51,100.29) .. controls (270.06,100.91) and (271.51,102.97) .. (273.03,105.13) .. controls (274.56,107.29) and (276.01,109.35) .. (275.56,109.96) .. controls (275.11,110.58) and (272.88,109.57) .. (270.55,108.52) .. controls (268.21,107.46) and (265.98,106.46) .. (265.53,107.07) .. controls (265.08,107.68) and (266.53,109.74) .. (268.06,111.9) .. controls (269.58,114.07) and (271.03,116.13) .. (270.58,116.74) .. controls (270.13,117.35) and (267.9,116.35) .. (265.57,115.29) .. controls (263.23,114.24) and (261,113.23) .. (260.55,113.84) .. controls (260.1,114.46) and (261.55,116.52) .. (263.08,118.68) .. controls (264.6,120.84) and (266.05,122.9) .. (265.6,123.52) .. controls (265.15,124.13) and (262.93,123.13) .. (260.59,122.07) .. controls (258.25,121.01) and (256.02,120.01) .. (255.57,120.62) .. controls (255.12,121.23) and (256.57,123.29) .. (258.1,125.46) .. controls (259.62,127.62) and (261.08,129.68) .. (260.62,130.29) .. controls (260.17,130.9) and (257.95,129.9) .. (255.61,128.84) .. controls (253.27,127.79) and (251.04,126.78) .. (250.59,127.4) .. controls (250.14,128.01) and (251.59,130.07) .. (253.12,132.23) .. controls (254.64,134.39) and (256.1,136.45) .. (255.65,137.07) .. controls (255.19,137.68) and (252.97,136.68) .. (250.63,135.62) .. controls (248.29,134.56) and (246.06,133.56) .. (245.61,134.17) .. controls (245.16,134.79) and (246.61,136.85) .. (248.14,139.01) .. controls (248.7,139.8) and (249.25,140.59) .. (249.7,141.29) ;
        
        \draw  [color={rgb, 255:red, 155; green, 155; blue, 155 }  ,draw opacity=1 ][fill={rgb, 255:red, 255; green, 255; blue, 255 }  ,fill opacity=1 ] (145,141.23) .. controls (145,135.91) and (149.45,131.6) .. (154.94,131.6) .. controls (160.43,131.6) and (164.88,135.91) .. (164.88,141.23) .. controls (164.88,146.55) and (160.43,150.87) .. (154.94,150.87) .. controls (149.45,150.87) and (145,146.55) .. (145,141.23) -- cycle ;
        
        \draw    (316.11,34.6) -- (286.1,87.2) ;
        
        \draw    (340.35,94.58) -- (286.1,87.2) ;
        
        \draw  [fill={rgb, 255:red, 0; green, 0; blue, 0 }  ,fill opacity=1 ] (284.36,87.2) .. controls (284.36,86.18) and (285.14,85.36) .. (286.1,85.36) .. controls (287.05,85.36) and (287.83,86.18) .. (287.83,87.2) .. controls (287.83,88.22) and (287.05,89.05) .. (286.1,89.05) .. controls (285.14,89.05) and (284.36,88.22) .. (284.36,87.2) -- cycle ;
        
        \draw  [fill={rgb, 255:red, 0; green, 0; blue, 0 }  ,fill opacity=1 ] (248,141.77) .. controls (248,140.75) and (248.78,139.93) .. (249.73,139.93) .. controls (250.69,139.93) and (251.46,140.75) .. (251.46,141.77) .. controls (251.46,142.79) and (250.69,143.61) .. (249.73,143.61) .. controls (248.78,143.61) and (248,142.79) .. (248,141.77) -- cycle ;
        \draw   (301.28,54.39) -- (305.63,52.62) -- (307.06,58.01) ;
        \draw   (316.17,95.02) -- (313.48,90.97) -- (318.02,88.16) ;
        
        \draw  [color={rgb, 255:red, 155; green, 155; blue, 155 }  ,draw opacity=1 ][fill={rgb, 255:red, 255; green, 255; blue, 255 }  ,fill opacity=1 ] (145,166.6) .. controls (145,161.28) and (149.45,156.96) .. (154.94,156.96) .. controls (160.43,156.96) and (164.88,161.28) .. (164.88,166.6) .. controls (164.88,171.92) and (160.43,176.23) .. (154.94,176.23) .. controls (149.45,176.23) and (145,171.92) .. (145,166.6) -- cycle ;
        
        \draw  [color={rgb, 255:red, 155; green, 155; blue, 155 }  ,draw opacity=1 ][fill={rgb, 255:red, 255; green, 255; blue, 255 }  ,fill opacity=1 ] (145,191.18) .. controls (145,185.86) and (149.45,181.54) .. (154.94,181.54) .. controls (160.43,181.54) and (164.88,185.86) .. (164.88,191.18) .. controls (164.88,196.5) and (160.43,200.81) .. (154.94,200.81) .. controls (149.45,200.81) and (145,196.5) .. (145,191.18) -- cycle ;
        
        \draw  [color={rgb, 255:red, 155; green, 155; blue, 155 }  ,draw opacity=1 ][fill={rgb, 255:red, 255; green, 255; blue, 255 }  ,fill opacity=1 ] (338.04,141.52) .. controls (338.04,136.2) and (342.49,131.89) .. (347.99,131.89) .. controls (353.48,131.89) and (357.93,136.2) .. (357.93,141.52) .. controls (357.93,146.84) and (353.48,151.16) .. (347.99,151.16) .. controls (342.49,151.16) and (338.04,146.84) .. (338.04,141.52) -- cycle ;
        
        \draw  [color={rgb, 255:red, 155; green, 155; blue, 155 }  ,draw opacity=1 ][fill={rgb, 255:red, 255; green, 255; blue, 255 }  ,fill opacity=1 ] (338.04,166.11) .. controls (338.04,160.79) and (342.49,156.47) .. (347.99,156.47) .. controls (353.48,156.47) and (357.93,160.79) .. (357.93,166.11) .. controls (357.93,171.43) and (353.48,175.74) .. (347.99,175.74) .. controls (342.49,175.74) and (338.04,171.43) .. (338.04,166.11) -- cycle ;
        
        \draw  [color={rgb, 255:red, 155; green, 155; blue, 155 }  ,draw opacity=1 ][fill={rgb, 255:red, 255; green, 255; blue, 255 }  ,fill opacity=1 ] (338.04,190.69) .. controls (338.04,185.37) and (342.49,181.05) .. (347.99,181.05) .. controls (353.48,181.05) and (357.93,185.37) .. (357.93,190.69) .. controls (357.93,196.01) and (353.48,200.32) .. (347.99,200.32) .. controls (342.49,200.32) and (338.04,196.01) .. (338.04,190.69) -- cycle ;

        \draw (148.8,186.83) node [anchor=north west][inner sep=0.75pt]  [font=\footnotesize] [align=left] {$\displaystyle u$};
        
        \draw (343.13,186.83) node [anchor=north west][inner sep=0.75pt]  [font=\footnotesize] [align=left] {$\displaystyle u$};
        
        \draw (149.96,136.67) node [anchor=north west][inner sep=0.75pt]  [font=\footnotesize] [align=left] {$\displaystyle c$};
        
        \draw (344.43,136.67) node [anchor=north west][inner sep=0.75pt]  [font=\footnotesize] [align=left] {$\displaystyle s$};
        
        \draw (149.96,162.25) node [anchor=north west][inner sep=0.75pt]  [font=\footnotesize] [align=left] {$\displaystyle c$};
        
        \draw (344.28,162.25) node [anchor=north west][inner sep=0.75pt]  [font=\footnotesize] [align=left] {$\displaystyle c$};
        
        \draw (102.55,154.14) node [anchor=north west][inner sep=0.75pt]   [align=left] {$\displaystyle \Xi _{cc}^{++}$};
        
        \draw (372.07,154.14) node [anchor=north west][inner sep=0.75pt]   [align=left] {$\displaystyle \Xi _{c}^{+}$};
        
        \draw (234.08,95.15) node [anchor=north west][inner sep=0.75pt]   [align=left] {$\displaystyle \mathbf{W}^{+}$};
        
        \draw (308.8,99.49) node [anchor=north west][inner sep=0.75pt] [align=left] {$\displaystyle \bar{\ell }$};
        
        \draw (309.11,52.67) node [anchor=north west][inner sep=0.75pt] [align=left] {$\displaystyle \nu_{\ell }$};
        
    \end{tikzpicture}
    \caption{Feynman diagram for $\Xi_{cc}^{++} \rightarrow \Xi_{c}^{+} \bar{\ell}\nu_{\ell}$ decay}
    \label{feynmandiagram}
\end{figure}
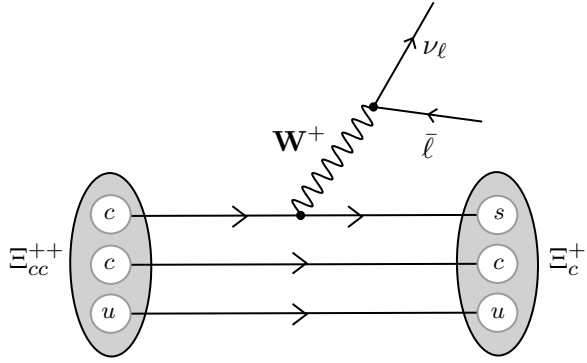

The decay amplitude for this process can be written as:
\begin{align}
    \label{eq:A}
    \mathcal{A} = \frac{G_F}{\sqrt{2}} \, V_{cs}^{*} \, L^\mu \, \langle \Xi_c^{+} | \bar{s} \gamma_\mu (1 - \gamma_5) c | \Xi_{cc}^{++} \rangle,
\end{align}
where $ L^\mu=\bar{\nu}_\ell \, \gamma^\mu (1 - \gamma^5) \, \ell$ is the leptonic current, $G_F$ is 
the Fermi constant, and $V_{cs}$ is the relevant CKM matrix element.

{The hadronic current has the standard $V-A$ structure. For a $\tfrac12\to\tfrac12$ baryon transition, Lorentz covariance allows the transition matrix element to be decomposed in terms of six independent form factors~\cite{Tousi:2024usi}.} The vector 
and axial vector currents become,

\begin{align*}
\langle \Xi_c^+(p',s') | V^\mu | \Xi_{cc}^{++}(p,s) \rangle &=
\bar{u}_{\Xi_c^+}(p',s')\left[
\mathcal{F}_1(q^2)\gamma^\mu + \mathcal{F}_2(q^2)\frac{p^\mu}{M_{\Xi_{cc}^{++}}} + \mathcal{F}_3(q^2)\frac{p'^\mu}{M_{\Xi_c^+}}
\right] u_{\Xi_{cc}^{++}}(p,s), \\
\langle \Xi_c^+(p',s') | A^\mu | \Xi_{cc}^{++}(p,s) \rangle &=
\bar{u}_{\Xi_c^+}(p',s')\left[
\mathcal{G}_1(q^2)\gamma^\mu + \mathcal{G}_2(q^2)\frac{p^\mu}{M_{\Xi_{cc}^{++}}} + \mathcal{G}_3(q^2)\frac{p'^\mu}{M_{\Xi_c^+}}
\right]\gamma_5 u_{\Xi_{cc}^{++}}(p,s),
\end{align*}
where $\mathcal{F}_1(q^2)$, $\mathcal{F}_2(q^2)$, and $\mathcal{F}_3(q^2)$ are the form factors for the vector current while $\mathcal{G}_1(q^2)$, $\mathcal{G}_2(q^2)$ and $\mathcal{G}_3(q^2)$ are the form factors for the axial vector current. 

{To discuss the polarization properties of the particles involved in the decay $\Xi_{cc}^{++}\to \Xi_c^+\bar\ell\nu_\ell$, we define a set of mutually orthogonal spin-quantization axes:
\begin{align}
    \hat{e}_L = 
    \frac{\vec{p}_{m}}{|\vec{p}_{m}|},
    \quad
    \hat{e}_N = 
    \frac{\vec{k_r} \times \vec{p}_{m}}{|\vec{k_r} \times \vec{p}_{m}|},
    \quad
    \hat{e}_T = 
    \hat{e}_N \times \hat{e}_L,
\end{align}
where labels $L, \ T, \ N$ correspond to longitudinal ($L$), transverse ($T$), and normal ($N$) polarization directions, respectively. Here $\vec p_m$ is the momentum of the particle $m$ under consideration ($m=\Xi_{cc}^{++}, \Xi_{c}^{+}, \bar\ell$), and $\vec{k}_r$ is a reference direction used to define the normal to the plane spanned by $\vec{p}_m$ and $\vec{k}_r$. While discussing the polarization of $\bar\ell$, we take the reference $\vec k_r$ to point along the momentum of $\Xi_{cc}^{++}$. In the case of a discussion of the polarization of either baryon, the momentum of $\bar\ell$ provides the reference direction. These directions are defined most conveniently in the rest frame of the particle under consideration. Using these axes, we can define the spin polarization 4-vectors:
\begin{align}
    S^\mu_{n} = (0,\hat{e}_n),
\end{align}
where $n \in \{L,T,N\}$. When boosted to the center of mass frame, the longitudinal polarization four-vector transforms to:
\begin{align}
    S_{L,\mathrm{CM}}^\mu =
    \left(
    \frac{|\vec{p}_{m}|}{M_m},
    \frac{E_m \vec{p}_{m}}{M_m |\vec{p}_{m}|}
    \right), 
\end{align}
while the normal and transverse polarization four-vectors remain unchanged. Here, $M_m$ is the mass of particle $m$ and $E_m$ is its energy.}

{The polarization structure of the decay is given by the differential branching ratio,
\begin{align}
    \mathcal{D}_k^{(m,n)}(q^2,\cos{\theta}) & \equiv \frac{d^2\mathcal{\tilde{D}}_k^{(m,n)}(q^2,\cos{\theta})}{dq^2d(\cos{\theta})} , \nonumber \\
    & =\frac{G_F^2 |V_{cs}|^2\tau_{\Xi_{cc}^{++}}}{2}\frac{q^2\frac{\sqrt{Q_+Q_-}}{2M_{\Xi_{cc}^{++}}}\left(1-\frac{m_\ell^2}{q^2}\right)^2}{192\pi^3M_{\Xi_{cc}^{++}}^2}|\mathcal{A}_k^{(m,n)}(q^2,\cos{\theta})|^2,
\end{align}
where $Q_+=(M_{\Xi_{cc}^{++}}+M_{\Xi_{c}^{+}})^2-q^2$ and $Q_-=(M_{\Xi_{cc}^{++}}-M_{\Xi_{c}^{+}})^2-q^2$. Moreover, $\mathcal{A}_k^{(m,n)}(q^2,\cos{\theta})$ is the polarization amplitude of particle $m$ ($m\in\{\ell,\Xi_{cc}^{++},\Xi_{c}^{+}\}$) polarized in the $n$ ($n\in\{L,T\}$) direction in the decay process. To obtain this amplitude, we apply the spin projection operator  $\tfrac{1}{2}(1 + \gamma_5 \slashed{S})$ on the spinors of the particle being polarized in Eq.~\ref{eq:A}, by making the replacement $(\slashed{p}_{m} + M_m) \to (\slashed{p}_{m} + M_m)\tfrac{1}{2}(1 + \gamma_5 \slashed{S})$. The index $k$ takes on the values $\pm$, referring to the positive and negative particle polarizations. The massive particles were polarized one at a time, and their results of differential branching ratio $\mathcal{D}_k^{(m,n)}$ were compared with the unpolarized differential branching ratio $\mathcal D$.} {Further, note that we do not discuss the polarization characteristics with reference to the normal direction $N$. This is because the contributions due to the positive and negative normal polarization states are equal to each other, and half that due to the unpolarized one. This is a purely kinematic effect, where the scalar product of $S^\mu_N$ with the momentum of each participating particle is zero because the normal direction is perpendicular to the decay plane.} 
The expression for the squared amplitude $|\mathcal{A}_k^{(m,n)}|^2$, after integrating over $\cos{\theta}$, for the case of polarized particles is presented in Appendix~\ref{app_ampl} while the unpolarized case yields the result,
{\allowdisplaybreaks
\begin{align}
    |\mathcal{A}|^2 & = \frac{8}{3q^4} (q^2-m_{\mu}^2) \Bigg\{  24q^4M_{\Xi_c^+} M_{\Xi_{cc}^{++}} ( -\mathcal{F}_1^2 + \mathcal{G}_2^2 ) + 4 ( \mathcal{F}_1^2 + \mathcal{G}_2^2 ) \bigl[ q^4 (M_{\Xi_c^+}^2+M_{\Xi_{cc}^{++}}^2 - m_{\mu}^2) \nonumber \\
    &- m_{\mu}^2q^2(M_{\Xi_c^+}^2+ M_{\Xi_{cc}^{++}}^2) + (2m_{\mu}^2 + q^2)(M_{\Xi_c^+}^2-M_{\Xi_{cc}^{++}}^2)^2 - 2q^6 \bigr] + \frac{1}{M_{\Xi_{cc}^{++}}^2}\bigg[\big[(M_{\Xi_c^+}^2 \nonumber \\
    & - M_{\Xi_{cc}^{++}}^2)^2 - 2M_{\Xi_c^+}^2q^2 + q^4 \big] (2m_{\mu}^2 + q^2) - 2M_{\Xi_{cc}^{++}}^2q^2(q^2-m_{\mu}^2) \bigg] \bigg[Q_+\mathcal{F}_2^2 + Q_- \mathcal{G}_2^2 \bigg] \nonumber \\
    & + \frac{1}{M_{\Xi_c^+}^2}\bigg[\Big[(M_{\Xi_c^+}^2 - M_{\Xi_{cc}^{++}}^2)^2 - 2M_{\Xi_{cc}^{++}}^2q^2 + q^4\Big](2m_{\mu}^2 + q^2)   - 2M_{\Xi_c^+}^2q^2(q^2-m_{\mu}^2) \bigg] \bigg[Q_+\mathcal{F}_3^2 \nonumber \\
    & + Q_-\mathcal{G}_3^2 \bigg] + 4 Q_+\mathcal{F}_1 \Bigg[ \left(\frac{\mathcal{F}_2}{M_{\Xi_{cc}^{++}}} + \frac{\mathcal{F}_3}{M_{\Xi_c^+}} \right) \big((M_{\Xi_c^+}-M_{\Xi_{cc}^{++}})^2(2m_{\mu}^2+q^2)- q^4 \big) (M_{\Xi_c^+} \nonumber \\
    & +M_{\Xi_{cc}^{++}}) + m_{\mu}^2q^2\bigg[ \frac{\mathcal{F}_2}{M_{\Xi_{cc}^{++}}}(M_{\Xi_{cc}^{++}}-2M_{\Xi_c^+}) + \frac{\mathcal{F}_3}{M_{\Xi_c^+}}(M_{\Xi_c^+}-2M_{\Xi_{cc}^{++}})\bigg] \Bigg] + 4 Q_-\mathcal{G}_2 \nonumber \\
    & \times \Bigg[ \bigg(-\frac{\mathcal{G}_2}{M_{\Xi_{cc}^{++}}} + \frac{\mathcal{G}_3}{M_{\Xi_c^+}}\bigg)\left((M_{\Xi_c^+}+M_{\Xi_{cc}^{++}})^2(2m_{\mu}^2+q^2)-q^4 \right)(M_{\Xi_c^+}-M_{\Xi_{cc}^{++}}) + m_{\mu}^2q^2 \nonumber \\
    & \times \bigg[ -\frac{\mathcal{G}_2}{M_{\Xi_{cc}^{++}}}(M_{\Xi_{cc}^{++}} +2M_{\Xi_c^+}) + \frac{\mathcal{G}_3}{M_{\Xi_c^+}}(M_{\Xi_c^+}+2M_{\Xi_{cc}^{++}})\bigg] \Bigg] + \frac{2}{M_{\Xi_c^+}M_{\Xi_{cc}^{++}}}\bigg[ -q^4\Big[2(M_{\Xi_c^+}^2 \nonumber \\
    & +M_{\Xi_{cc}^{++}}^2) + m_{\mu}^2\Big] + (2m_{\mu}^2+q^2)(M_{\Xi_c^+}^2-M_{\Xi_{cc}^{++}}^2)^2- m_{\mu}^2q^2(M_{\Xi_c^+}^2+M_{\Xi_{cc}^{++}}^2) + q^6 \bigg] \Big[Q_+\mathcal{F}_2\mathcal{F}_3 \nonumber \\
    & + Q_-\mathcal{G}_2\mathcal{G}_3\Big] \Bigg\}.
\end{align}
}
The forward-backward asymmetry (FBA) is defined as,
\begin{align}
    \text{FBA}_k^{(m,n)}(q^2)=\frac{\mathcal{N}^{F(m,n)}_k-\mathcal{N}^{B(m,n)}_k}{\mathcal{N}^{F(m,n)}_k+\mathcal{N}^{B(m,n)}_k},
\end{align}
where $ \mathcal{N}^{F(m,n)}_k\left(\mathcal{N}^{B(m,n)}_k\right) $ is the probability of the lepton $\ell=\mu$ moving in
the forward (backward) direction. In terms of the differential decay rate, these probabilities can computed using,
\begin{align}
\mathcal{N}^{F(m,n)}_k(q^2)=\int_{0}^{1}d(\cos\theta) \mathcal{D}^{(m,n)}_k(q^2,\cos{\theta}) \ ,\quad
\mathcal{N}^{B(m,n)}_k(q^2)=\int_{-1}^{0}d(\cos\theta) \mathcal{D}^{(m,n)}_k(q^2,\cos{\theta}).
\end{align}
It is useful to study the polarization asymmetries $P^{(m,n)}$ defined via,
\begin{align}\label{eqn.pol_asymm}
    P^{(m,n)}(q^2)=\frac{\bigintsss_{-1}^{1}d(\cos\theta) \mathcal{D}^{(m,n)}_+(q^2,\cos{\theta}) - \bigintsss_{-1}^{1}d(\cos\theta) \mathcal{D}^{(m,n)}_-(q^2,\cos{\theta})}{\bigintsss_{-1}^{1}d(\cos\theta) \mathcal{D}^{(m,n)}_+(q^2,\cos{\theta}) + \bigintsss_{-1}^{1}d(\cos\theta) \mathcal{D}^{(m,n)}_-(q^2,\cos{\theta})}.
\end{align}
{The quantity $P^{(m,n)}(q^{2})$ measures the polarization asymmetry of particle $m$ along the direction $n$ through the normalized difference between the decay rates of its positive and negative polarization states.} We go on to also compute the $q^2$-averaged values of the $P^{(m,n)}$. 

We also consider and define the ratios for the polarizations of the parent baryon $\Xi_{cc}^{++}$, the daughter baryon $\Xi_c^+$, and the muon $\ell$, defined via the following expression:
\begin{equation}
    \label{ratio_equation}
    \mathcal{R}_{nn'}^{(m,m')}[j] \equiv \frac{\bigintsss_{-1}^{1}d(\cos\theta) \mathcal{D}_j^{(m,n)}(q^2,\cos{\theta})}{\bigintsss_{-1}^{1}d(\cos\theta) \mathcal{D}_{j}^{(m',n')}(q^2,\cos{\theta})},
\end{equation}
where $j=\pm$ refers to the polarization\footnote{One may argue that a more general ratio, for example $\mathcal{R}_{TL}^{(\Xi_{cc}^{++},\ell)}[j,k]\equiv{\mathcal{D}{_j^{(\Xi_{cc}^{++},T)}}}/{\mathcal{D}{_k^{(\ell,L)}}}$, 
could be considered. We did try looking at such quantities but found the description 
too cumbersome without any clear phenomenological advantage and, therefore, chose to settle with the simpler quantities with $j=k$.}. The ratio $\mathcal{R}^{(m,m')}_{nn'}[j]$ provides a measure for 
comparing the `numbers' of $m$-type particles polarized with helicity $j$ in the $n$ direction with the $m'$-type particles polarized with helicity $j$ in the $n'$ direction.  
It turns out that we can define polarization asymmetry ratios building on Eq.~\ref{ratio_equation} that may be more robust against form factor uncertainties:
\begin{align}
    \label{pol_asymm_ratio_equation}
    \mathcal{R}_{P_{nn'}^{(m,m')}} &\equiv \frac{\mathcal{R}_{nn'}^{(m,m')}[+]\,-\,\mathcal{R}_{nn'}^{(m,m')}[-]}{\mathcal{R}_{nn'}^{(m,m')}[+]\,+\,\mathcal{R}_{nn'}^{(m,m')}[-]}.
\end{align}

\section{Phenomenological study}\label{PS}

\subsection{ Input parameters}
We show, in Table~\ref{tab:input_parameters}, various parameters (masses, Fermi constant, the relevant CKM matrix element and lifetime of the $\Xi_{cc}^{++}$) used in our analysis~\cite{ParticleDataGroup:2024cfk}. We note that we have used the central values given in Table~\ref{tab:input_parameters}. 
\begin{table}[H]
\centering
\fbox{
\begin{minipage}{1\textwidth}
\begin{center}
\resizebox{1\textwidth}{!}{
\begin{tabular}{lll}
$m_{c} = (1.27 \pm 0.02)\,\mathrm{GeV}$ &
$m_{e} = 0.51\,\mathrm{MeV}$ &
$m_{\mu} = 105\,\mathrm{MeV}$ \\
$M_{\Xi_{cc}^{++}} = (3.62 \pm 0.0015)\,\mathrm{GeV}$ &
$M_{\Xi_{c}^{+}} = (2.46 \pm 0.00023)\,\mathrm{GeV}$ &
$G_{F} = 1.17 \times 10^{-5}\,\mathrm{GeV}^{-2}$ \\
$|V_{sc}| = (0.974 \pm 0.006)$ &
$\tau_{\Xi_{cc}^{++}} = (2.56 \pm 0.27)\times 10^{-13}\,\mathrm{s}$ &
\\
\end{tabular}}
\end{center}
\end{minipage}
}
\caption{Numerical values of various input parameters used in numerical analysis~\cite{ParticleDataGroup:2024cfk}.}
\label{tab:input_parameters}
\end{table}
The hadronization of quarks and gluons is described in terms of form factors that are a source 
of theoretical uncertainties. The form factors that we use for $\Xi_{cc}^{++} \rightarrow \Xi_c^{+}$ are computed using QCD sum rules and taken from K. Azizi \textit{et al}~\cite{Tousi:2024usi}, with
\begin{equation}
{F}(q^2) = \frac{{F}(0)}{
    \left(
        1 - \alpha_1 \left(\frac{q}{M_{\Xi_{cc}^{++}}}\right)^2
          + \alpha_2 \left(\frac{q}{M_{\Xi_{cc}^{++}}}\right)^4
          + \alpha_3 \left(\frac{q}{M_{\Xi_{cc}^{++}}}\right)^6
          + \alpha_4 \left(\frac{q}{M_{\Xi_{cc}^{++}}}\right)^8
    \right)
},
\end{equation}
describing the $q^2$ (momentum transfer) dependence of the form factors $\mathcal{F}_i$ and $\mathcal{G}_i$. 
The values of $F(0)$, $\alpha_1$, $\alpha_2$, $\alpha_3$ and $\alpha_4$ corresponding to the various form factors are given in Table~\ref{FF}.

\begin{table}[H]
  \centering
  \label{tab:fit_function_parameters}
  \resizebox{1\textwidth}{!}{
  \begin{tabular}{|c|c|c|c|c|c|c|}
    \hline
    & $\mathcal{F}_{1}$ & $\mathcal{F}_{2}$ & $\mathcal{F}_{3}$ & $\mathcal{G}_{1}$ & $\mathcal{G}_{2}$ & $\mathcal{G}_{3}$ \\
    \hline
    $F(0)$ 
      & $-0.37 \pm 0.13$ 
      & $1.35 \pm 0.43$ 
      & $0.16 \pm 0.06$ 
      & $0.20 \pm 0.06$ 
      & $1.34 \pm 0.43$ 
      & $-1.86 \pm 0.65$ \\
   
    $\alpha_{1}$ 
      & $1.54$ 
      & $-0.05$ 
      & $1.99$ 
      & $2.79$ 
      & $1.07$ 
      & $2.15$ \\
  
    $\alpha_{2}$ 
      & $-23.84$ 
      & $-111.38$ 
      & $17.17$ 
      & $35.82$ 
      & $-78.83$ 
      & $-27.91$ \\

    $\alpha_{3}$ 
      & $268.50$ 
      & $1329.88$ 
      & $-442.35$ 
      & $-541.92$ 
      & $935.73$ 
      & $449.20$ \\

    $\alpha_{4}$ 
      & $-1174.79$ 
      & $-5206.92$ 
      & $2372.78$ 
      & $2240.25$ 
      & $-3550.64$ 
      & $-2702.99$ \\
    \hline
  \end{tabular}}
  \caption{ Form factors of $\Xi_{cc}^{++} \rightarrow \Xi_c^{+}$ decays which are calculated using QCD sum rules~\cite{Tousi:2024usi}}
  \label{FF}
\end{table}

\subsection{Observables focusing on one particle at a time}
In this section, we present the phenomenological analysis of various unpolarized and polarized observables from the corresponding differential decay distributions that focus on one 
particle at a time. In particular, we discuss the branching ratio $\mathcal{D}_k^{(m,n)}$, forward backward asymmetry $\text{FBA}_k^{(m,n)}$, and polarization asymmetries $P^{(m,n)}$. We have used a consistent color scheme for all the figures: black is used for the unpolarized results, green corresponds to the $k = +$ polarized state, and red denotes the $k = -$ polarized state. We use dashed lines to indicate the form factor uncertainties.

\subsubsection{Polarized $(m,n)$ cases}

We begin presenting our results by discussing the branching ratio 
$\mathcal{D}_k^{(m,L)}$ and $\text{FBA}_k^{(m,L)}$ as a function of $q^2$ for the cases of $m\in\left\{\Xi_{cc}^{++}, \Xi_{c}^{+}, \mu^+\right\}$ 
with the longitudinal direction chosen for polarization. As 
described earlier, the case of $k=+$ is presented in green and $k=-$ in red with the unpolarized result in black.

\begin{figure}[H]
    \centering
    \begin{subfigure}[b]{0.4\textwidth}
        \centering
        \includegraphics[width=\textwidth]{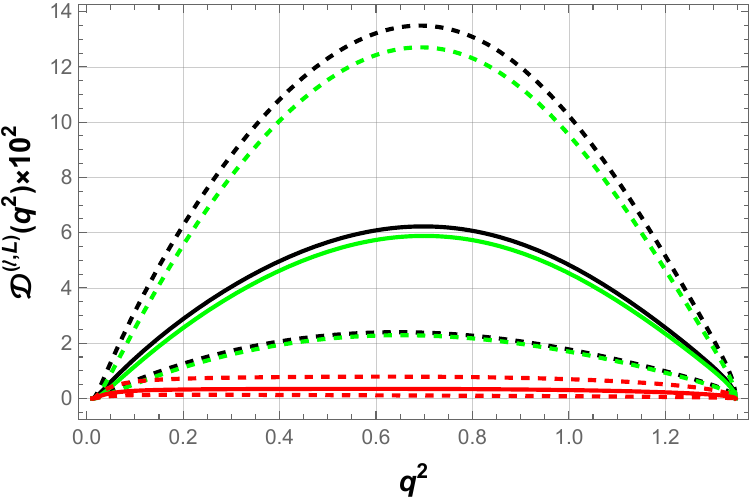}
        \caption{}
        \label{br_lepton}
    \end{subfigure}
    \begin{subfigure}[b]{0.4\textwidth}
        \centering
        \includegraphics[width=\textwidth]{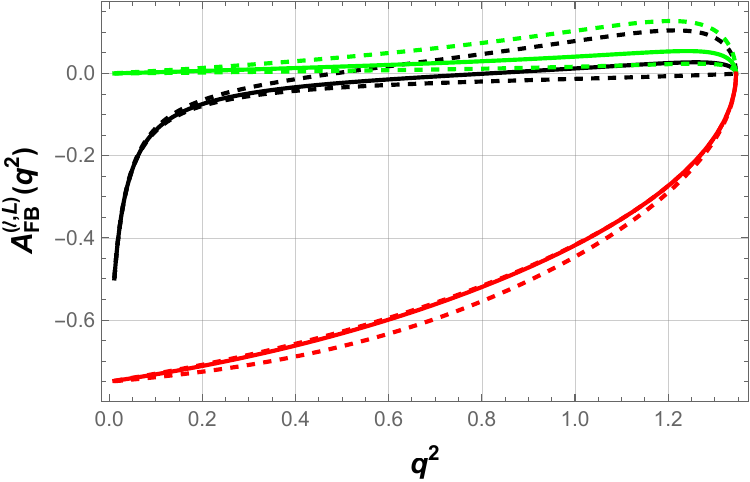}
        \caption{}
        \label{afb_lepton}
    \end{subfigure}
    \caption{Differential branching ratios $\mathcal{D}_k^{(m,n)}(q^2)$, integrated over $\cos{\theta}$,  and forward-backward asymmetry FBA$_k^{(m,n)}(q^2)$ as functions of $q^{2}$ for the SM unpolarized (black) and the longitudinally ($L$) polarized muon ($\ell$) cases. The $k=+$ longitudinally polarized state is in green and the $k=-$ one is in red. The dotted region represents uncertainty due to form factors.}
    \label{longi_lepton}
\end{figure}

In Fig.~\ref{br_lepton}, we present the branching ratio for the longitudinal direction in the case of the muon.
The peak height and position of the green and black curves are effectively the same, while the red curve is an order 
of magnitude lower 
exhibiting a clear hierarchy $\mathcal{D}(q^2)\sim \mathcal{D}_+^{(\ell,L)}(q^2)\gg \mathcal{D}_-^{(\ell,L)}(q^2)$. 
Near the endpoint of the distribution, the projector continues to favor $k = +$, while 
the $k = -$ rate is suppressed due to kinematic and spin effects\footnote{This suppression of the $k=-$ 
state relative to the $k=+$ state
at the end-point of the distribution is not obvious in Fig.~\ref{br_lepton}}. 

In Fig.~\ref{afb_lepton}, we show the forward-backward asymmetry $\text{FBA}_k^{(\ell,L)}$ as a function of $q^2$. It can be seen that for the case of the muon, the longitudinal polarization effects are quite distinct from each other in the low $q^2$ bin but do decrease towards the high $q^2$ bin. {Since the branching ratio $\mathcal{D}(q^2)\sim \mathcal{D}_+^{(\ell,L)}(q^2)$, the forward-backward asymmetry in the unpolarized (black) case generally tracks the $k=+$ (green) case, except in the low $q^2$ region. This is where the branching ratio is small overall, and there is a large forward-backward asymmetry in the $k=-$ (red) state.} The black curve exhibits a zero crossing at $q^2\sim 0.9\,{\rm GeV}^2$, while the red (green) curve remains negative (positive) throughout the full range of $q^2$.

\begin{figure}[H]
    \centering
    \begin{subfigure}[b]{0.4\textwidth}
        \centering
        \includegraphics[width=\textwidth]{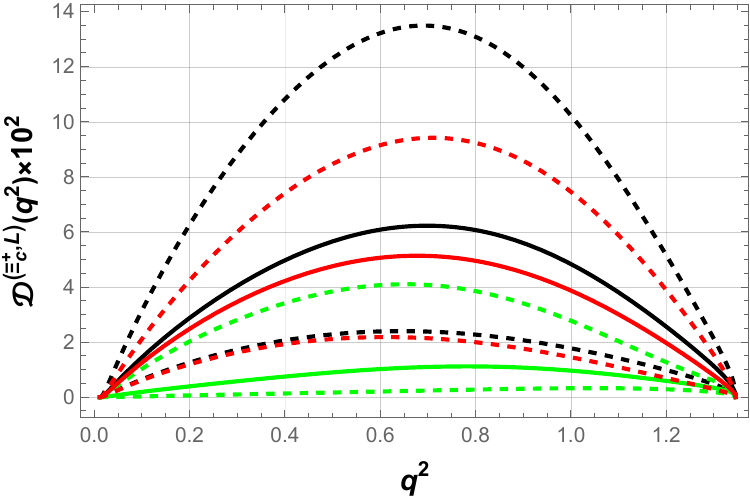}
        \caption{}
        \label{br_daughter}
    \end{subfigure}
    \begin{subfigure}[b]{0.4\textwidth}
        \centering
        \includegraphics[width=\textwidth]{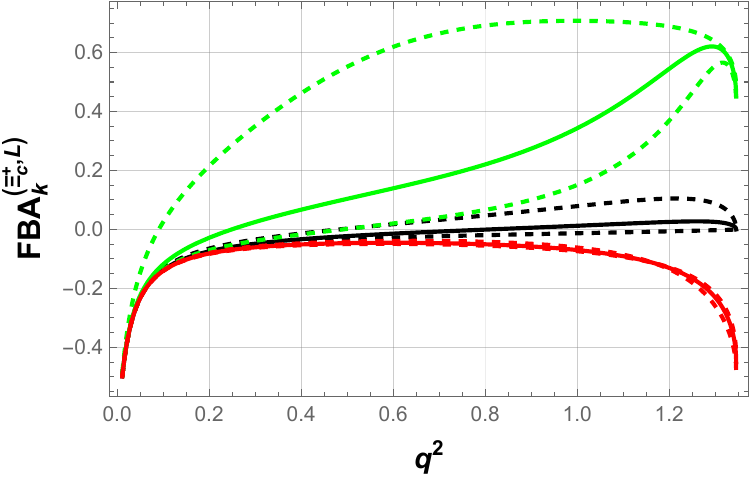}
        \caption{}
        \label{afb_daughter}
    \end{subfigure}
    \caption{Differential branching ratios $\mathcal{D}_k^{(m,n)}(q^2)$, integrated over $\cos{\theta}$, and forward–backward asymmetry FBA$_k^{(m,n)}(q^2)$ as functions of $q^{2}$ for the SM unpolarized (black) and the longitudinally ($L$) polarized daughter baryon $\Xi_c^+$ cases. Color coding is the same as in Fig.~\ref{longi_lepton}.}
\end{figure}

In Fig.~\ref{br_daughter}, we examine the case of the longitudinally polarized daughter baryon $\Xi_c^+$. The corresponding observable is $\mathcal{D}_k^{(\Xi_c^+,L)}(q^2)$. We observe that the $\Xi_c^+$ is predominantly produced in the $k=-$ (red) state, while the $k=+$ (green) state is strongly suppressed. Most of the total BR comes from the $k=-$ state. In Fig.~\ref{afb_daughter}, we show the $\text{FBA}_k^{(\Xi_c^+,L)}(q^2)$. At low $q^2$, near the threshold, the curve starts at negative values for the unpolarized as well as the polarized cases. The forward-backward asymmetry 
increases in the low $q^2$ region with both the $k=+$ longitudinally polarized (green) as well as the unpolarized (black) values exhibiting a zero crossing. In particular, at large $q^2$, the $k=+$ longitudinally polarized $\Xi_c^+$ particle 
exhibits a large forward-backward asymmetry. Since the decay is predominantly into the $k=-$ state, the 
unpolarized forward-backward asymmetry mostly tracks this state for most of the dynamical range ($q^2\lesssim 1.0~ \rm{Gev}^2$). Overall, the $\Xi_c^+$ with $k=+$ (green) favors the forward direction, while $k=-$ (red) maintains a backward tendency. Near $q^2_{\text{max}}$, the green curve rises peaks giving a forward-backward asymmetry $\sim 0.62$, whereas the red curve drops to a negative value $\sim -0.5$.

\begin{figure}[h!]
    \centering
     \begin{subfigure}[b]{0.4\textwidth}
        \centering
        \includegraphics[width=\textwidth]{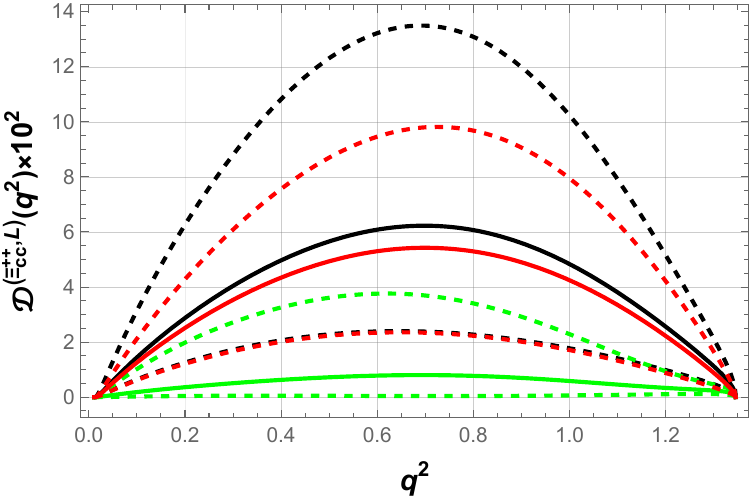}
        \caption{}
        \label{br_parent}
    \end{subfigure}
     \begin{subfigure}[b]{0.4\textwidth}
        \centering
        \includegraphics[width=\textwidth]{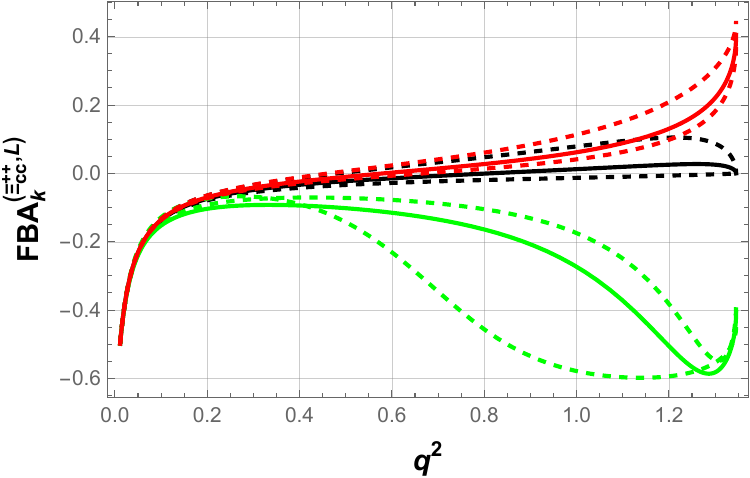}
        \caption{}
        \label{afb_parent}
    \end{subfigure}
    \caption{Differential branching ratios $\mathcal{D}_k^{(m,n)}(q^2)$, integrated over $\cos{\theta}$, and forward–backward asymmetry FBA$_k^{(m,n)}(q^2)$, integrated over $\cos{\theta}$, as functions of $q^{2}$ for the SM unpolarized (black) and the longitudinally ($L$) polarized parent baryon $\Xi_{cc}^{++}$ cases. Color coding is the same as in Fig.~\ref{longi_lepton}.}
\end{figure}

In Fig.~\ref{br_parent}, the parent baryon $\Xi_{cc}^{++}$ is longitudinally polarized, and we probe the effect of the longitudinal spin orientation of the initial state on the decay dynamics. The hierarchy follows the pattern 
$\mathcal{D}^{(\Xi_{cc}^{++},L)}~({\rm black})\gtrsim \mathcal{D}_-^{(\Xi_{cc}^{++},L)}~({\rm red})\gg \mathcal{D}_+^{(\Xi_{cc}^{++},L)}({\rm green})$, with the maximum of the green curve ($k=+$) an order of magnitude smaller than the maximum of the red ($k=-$). In Fig.~\ref{afb_parent}, we present $\text{FBA}_k^{(\Xi{cc}^{++},L)}$ as  function 
of the momentum transfer. There is a zero crossing at $q^2\sim 0.5\,{\rm GeV}^2$ for the black (unpolarized) and red ($k=-$) curves, whereas the green ($k=+$) curve remains negative. Since the decay amplitude for the $k=-$ state is much larger than the $k=+$ state, the forward-backward asymmetry for the unpolarized case mainly tracks the red ($k=-$) curve. The distinction between the two polarization states is obvious in the high $q^2$ bin, which forces the unpolarized (black) forward-backward asymmetry to deviate somewhat from the red ($k=-$) curve. At low $q^2$, we observe a strong overall backward preference. 

\begin{figure}[H]
    \centering
    \begin{subfigure}[b]{0.4\textwidth}
        \centering
        \includegraphics[width=\textwidth]{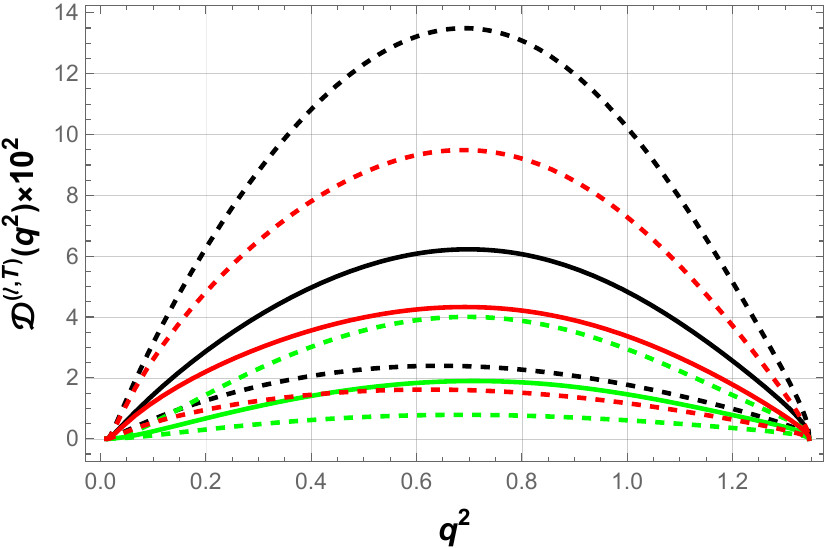}
        \caption{}
        \label{br_trans_lepton}
    \end{subfigure}
     \begin{subfigure}[b]{0.4\textwidth}
        \centering
        \includegraphics[width=\textwidth]{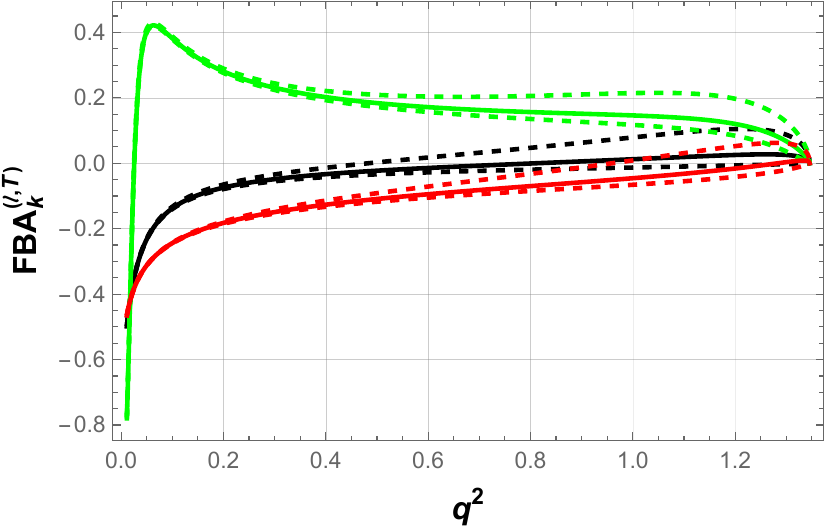}
        \caption{}
        \label{afb_trans_lepton}
    \end{subfigure}
    \caption{Differential branching ratios $\mathcal{D}_k^{(m,n)}(q^2)$, integrated over $\cos{\theta}$, and forward–backward asymmetry FBA$_k^{(m,n)}(q^2)$, integrated over $\cos{\theta}$, as functions of $q^{2}$ for the SM unpolarized and the transversely ($T$) polarized muon $\ell$ cases. Color coding is the same as in Fig.~\ref{longi_lepton}.}
\end{figure}

We now move on to present the differential branching ratio and the forward-backward asymmetry for the case of transversely polarized ($n = T$) particles.
In Fig.~\ref{br_trans_lepton}, we can see that for the transverse polarized muon, the hierarchy at all $q^2$ is $\mathcal{D}(q^2) \,(\textrm{black})> \mathcal{D}_-^{(\ell,T)}(q^2) \,(\textrm{red})> \mathcal{D}_+^{(\ell,T)}(q^2)\,(\textrm{green})$. The $\text{FBA}_k^{(\ell,T)}(q^2)$ for the transverse polarized muon presented in Fig.~\ref{afb_trans_lepton} clearly shows the distinction between the $k=+$ state (green) as having a forward preference with the $k=-$ (red) having a backward one. The unpolarized forward-backward asymmetry sits between the two and shows very little forward-backward asymmetry in the mid to large $q^2$ regime. At low $q^2$, all curves start negative, but the green one ($k=+$) increases rapidly and becomes positive ($\sim0.01\,\textrm{GeV}$) almost immediately. The unpolarized forward-backward asymmetry is more influenced by the $k=-$ (red) state.  

\begin{figure}[H]
    \centering
    \begin{subfigure}[b]{0.4\textwidth}
        \centering
        \includegraphics[width=\textwidth]{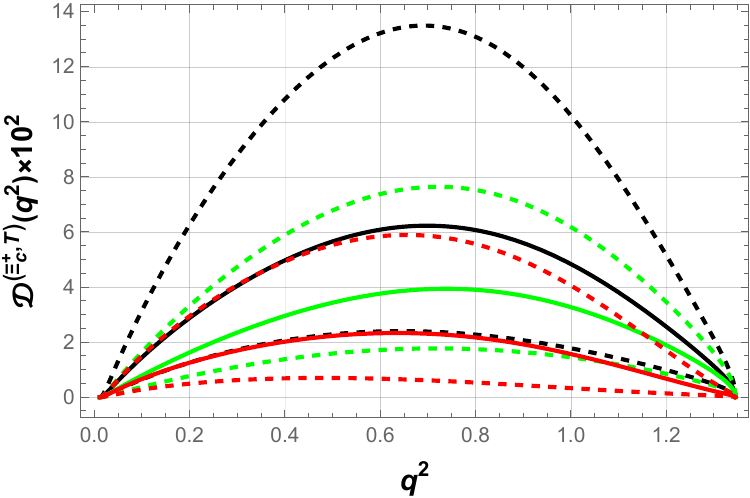}
        \caption{}
        \label{br_trans_daughter}
    \end{subfigure}
    \begin{subfigure}[b]{0.4\textwidth}
        \centering
        \includegraphics[width=\textwidth]{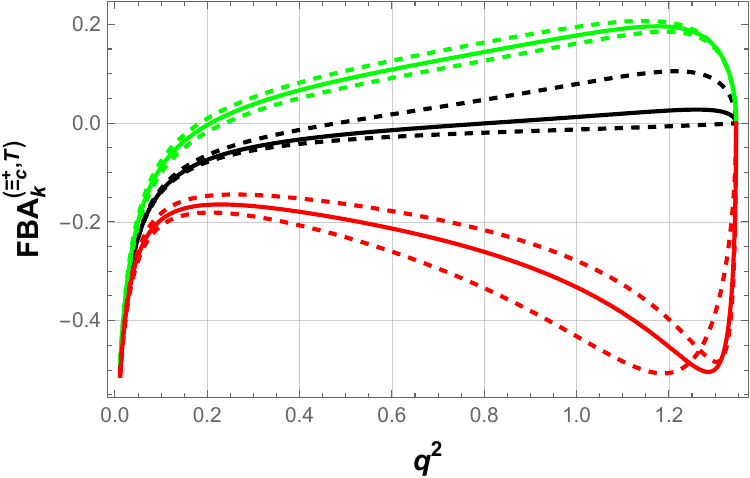}
        \caption{}
        \label{afb_trans_daugter}
    \end{subfigure}
    \caption{Differential branching ratios $\mathcal{D}_k^{(m,n)}(q^2)$, integrated over $\cos{\theta}$, and forward–backward asymmetry FBA$_k^{(m,n)}(q^2)$, integrated over $\cos{\theta}$, as functions of $q^{2}$ for the SM unpolarized  and the transversely ($T$) polarized daughter $\Xi_{c}^{+}$ cases. Color coding is the same as in Fig.~\ref{longi_lepton}.}
\end{figure}

In Fig.~\ref{br_trans_daughter} and Fig.~\ref{afb_trans_daugter}, we present the branching ratio and the forward-backward asymmetry for the case of the 
transverse polarized $\Xi_c^+$, respectively. We can see that the splitting between the branching ratio in the $k=+$ (green) and $k=-$ (red) states is modest in the low and high $q^2$ regions, and peaks for the mid $q^2$ allowed region. {The transverse projection reveals that $\Xi_c^+$ is produced slightly more in the $k=+$ (green) state.} The $\text{FBA}_k^{(\Xi{c}^{+},T)}(q^2)$ reveals that the $k=-$ (red) transverse polarized daughter $\Xi_c^+$ has a clear backward tendency as opposed to the $k=+$ (green) state. The unpolarized $\text{FBA}_k^{(\Xi{c}^{+},T)}(q^2)$ resultantly hovers around the zero mark for most of the $q^2$ region. 

\begin{figure}[H]
    \centering
     \begin{subfigure}[b]{0.4\textwidth}
        \centering
        \includegraphics[width=\textwidth]{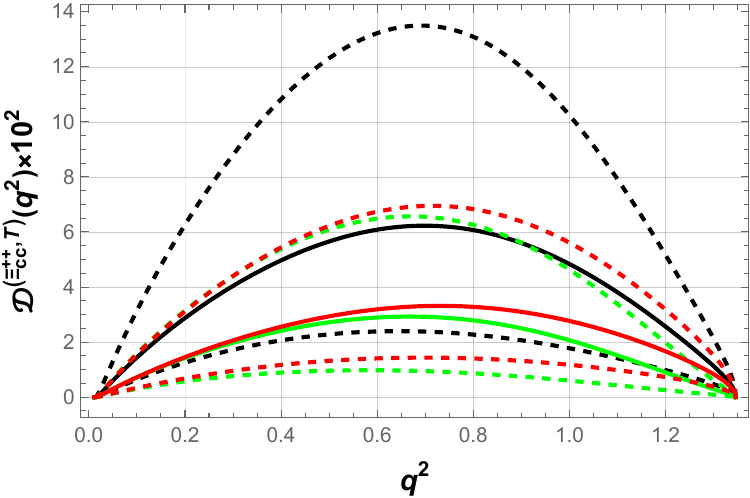}
        \caption{}
        \label{br_trans_parent}
    \end{subfigure}
     \begin{subfigure}[b]{0.4\textwidth}
        \centering
        \includegraphics[width=\textwidth]{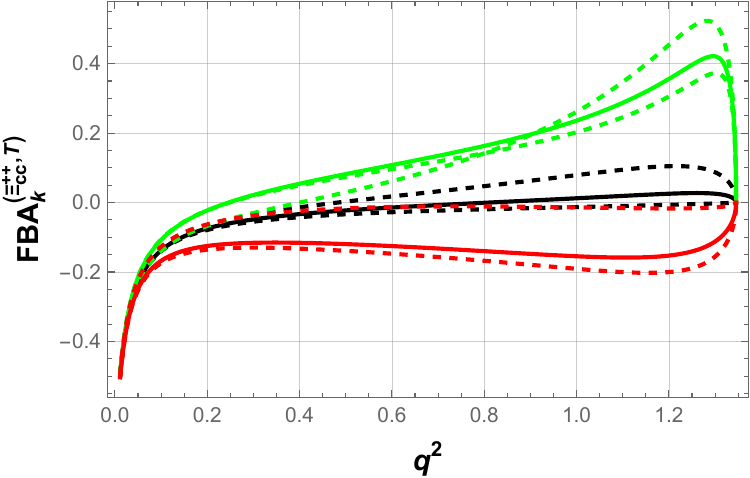}
        \caption{}
        \label{afb_trans_parent}
    \end{subfigure}
    \caption{Differential branching ratios $\mathcal{D}_k^{(m,n)}(q^2)$, integrated over $\cos{\theta}$, and forward–backward asymmetry FBA$_k^{(m,n)}(q^2)$, integrated over $\cos{\theta}$, as functions of $q^{2}$ for the SM unpolarized and the transversely ($T$) polarized parent $\Xi_{cc}^{++}$ cases. Color coding is the same as in Fig.~\ref{longi_lepton}.}
\end{figure}

In Fig.~\ref{br_trans_parent} and Fig.~\ref{afb_trans_parent}, we finally consider the transverse polarization effects of the $\Xi_{cc}^{++}$. 
It can be seen that throughout the dynamical range, $\mathcal{D}(q^2) \,(\textrm{black})> \mathcal{D}_-^{(\Xi_{cc}^{++},T)}(q^2) \,(\textrm{red})\sim \mathcal{D}_+^{(\Xi_{cc}^{++},T)}(q^2)\,(\textrm{green})$. 
At higher $q^2\gtrsim 0.5\,\textrm{GeV}^2$, we can see a somewhat higher value for $\mathcal{D}_-^{(\Xi_{cc}^{++},T)}(q^2)\,\textrm{(red)}$ as compared with $\mathcal{D}_+^{(\Xi_{cc}^{++},T)}(q^2)\,\textrm{(green)}$.  
In Fig.~\ref{afb_trans_parent}, we can see that $\text{FBA}_k^{(\Xi_{cc}^{++},T)}(q^2)$ starts off as negative. However, for $q^2\gtrsim 0.2\,\textrm{GeV}^2$, the $k=+$ (green) polarized $\Xi_{cc}^{++}$ shows a positive forward-backward asymmetry while the $k=-$ state continues to prefer the backward direction. Given that the branching ratios of the two states are similar, we can correspondingly see very little forward-backward asymmetry in the unpolarized case for the $\Xi_{cc}^{++}$.

\begin{figure}[H]
    \centering
    \begin{subfigure}[b]{0.49\textwidth}
        \centering
        \includegraphics[width=\textwidth]{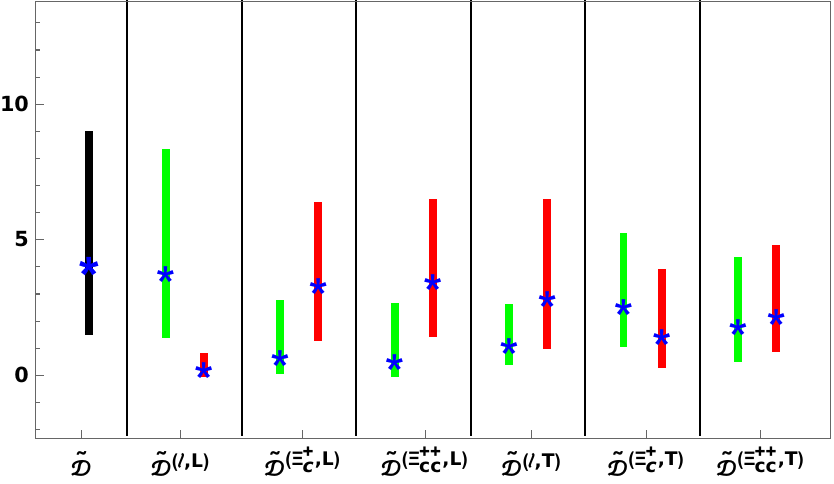}
        \caption{$\tilde{\mathcal{D}}_k^{(m,n)}$ for $q^2=[0.0112$-$1.346]$ GeV$^2$}
        \label{bar_a}
    \end{subfigure}
    \begin{subfigure}[b]{0.49\textwidth}
        \centering
        \includegraphics[width=\textwidth]{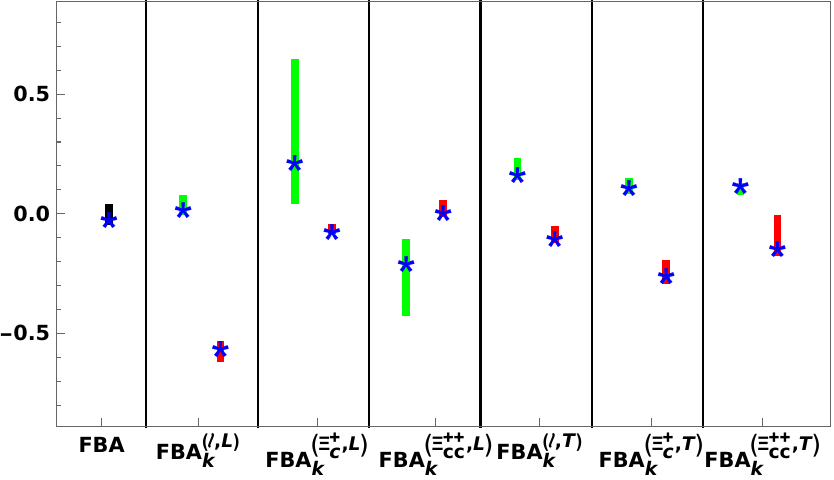}
        \caption{$\text{FBA}_k^{(m,n)}$ for $q^2=[0.0112$-$1.346]$ GeV$^2$}
        \label{bar_b}
    \end{subfigure}
        \begin{subfigure}[b]{0.49\textwidth}
        \centering
        \includegraphics[width=\textwidth]{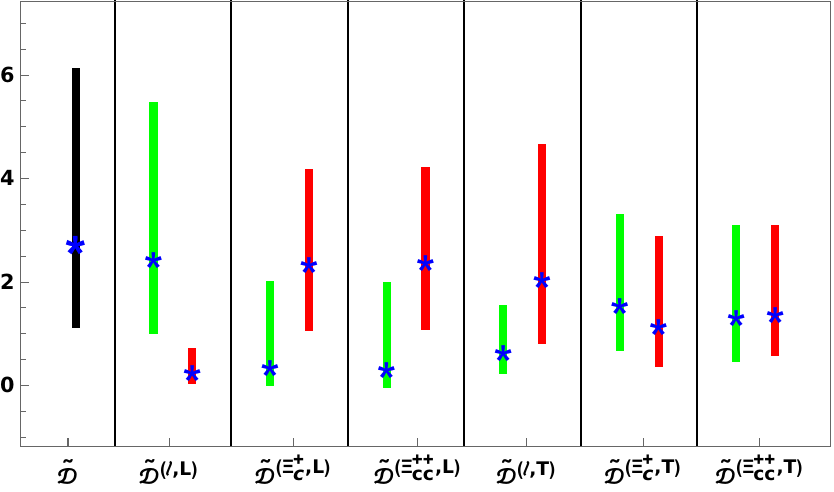}
        \caption{$\tilde{\mathcal{D}}_k^{(m,n)}$ for $q^2=[0.0112$-$0.4]$ GeV$^2$}
        \label{bar_c}
    \end{subfigure}
    \begin{subfigure}[b]{0.49\textwidth}
        \centering
        \includegraphics[width=\textwidth]{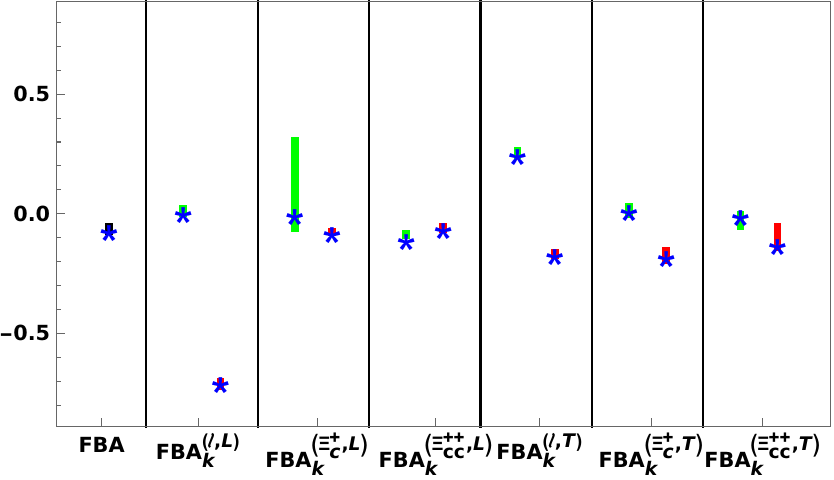}
        \caption{$\text{FBA}_k^{(m,n)}$ for $q^2=[0.0112$-$0.4]$ GeV$^2$}
        \label{bar_d}
    \end{subfigure}
        \begin{subfigure}[b]{0.49\textwidth}
        \centering
        \includegraphics[width=\textwidth]{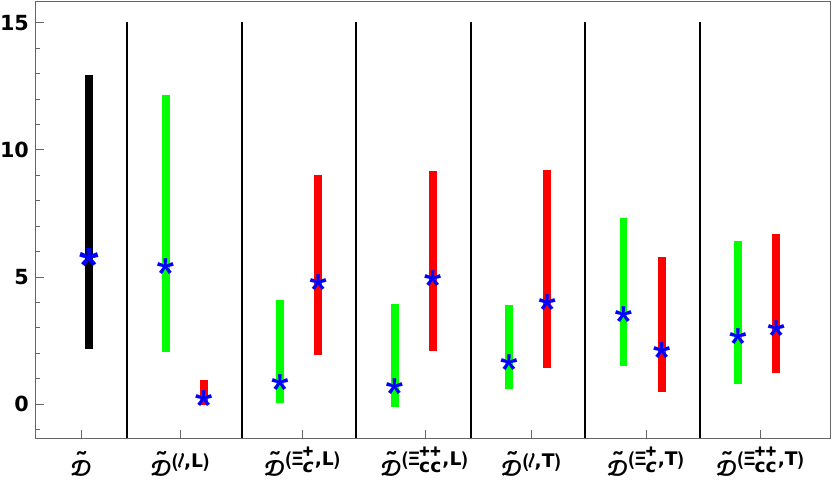}
        \caption{$\tilde{\mathcal{D}}_k^{(m,n)}$ for $q^2=[0.4$-$0.8]$ GeV$^2$}
        \label{bar_e}
    \end{subfigure}
    \begin{subfigure}[b]{0.49\textwidth}
        \centering
        \includegraphics[width=\textwidth]{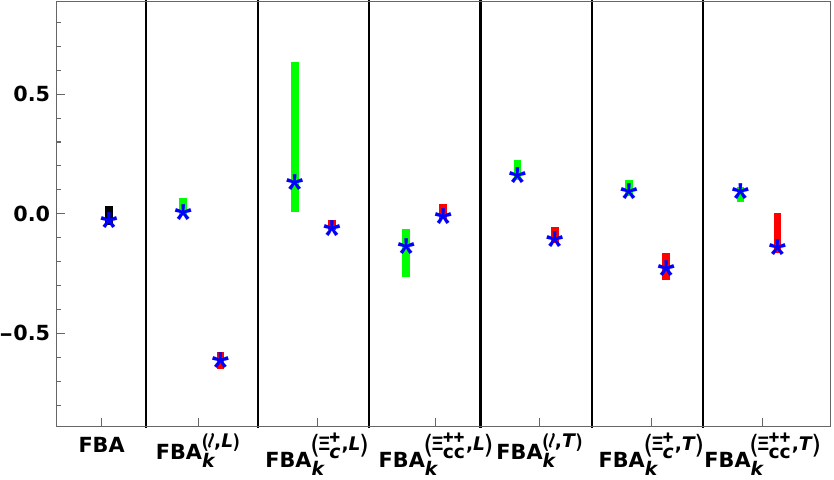}
        \caption{$\text{FBA}_k^{(m,n)}$ for $q^2=[0.4$-$0.8]$ GeV$^2$}
        \label{bar_f}
    \end{subfigure}
        \begin{subfigure}[b]{0.49\textwidth}
        \centering
        \includegraphics[width=\textwidth]{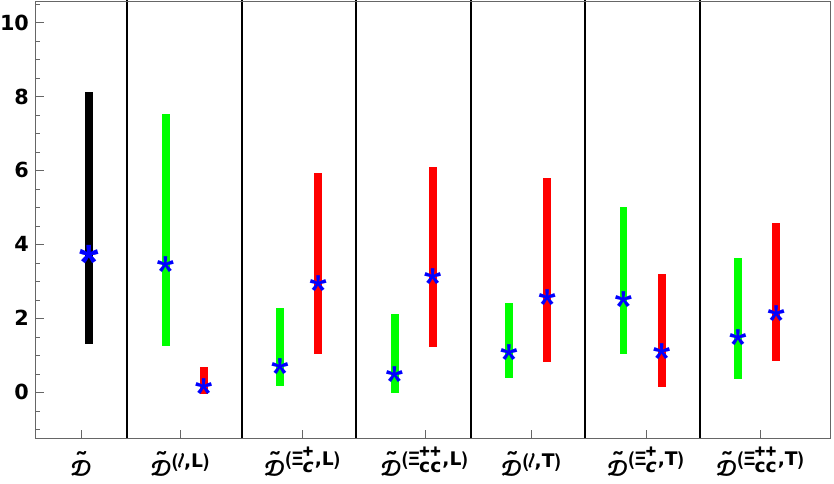}
        \caption{$\tilde{\mathcal{D}}_k^{(m,n)}$ for $q^2=[0.8$-$1.346]$ GeV$^2$}
        \label{bar_g}
    \end{subfigure}
    \begin{subfigure}[b]{0.49\textwidth}
        \centering
        \includegraphics[width=\textwidth]{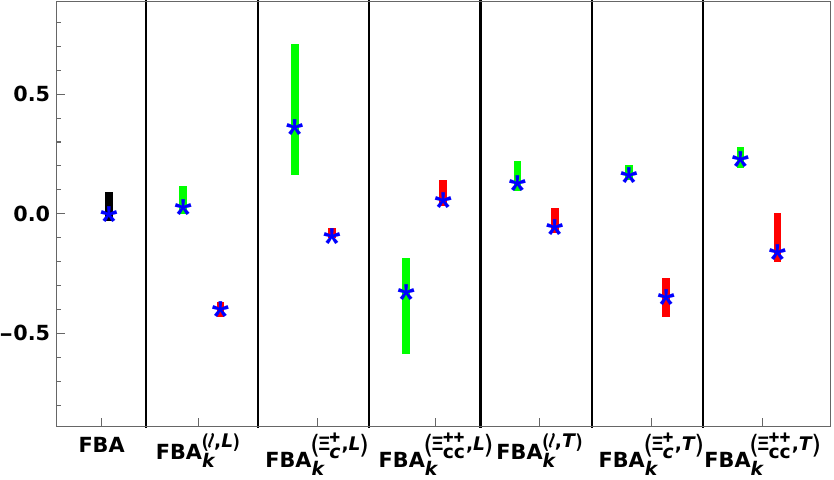}
        \caption{$\text{FBA}_k^{(m,n)}$ for $q^2=[0.8$-$1.346]$ GeV$^2$}
        \label{bar_h}
    \end{subfigure}
    \caption{Variation in the magnitudes of the branching ratio $\tilde{\mathcal{D}}_k^{(m,n)}$ and the forward–backward asymmetry $\text{FBA}_k^{(m,n)}$ shown in left and right panels, respectively, after integrating over four different $q^2$ bins: [$0.0112$-$1.346$] GeV$^2$ in Figs.~\ref{bar_a} and \ref{bar_b}, [$0.0112$-$0.4$] GeV$^2$ in Figs.~\ref{bar_c} and \ref{bar_d}, [$0.4$-$0.8$] GeV$^2$ in Figs.~\ref{bar_e} and \ref{bar_f}, and [$0.8$-$1.346$] GeV$^2$ in Figs.~\ref{bar_g} and \ref{bar_h}. The blue asterisk `$\ast{}$' marks the central value of the branching ratio or the forward-backward asymmetry in the respective bins.}
    \label{barplots}
\end{figure}

The variations due to form factor uncertainties in the binned branching ratios and forward-backward asymmetry for each case are respectively shown in the left and right panels of Fig.~\ref{barplots}. The available $q^2$ range in Fig.~\ref{barplots} is divided into four bins: low $q^2\in [0.0112,0.4]$ GeV$^2$ (row 2), mid $q^2\in[0.4,0.8]$ GeV$^2$ (row 3), high $q^2\in[0.8,1.346]$ GeV$^2$ (row 4), with the fourth bin corresponding to the whole dynamical range $q^2\in[0.0112,1.346]$ GeV$^2$ (row 1). {For each polarization state ($k = 0, +, -$), the differential branching ratio $\mathcal{D}_k^{(m,n)}(q^2)$ is integrated over each bin via $\int_{\text{bin}} dq^2\, \mathcal{D}_k^{(m,n)}(q^2)/\int_{\text{bin}} dq^2$, whereas the forward-backward asymmetry FBA$_k^{(m,n)}(q^2)$ is integrated over each bin via $\int_{\text{bin}} dq^2 (\mathcal{N}^{F(m,n)}_k-\mathcal{N}^{B(m,n)}_k)/\int_{\text{bin}} dq^2 (\mathcal{N}^{F(m,n)}_k+\mathcal{N}^{B(m,n)}_k)$.} The error bars for each bin are colored-coded the same way as earlier stated, with black, green, and red, corresponding to the bin averaged values of $\mathcal{D}^{(m,n)}(q^2)$, $\mathcal{D}_+^{(m,n)}(q^2)$, and $\mathcal{D}_-^{(m,n)}(q^2)$ $\Big($FBA$^{(m,n)}(q^2)$, FBA$_+^{(m,n)}(q^2)$, and FBA$_-^{(m,n)}(q^2)$$\Big)$, respectively. The blue asterisk in each graph is the central value of the respective observable. The values for both branching ratios and forward-backward asymmetry for each bin are mentioned in Table~\ref{BRbarplotsdatasetfull},~\ref{BRbarplotsdataset1},~\ref{BRbarplotsdataset2}, and~\ref{BRbarplotsdataset3} in Appendix~\ref{Tables}.

Generally, the branching ratios $\mathcal{D}_k^{(m,n)}$ show a larger magnitude variation in the mid $q^2$ [0.4-0.8] GeV$^2$ (row 3) and high $q^2$ [0.8-1.346] GeV$^2$ (row 4) bins, shown in Figs.~\ref{bar_e} and~\ref{bar_g} respectively, than in the low $q^2$ [0.0112-0.4] GeV$^2$ bin, shown in Fig.~\ref{bar_c}. The exception to this behavior is the negatively polarized muon $\ell$ in the longitudinal direction, where the variation across all bins is roughly the same.

The effect of the form factor uncertainties on the forward-backward asymmetry $\text{FBA}_k^{(m,n)}$ for each $m, \ n$ case is less apparent compared to that on their respective branching ratios, as shown on the right panels of Fig.~\ref{barplots}. {However, that is not the case for the positively polarized daughter $\Xi_c^+$ and the parent $\Xi_{cc}^{++}$ baryon in the longitudinal direction, where the error bars are significant in each bin.} 

{While the central value of the forward-backward asymmetry for most cases lie within their variation from the chosen combination of form factor uncertainties, the one for the positively polarized parent baryon $\Xi_{cc}^{++}$ in the transverse direction is beyond the variation shown in Fig.~\ref{bar_b}. This is because the `upper' and `lower' limits of all the FBA$_k$ plots just correspond to the results obtained by considering the upper and lower form factor values in Table~\ref{FF}. This only gives an idea of the order of magnitude of the form factor uncertainty, as there are interference effects between different form factors in the amplitude.} As a result, the true variation in the branching ratio, as well as the FBA$_k$ as a result of form factor uncertainties is expected to be larger than the ones shown in Fig.~\ref{barplots}.

\begin{figure}[H]
    \centering
    \includegraphics[width=0.45\linewidth]{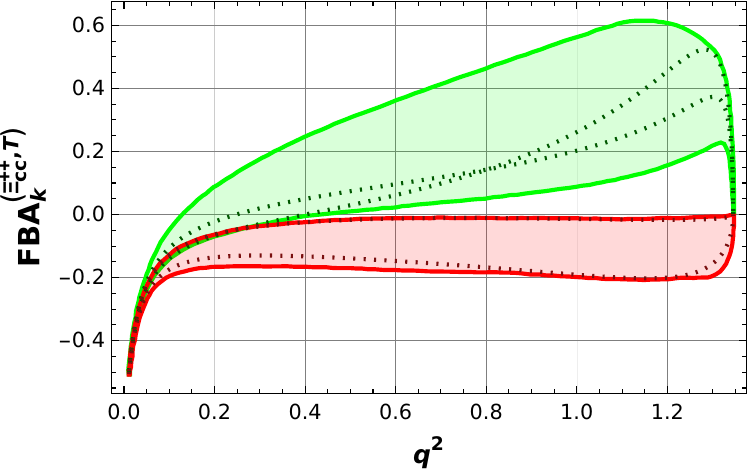}
    \caption{A detailed analysis of the form factor uncertainty of the forward-backward asymmetry FBA$_k^{m,n}(q^2)$, integrated over $\cos{\theta}$, of the transversely ($T$) polarized parent baryon $\Xi_{cc}^{++}$ as a function of $q^2$, obtained by varying form factor uncertainties within their $\pm 1\sigma$ allowed regions. The green envelope indicates the total variation by the positive polarization, whereas the red envelope indicates that by the negative one. The dotted dark green and dark red lines correspond to the uncertainty reported in Fig.~\ref{afb_trans_parent}.}
    \label{afb_trans_parent_64}
\end{figure}

{In order to analyze this properly, we did an exhaustive computation of the forward-backward asymmetry due to all combinations of uncertainties caused by form factors for one case. This case was that of the transversely polarized parent baryon $\Xi_{cc}^{++}$, and the results are shown in Fig.~\ref{afb_trans_parent_64}.} In Fig.~\ref{afb_trans_parent_64}, the true uncertainty 
envelope for the forward backward asymmetry due to form factors is shown as shaded green (for the $k=+$ transverse polarized parent baryon) and red (for the $k=-$ transverse polarized parent baryon) regions. We also show, as dotted lines, the uncertainty regions that were obtained for the same observable in Fig.~\ref{afb_trans_parent}. This analysis confirms the interference effects in the amplitude and show that the expected picture emerges once all uncertainties are accounted for correctly.

\subsection{Lepton-Flavor Universality}

\begin{figure}[H]
    \centering
    \includegraphics[width=0.5\linewidth]{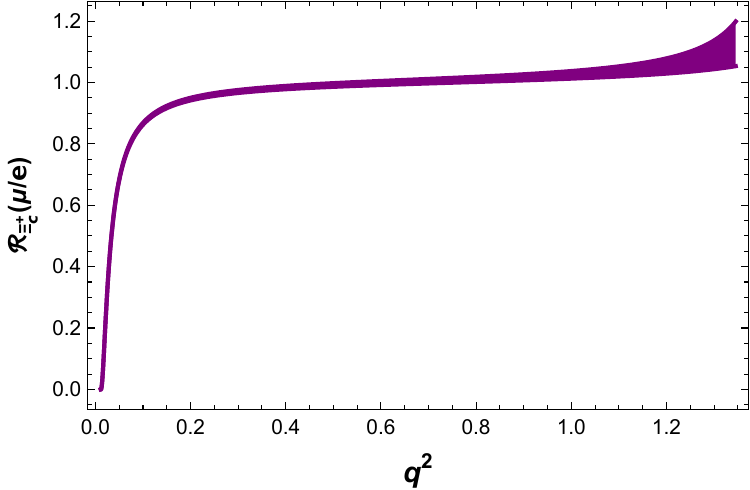}
    \caption{The lepton flavor universality ratio $\mathcal{R}_{\Xi_c^+}(\mu/e)$ as a function of $q^{2}$.}
    \label{unpol_mu_e}
\end{figure}

The SM is based on a lepton-flavor universal structure of the gauge interactions, and recent measurements of lepton-flavor universality in $\mathcal{R}_{K^*}$ are consistent with SM predictions. However, anomalies have been reported in the $\mathcal{R}_{D^{(*)}}$~\cite{Belle-II:2025yjp}. At any rate, the lepton flavor universality ratios provide a robust test of the SM. We have also calculated the LFU ratio, defined via $\mathcal{R}_{\Xi_c^+}(\mu/e) \equiv \mathcal{D}(\Xi_{cc}^{++}\to \Xi_c^+\mu^+\nu_\mu)/\mathcal{D}(\Xi_{cc}^{++}\to \Xi_c^+e^+\nu_e)$. The result of this LFU ratio across the entire $q^2$ range is shown in Fig.~\ref{unpol_mu_e}. By averaging over the entire $q^2$ region, we obtain the value $\mathcal{R}_{\Xi_c^+}(\mu/e)= 1.00229_{-0.01551}^{+0.00374}$. It is also interesting to note that $\mathcal{R}_{\Xi_c^+}(\mu/e)$ has the values $0.942998_{-0.008187}^{+0.002094}$, $1.00224_{{-0.01070}}^{+0.00296}$, $1.03615_{-0.02149}^{+0.00710}$ in the low, mid and high $q^2$ regions, respectively.

\subsection{Polarization Asymmetry Analysis}

We now focus our attention to an analysis of the polarization asymmetry in the decay $\Xi_{cc}^{++} \rightarrow \Xi_{c}^{+} \bar{\ell}\nu_{\ell}$. We describe the polarization asymmetry $P^{(m,n)}$, the polarization ratio $\mathcal{R}_{nn'}^{(m,m')}[j]$, and the polarization asymmetry ratio $\mathcal{R}_{P_{nn'}^{(m,m')}}$ as defined in Eq.~\ref{eqn.pol_asymm}, Eq.~\ref{ratio_equation}, and Eq.~\ref{pol_asymm_ratio_equation}, respectively.

\subsubsection{Polarization Asymmetry $P^{(m,n)}$}

\begin{figure}[H]
    \centering
    \begin{subfigure}[b]{0.32\textwidth}
        \centering
        \includegraphics[width=\textwidth]{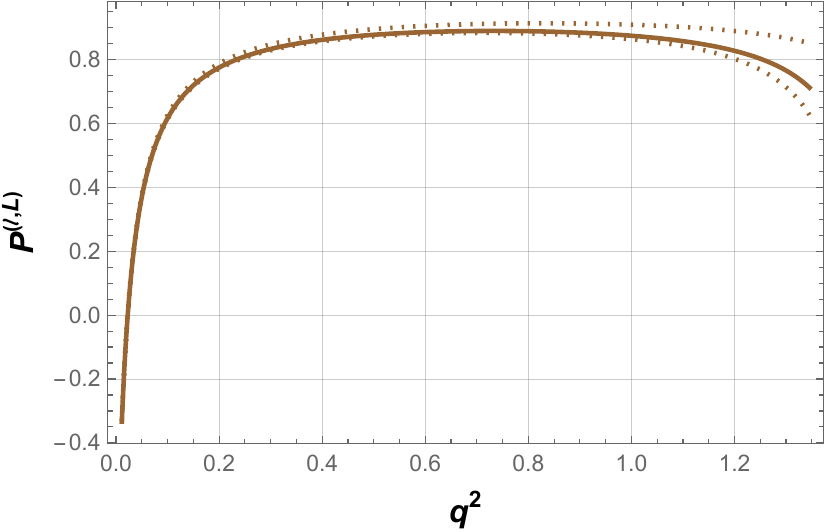}
        \caption{}
        \label{lepton_longi_asymm}
    \end{subfigure}
     \begin{subfigure}[b]{0.32\textwidth}
        \centering
        \includegraphics[width=\textwidth]{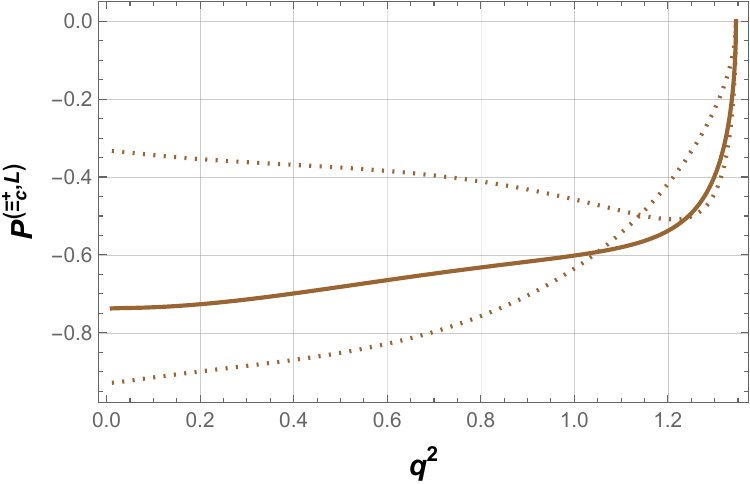}
        \caption{}
        \label{Ec_longi_asymm}
    \end{subfigure}
    \begin{subfigure}[b]{0.32\textwidth}
        \centering
        \includegraphics[width=\textwidth]{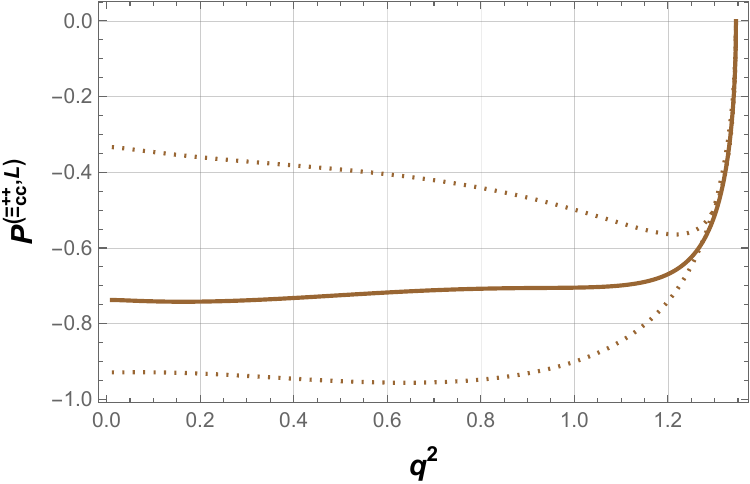}
        \caption{}
        \label{Ecc_longi_asymm}
    \end{subfigure}
    \caption{SM predictions for polarization asymmetries $P^{(m,n)}$ for particles $m$ ($m\in\{\ell,\Xi_{cc}^{++},\Xi_{c}^{+}\}$) in the longitudinal ($n=L$) direction.}
    \label{polatization_longi_asymm}
\end{figure}

\begin{figure}[H]
    \centering
    \begin{subfigure}[b]{0.32\textwidth}
        \centering
        \includegraphics[width=\textwidth]{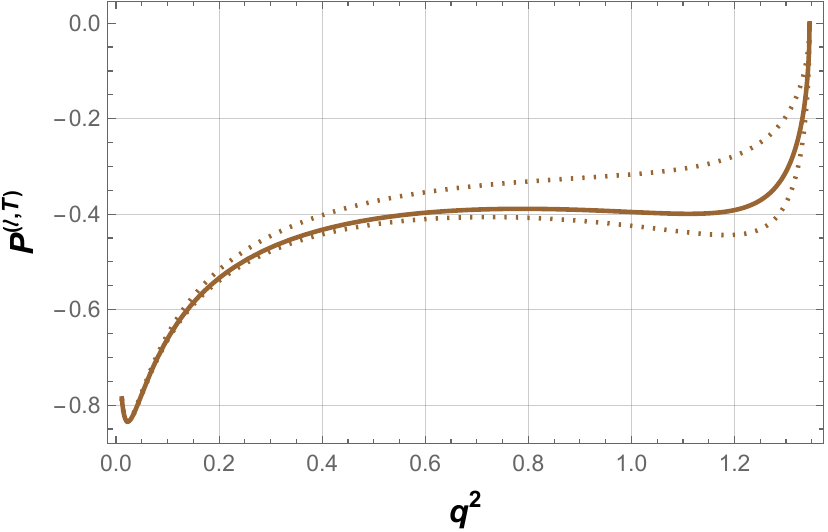}
        \caption{}
        \label{lepton_trans_asymm}
    \end{subfigure}
    \begin{subfigure}[b]{0.32\textwidth}
        \centering
        \includegraphics[width=\textwidth]{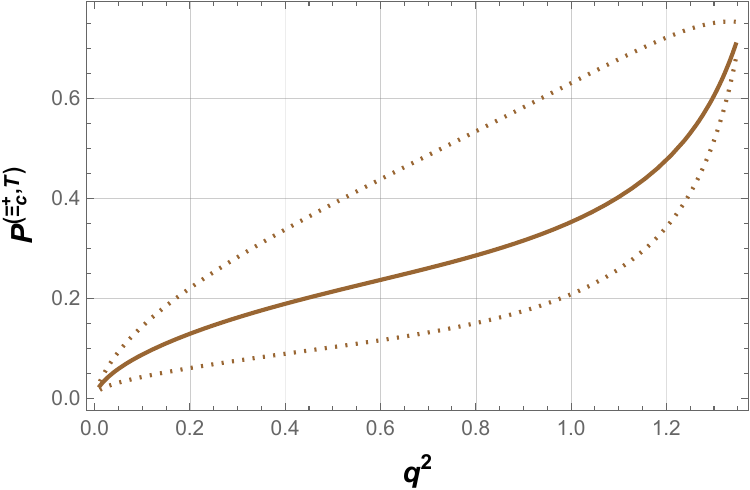}
        \caption{}
        \label{Ec_trans_asymm}
    \end{subfigure}
    \begin{subfigure}[b]{0.32\textwidth}
        \centering
        \includegraphics[width=\textwidth]{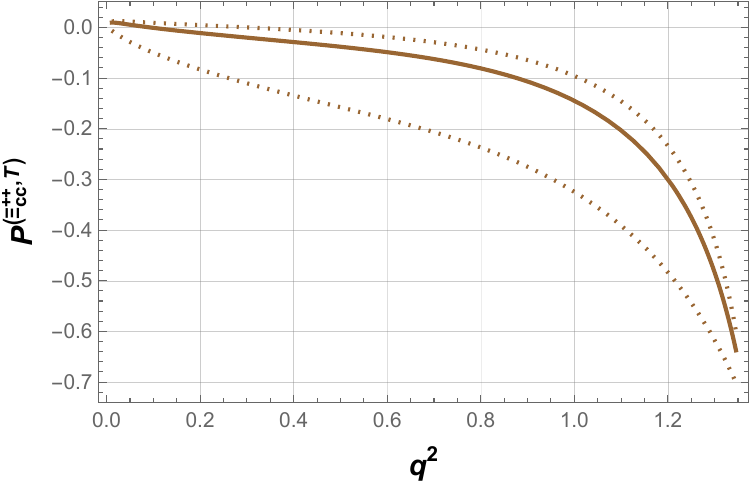}
        \caption{}
        \label{Ecc_trans_asymm}
    \end{subfigure}
    \caption{SM predictions for polarization asymmetries $P^{(m,n)}$ for particles $m$ ($m\in\{\ell,\Xi_{cc}^{++},\Xi_{c}^{+}\}$) in the transverse ($n=T$) direction.}
    \label{polatization_trans_asymm}
\end{figure}

In this section, we analyze the spin polarization asymmetries $P^{(m,n)}$ by plotting those as functions of $q^2$ in Figs.~\ref{polatization_longi_asymm} and~\ref{polatization_trans_asymm}. The $P^{(m,n)}$ quantifies the differences in the branching ratios (after integrating over $\cos\theta$) of the positive ($k=+$) and negative ($k=-$) polarization states of the particle $m$ ($m\in\{\ell,\Xi_{cc}^{++},\Xi_{c}^{+}\}$) in the $n$ ($n\in\{L,T\}$) direction. The brown curve represents the central value of $P^{(m,n)}$, while the dotted curves indicate variation due to uncertainties in the form factors. 

In Fig.~\ref{polatization_longi_asymm}, we present the spin polarization asymmetries $P^{(m,L)}$ for the longitudinally polarized muon $\ell$, daughter baryon $\Xi_c^+$, and parent baryon $\Xi_{cc}^{++}$. For the muon, Fig.~\ref{lepton_longi_asymm} shows that the observable $P^{(\ell,L)}$ starts from a negative value at very low $q^2$, rises rapidly toward positive values, and quickly becomes relatively flat. It has a maximum of $\sim 0.9$ in the mid $q^2$ region, followed by a slight decline at high $q^2$. Similarly, the observable $P^{(\Xi_c^+,L)}$ for the longitudinally polarized daughter baryon in Fig.~\ref{Ec_longi_asymm} starts negative and gradually increases towards the high $q^2$ region, after which it rapidly increases towards zero. It also has a large uncertainty deviation in the low $q^2$ region, which decreases towards the higher $q^2$ values. {The crossing of the central and indicated uncertainty plots in Fig.~\ref{Ec_longi_asymm} reminds us of the interference between the various form factor terms in the amplitude. Accordingly, as explained earlier, a more thorough analysis of form factor uncertainties is warranted with the uncertainty region indicated by our analysis. Similar behaviors for both the central and uncertainty plots are also seen for the parent baryon in the longitudinal direction in Fig.~\ref{Ecc_longi_asymm}. The only minor difference is that the central $P^{(\Xi_{cc}^{++},L)}$ plot is more or less constant for the most part and only rapidly increases towards zero at the tail of the distribution ($q^2\gtrsim 1.3 \textrm{GeV}^2$).}

In Fig.~\ref{polatization_trans_asymm}, we present the polarization asymmetries $P^{(m,T)}$ of the transversely polarized muon $\ell$, daughter baryon $\Xi_c^+$, and parent baryon $\Xi_{cc}^{++}$ as a function of $q^2$. For the muon $\ell$, $P^{(\ell,T)}$ in Fig.~\ref{lepton_trans_asymm} remains negative throughout the $q^2$ range, starting near {$-0.8$} and gradually increasing towards zero at higher $q^2$. Its uncertainty deviates slightly from the central plot in the mid-to-high $q^2$ region, after which this gap closes in at high $q^2$. Furthermore, for both the daughter and the parent baryons as shown in Figs.~\ref{Ec_trans_asymm} and~\ref{Ecc_trans_asymm}, their respective observables  $P^{(\Xi_c^+,T)}$ and $P^{(\Xi_{cc}^{++},T)}$ start from zero. However, the former remains positive and monotonically increases throughout the dynamical $q^2$ range, whereas the latter is negative and monotonically decreases throughout the same range. Both their uncertainty plots deviate the most in the mid $q^2$ region, with this deviation only slightly decreasing in the high $q^2$ region.

We present the averaged values of the $P^{(m,L)}$ and $P^{(m,T)}$ over the $q^2$ bins: $[0.0112$-$0.4]$ GeV$^2$, $[0.4$-$0.8]$ GeV$^2$, $[0.8$-$0.1346]$ GeV$^2$, and $[0.0112$-$0.1346]$ GeV$^2$ in Table~\ref{Table_asymmetries} of Appendix~\ref{Tables}.

\subsubsection{Polarization Ratios $\mathcal{R}_{nn'}^{(m,m')}[j]$}

In order to further analyze the process, we now discuss the polarization ratios $\mathcal{R}_{nn'}^{(m,m')}[j]$. {As described earlier, the ratio $\mathcal{R}^{(m,m')}_{nn'}[j]$ is a convenient measure for comparing the differential branching ratio of $m$-type particles polarized with helicity $j$ in the $n$ direction with that of the $m'$-type particles polarized with helicity $j$ in the $n'$ direction.}

\begin{figure}[H]
    \centering
    \begin{subfigure}{0.32\textwidth}\includegraphics[width=1.9in,height=1.25in]{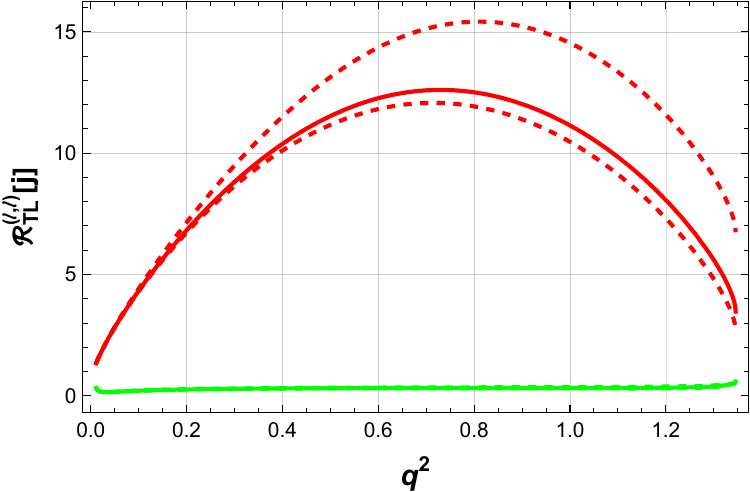}\caption{}\label{fig:Rl}
    \end{subfigure}
    \begin{subfigure}{0.32\textwidth}\includegraphics[width=1.9in,height=1.25in]{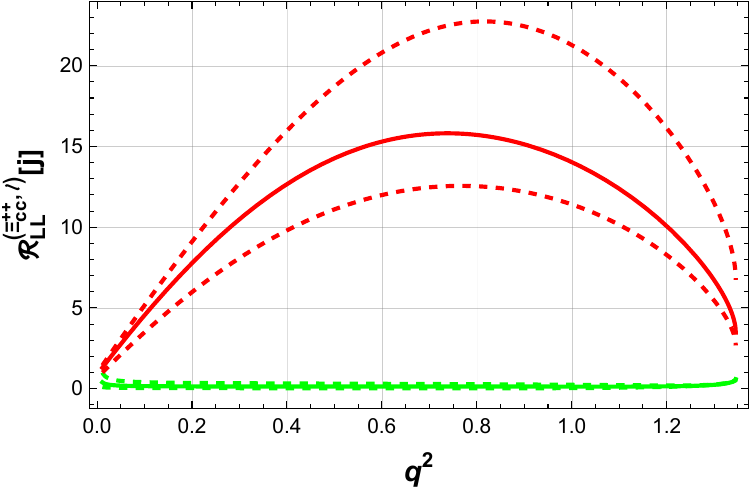}\caption{}\label{fig:REccl_LL}\end{subfigure}
    \begin{subfigure}{0.32\textwidth}\includegraphics[width=1.9in,height=1.25in]{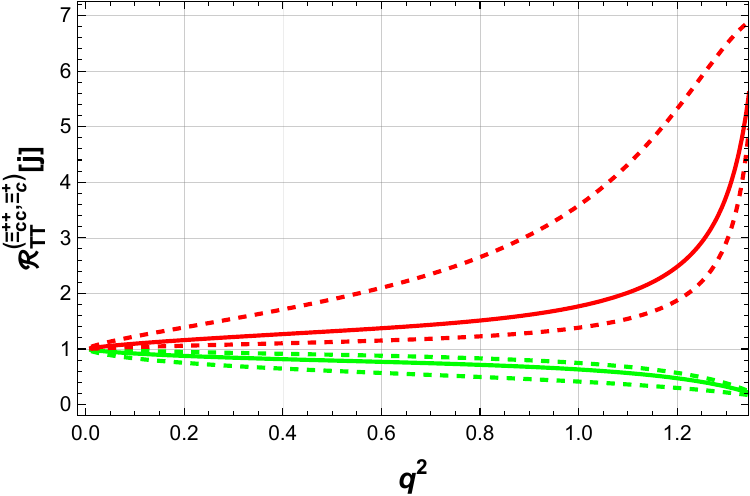}\caption{}\label{fig:REccEc_TT}\end{subfigure}
    \begin{subfigure}{0.32\textwidth} \includegraphics[width=1.9in,height=1.25in]{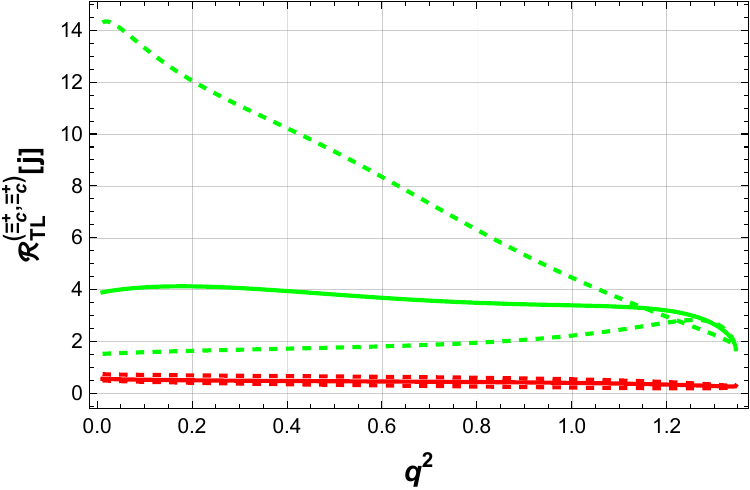}\caption{}\label{fig:REc}\end{subfigure}
    \begin{subfigure}{0.32\textwidth}\includegraphics[width=1.9in,height=1.25in]{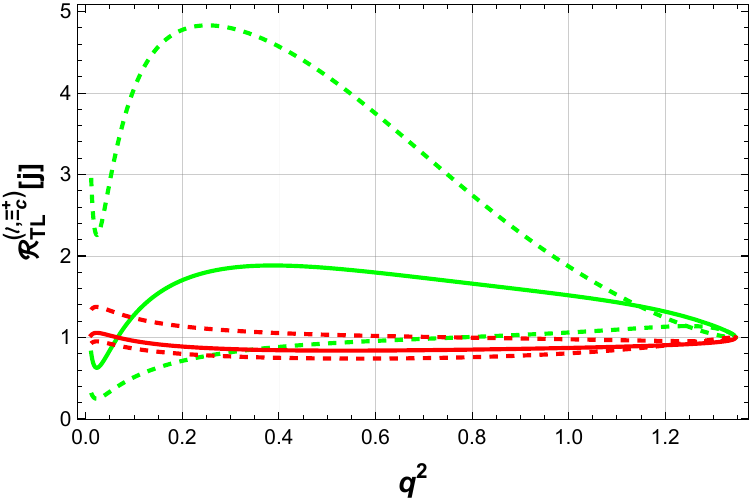}\caption{}\label{fig:RlEc}\end{subfigure}
    \begin{subfigure}{0.32\textwidth}\includegraphics[width=1.9in,height=1.25in]{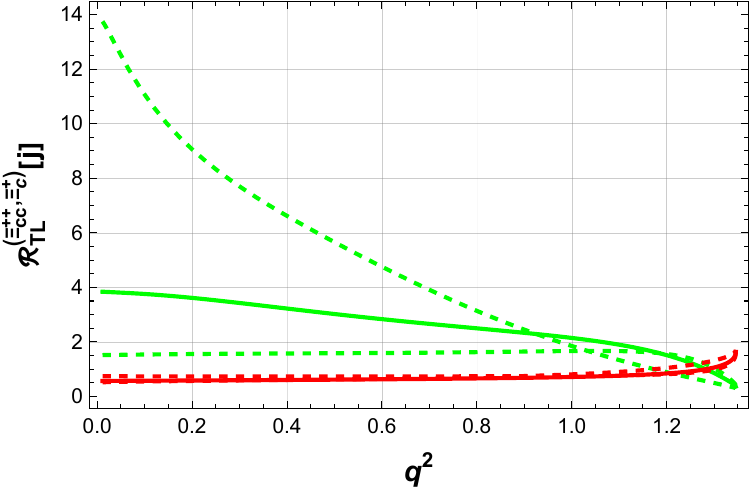}\caption{}\label{fig:REccEc}\end{subfigure}
    \begin{subfigure}{0.32\textwidth}\includegraphics[width=1.9in,height=1.25in]{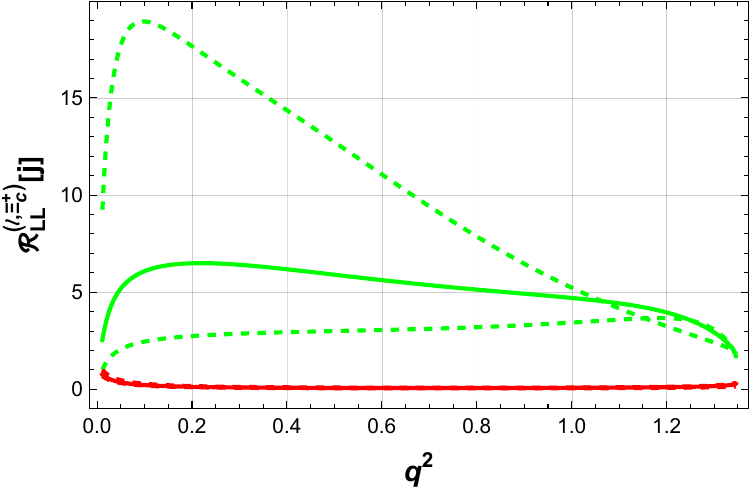}\caption{}\label{fig:RlEc_LL}\end{subfigure}
    \begin{subfigure}{0.32\textwidth}\includegraphics[width=1.9in,height=1.25in]{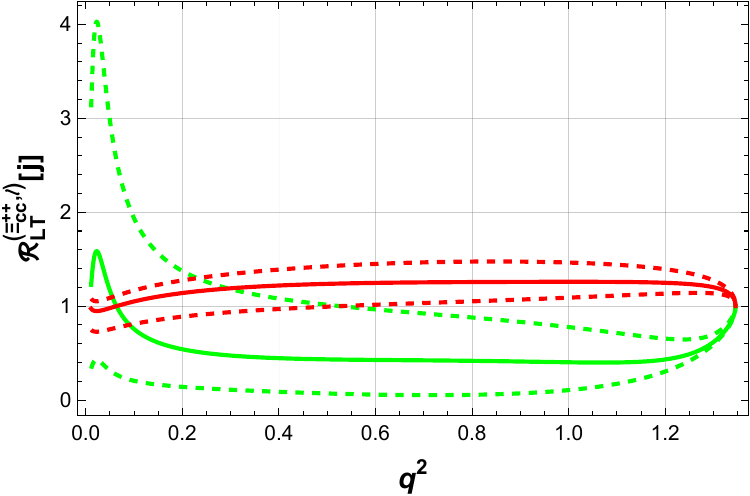}\caption{}\label{fig:RlEcc}\end{subfigure}
    \caption{SM predictions for polarization ratios $\mathcal{R}_{nn'}^{(m,m')}[j]$ for different combinations of longitudinal and transverse polarizations as a function of $q^{2}$, where there is a significant uncertainty in one of the polarization states.}
    \label{Ratios1}
\end{figure}

\begin{figure}[h!]
    \centering
    \begin{subfigure}{0.32\textwidth}\includegraphics[width=1.9in,height=1.25in]{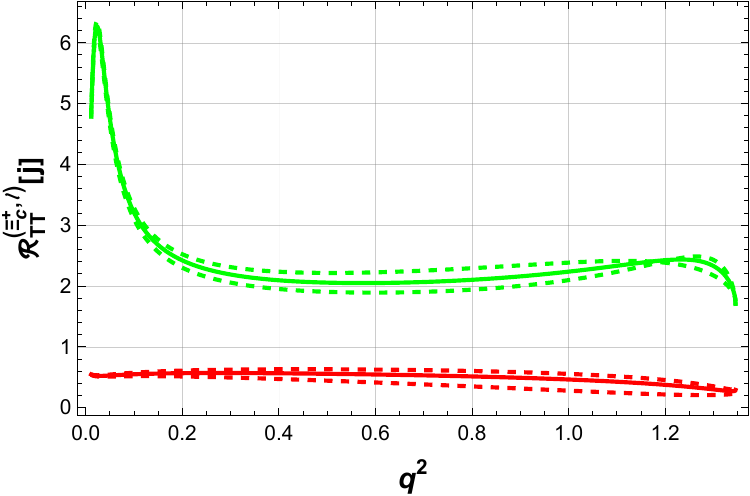}\caption{}\label{fig:REcl_TT}\end{subfigure}
    \begin{subfigure}{0.32\textwidth}\includegraphics[width=1.9in,height=1.25in]{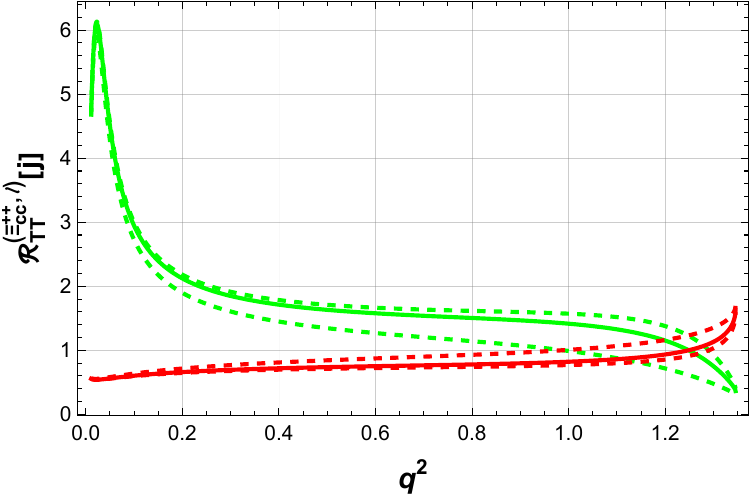}\caption{}\label{fig:REccl_TT}\end{subfigure}
    \begin{subfigure}{0.32\textwidth}\includegraphics[width=1.9in,height=1.25in]{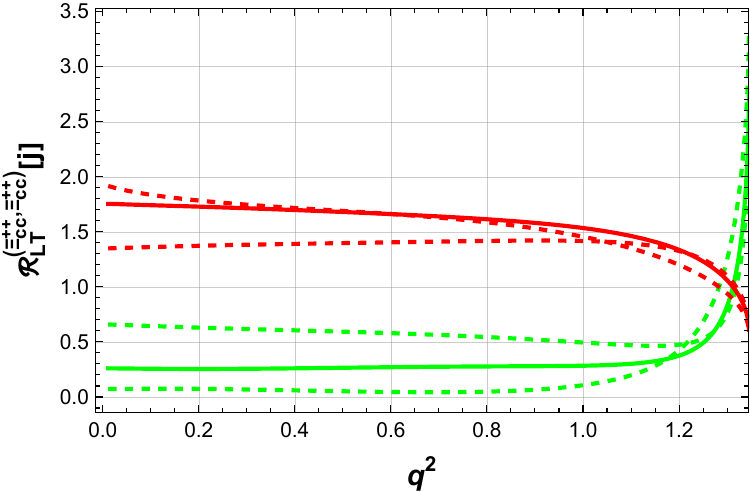}\caption{}\label{fig:REcc}\end{subfigure}
    \begin{subfigure}{0.32\textwidth}\includegraphics[width=1.9in,height=1.25in]{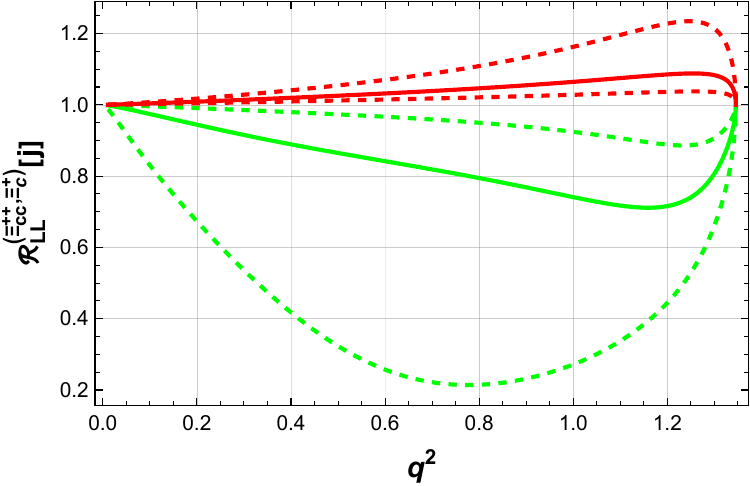}\caption{}\label{fig:REccEc_LL}\end{subfigure}
    \caption{SM predictions for polarization ratios $\mathcal{R}_{nn'}^{(m,m')}[j]$ for different combinations of longitudinal and transverse polarizations as a function of $q^{2}$, where there is minimal uncertainty for both polarization states.}
    \label{Ratios2}
\end{figure}

We present the polarization ratios $\mathcal{R}_{nn'}^{(m,m')}[j]$ as a function of $q^2$  for different combinations of longitudinal and transverse polarizations in Figs.~\ref{Ratios1} and~\ref{Ratios2}. Additionally, the results for various combinations of $\mathcal{R}_{nn'}^{(m,m')}[j]$ over the full dynamical $q^2$ range are provided in Table~\ref{TableBRbarplotsratio} of Appendix~\ref{Tables}.

In the ratios shown in Fig.~\ref{Ratios1}, we observe that typically one helicity state has a larger value for the polarization ratio $\mathcal{R}_{nn'}^{(m,m')}[j]$. We can also see different combinations will have varying sensitivity to form factor uncertainties. Noticeably, in each ratio, one polarization case has more variation from their central plot than the other, where the latter is generated using the central values of the form factors. {Figs.~\ref{fig:Rl},~\ref{fig:REccl_LL}, and~\ref{fig:REccEc_TT} have significantly more such deviation in the negative (red) polarization state than in the positive (green) one, whereas in Figs.~\ref{fig:REc},~\ref{fig:RlEc}, ~\ref{fig:REccEc}, ~\ref{fig:RlEc_LL}, and~\ref{fig:RlEcc} more uncertainty is manifest in the positive (green) polarization state.}

The ratios in Fig.~\ref{Ratios2} highlight several ``clean'' observables, as both polarization cases have little deviation from the central plot. This suggests that polarization, which can be experimentally reconstructed from decay products in future experiments, may provide valuable insights into physics beyond the SM.

\begin{figure}[h!]
    \centering
    \includegraphics[width=0.65\linewidth]{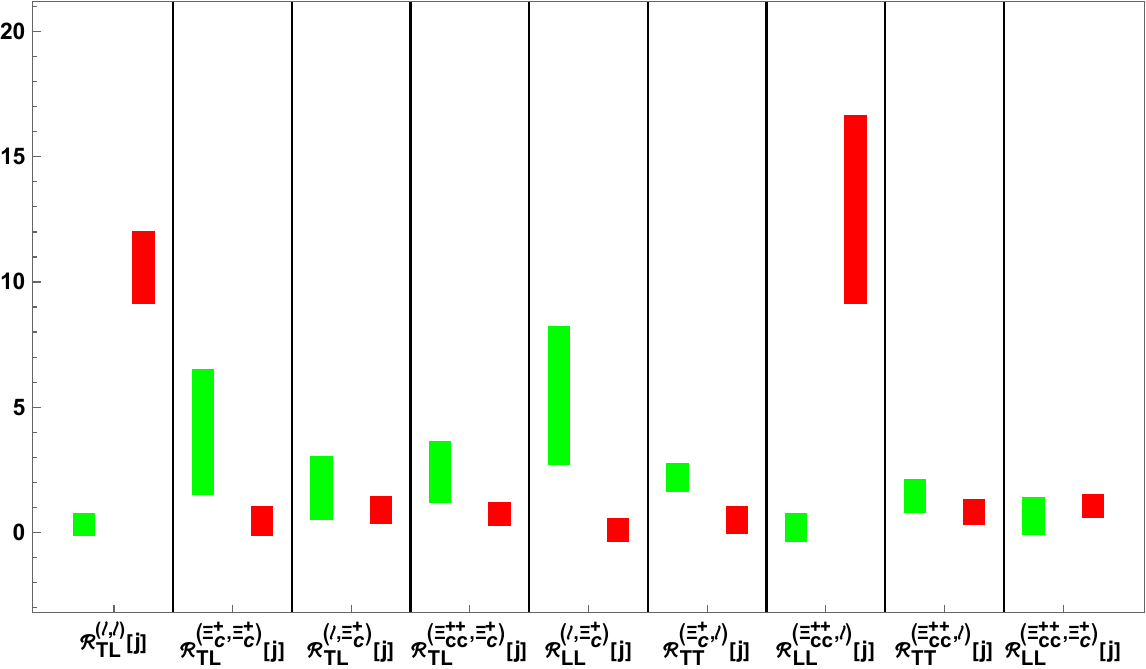}
    \caption{The variation in the magnitudes of the ratio $\mathcal{R}_{nn'}^{(m,m')}[j]$ for different polarization combinations of the particles $m,m'$ ($m,m' \in \{\ell, \Xi_{cc}^{++}, \Xi_{c}^{+}\}$) in their respective direction $n,n'$ ($n,n' \in \{L, T\}$) is shown. The index $j=\pm$ denotes the two polarization states, where the green bars correspond to $j=+$ and the red bars correspond to $j=-$. }
    \label{bar_ratios}
\end{figure}

The variation in the magnitude of these ratios due to form factor uncertainties are shown in Fig.~\ref{bar_ratios}. Generally, we observe a smaller magnitude variation for most cases, except for the postively polarized cases of $\mathcal{R}_{TL}^{(\Xi_{c}^{+},\Xi_{c}^{+})}$, $\mathcal{R}_{TL}^{(\ell,\Xi_{c}^{+})}$, $\mathcal{R}_{TL}^{(\Xi_{cc}^{++},\Xi_{c}^{+})}$, and $\mathcal{R}_{LL}^{(\ell,\Xi_{c}^{+})}$ and the negative polarized cases of $\mathcal{R}_{TL}^{(\ell,\ell)}$ and $\mathcal{R}_{LL}^{(\Xi_{cc}^{++},\ell)}$ where a larger variation is seen.

\subsubsection{Polarization Asymmetry Ratios $\mathcal{R}_{P_{nn'}^{(m,m')}}$}

\begin{figure}[H]
    \centering
    \begin{subfigure}{0.325\textwidth}\includegraphics[width=1.8in,height=1.17in]{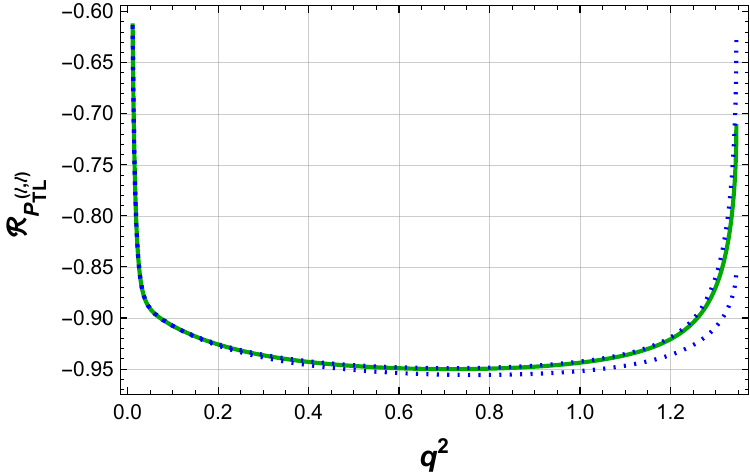}\caption{}\label{fig:PlTL}\end{subfigure}
    \begin{subfigure}{0.325\textwidth}\includegraphics[width=1.8in,height=1.17in]{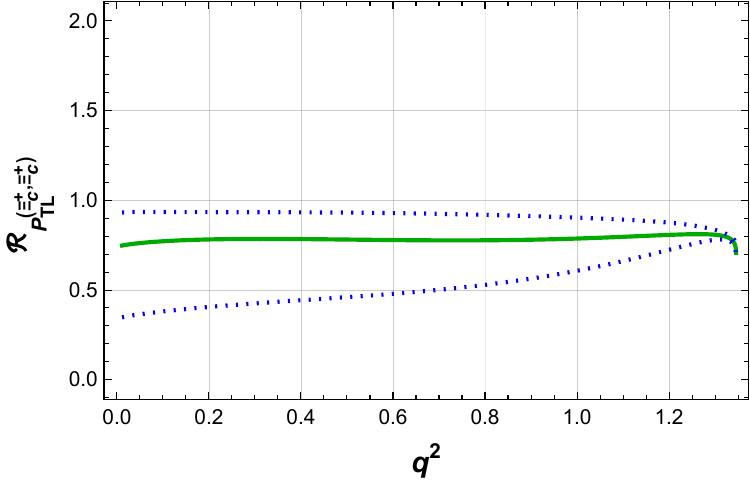}\caption{}\label{fig:PEcTL}\end{subfigure}
    \begin{subfigure}{0.325\textwidth}\includegraphics[width=1.8in,height=1.17in]{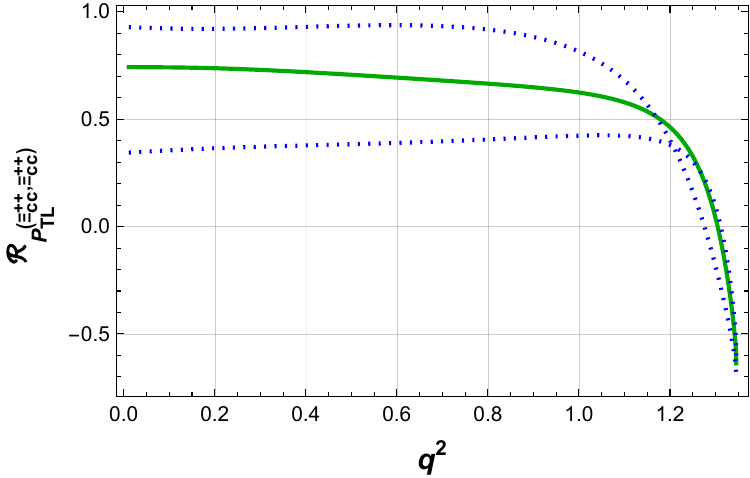}\caption{}\label{fig:PEccTL}\end{subfigure}
    \begin{subfigure}{0.325\textwidth}\includegraphics[width=1.8in,height=1.17in]{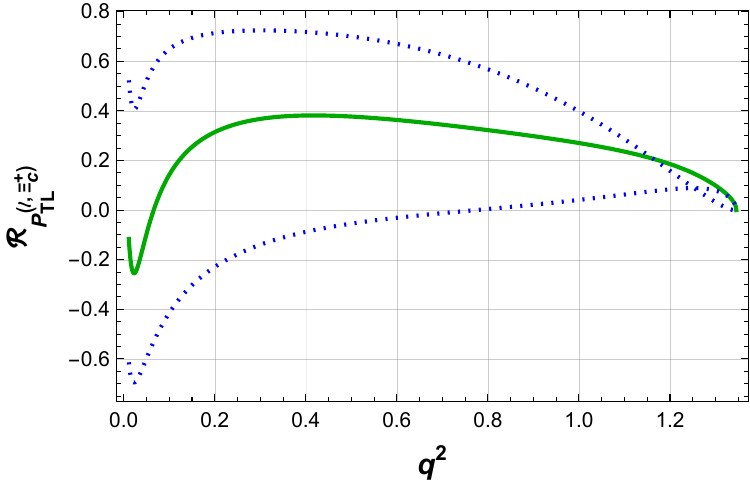}\caption{}\label{fig:PlEcTL}\end{subfigure}
    \begin{subfigure}{0.325\textwidth}\includegraphics[width=1.8in,height=1.17in]{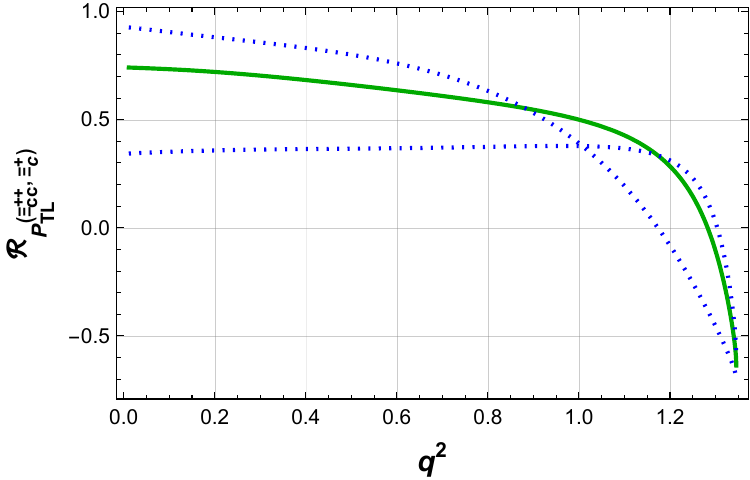}\caption{}\label{fig:PEccEcTL}\end{subfigure}
    \begin{subfigure}{0.325\textwidth}\includegraphics[width=1.8in,height=1.17in]{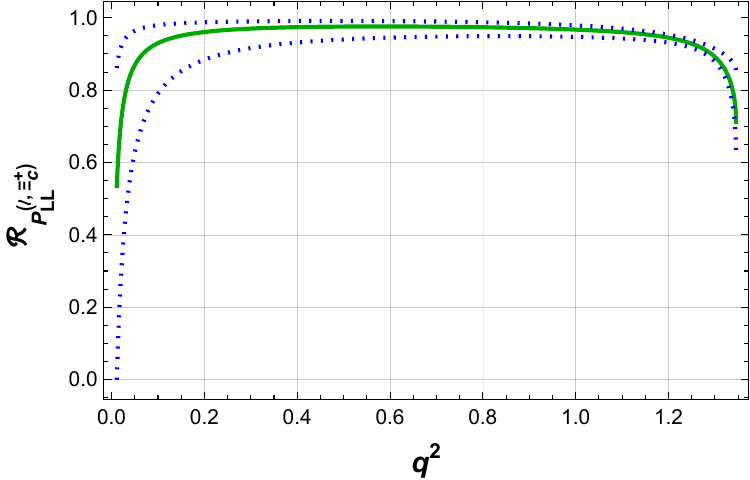}\caption{}\label{fig:PlEcLL}\end{subfigure}
    \begin{subfigure}{0.325\textwidth}\includegraphics[width=1.8in,height=1.17in]{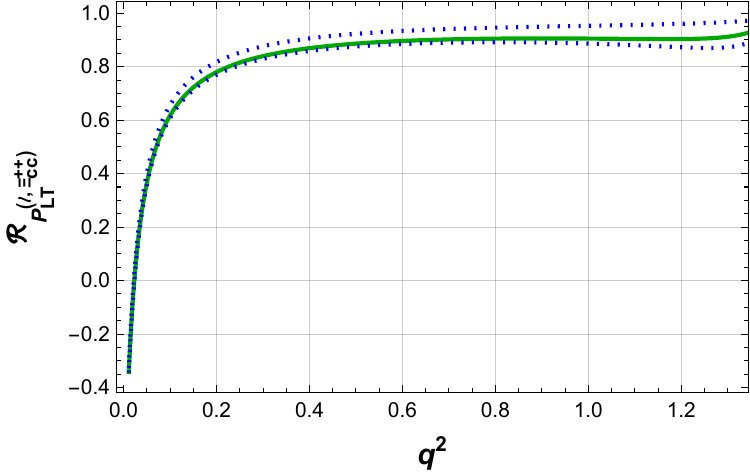}\caption{}\label{fig:PlEccTL}\end{subfigure}
    \begin{subfigure}{0.325\textwidth}\includegraphics[width=1.8in,height=1.17in]{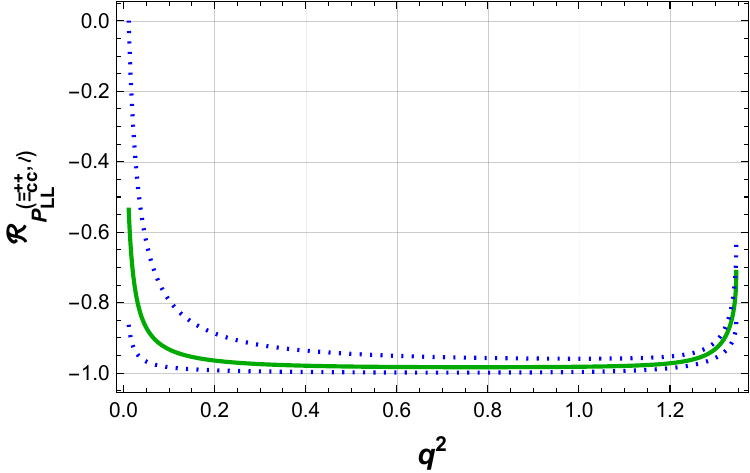}\caption{}\label{fig:PEcclLL}\end{subfigure}
    \begin{subfigure}{0.325\textwidth}\includegraphics[width=1.8in,height=1.17in]{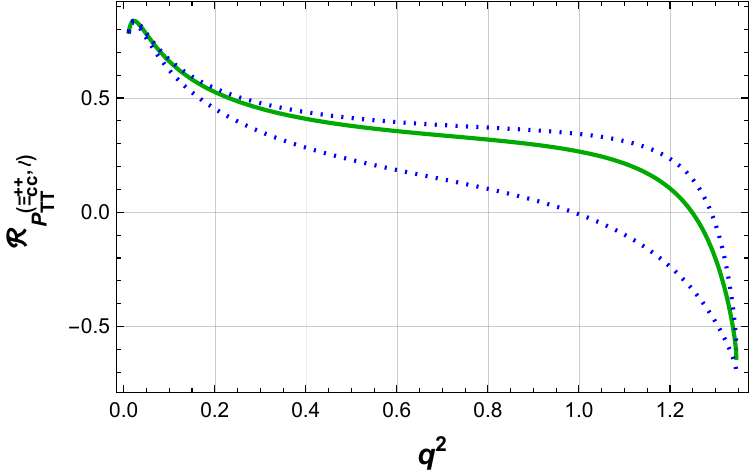}\caption{}\label{fig:PEcclTT}\end{subfigure}
    \caption{SM predictions for polarization asymmetries $\mathcal{R}_{P_{nn'}^{(m,m')}}$ for particles $m,m'$ ($m,m'\in\{\ell,\Xi_{cc}^{++},\Xi_{c}^{+}\}$) in the direction $n,n'$ ($n,n'\in\{L,T\}$).}
    \label{Ratios_asymmetries}
\end{figure}

We present the polarization asymmetry ratios $\mathcal{R}_{P_{nn'}^{(m,m')}}$ as a function of $q^2$ for some of the polarization ratios $\mathcal{R}_{nn'}^{(m,m')}$ found in the previous part in Fig.~\ref{Ratios_asymmetries} as predicted by the SM. These indicate the differences in the positive ($k=+$) and negative ($k=-$) result of a particular polarization ratio. Their expressions are constructed in a way such that the deviations produced by the form factors uncertainties are minimized. The results are also averaged over the $q^2$ bins: $[0.0112$-$0.4]$ GeV$^2$, $[0.4$-$0.8]$ GeV$^2$, $[0.8$-$0.1346]$ GeV$^2$, and $[0.0112$-$0.1346]$ GeV$^2$ and their values along with related uncertainties are mentioned in Table~\ref{Table_asymmetries_ratios} of Appendix~\ref{Tables}.

It can be evidently seen in all the plots in Fig.~\ref{Ratios_asymmetries} that the uncertainty graphs do not deviate too much from their central plots. We can note that plots in Figs.~\ref{fig:PEcTL},~\ref{fig:PEccTL},~\ref{fig:PlEcTL}, and~\ref{fig:PEccEcTL} differ somewhat more than the others in the low to mid $q^2$ range. This is because the plots of their corresponding polarization ratios - $\mathcal{R}_{TL}^{(\Xi_{c}^{+},\Xi_{c}^{+})}$, $\mathcal{R}_{TL}^{(\Xi_{cc}^{++},\Xi_{cc}^{++})}$, $\mathcal{R}_{TL}^{(\ell,\Xi_{c}^{+})}$, and $\mathcal{R}_{TL}^{(\Xi_{cc}^{++},\Xi_{c}^{+})}$ - have a large variation between their respective uncertainty graphs for the positively polarized case in the low to mid $q^2$ range. Nevertheless, the relatively minimal uncertainty in these ratios throughout the $q^2$ range means that these ratios are good observables to test the SM and perhaps 
look for physics beyond the SM. 

\subsection{Correlation Analysis of Polarized Observables}

\begin{figure}[h!]
\centering
\includegraphics[width=1.8in,height=1.25in]{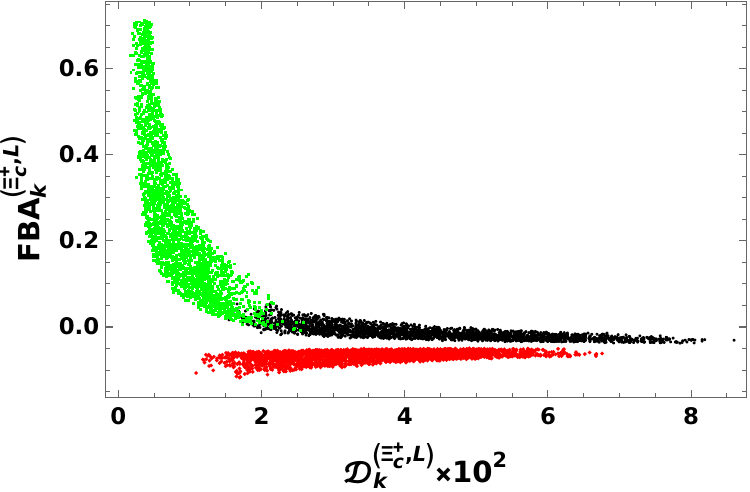}
\includegraphics[width=1.8in,height=1.25in]{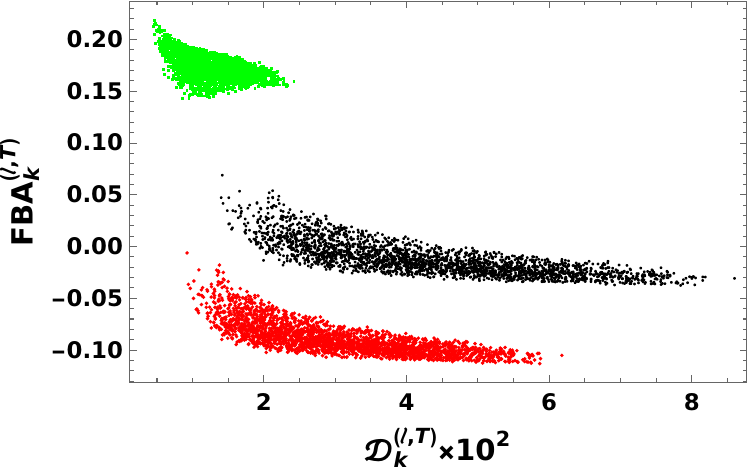}
\includegraphics[width=1.8in,height=1.25in]{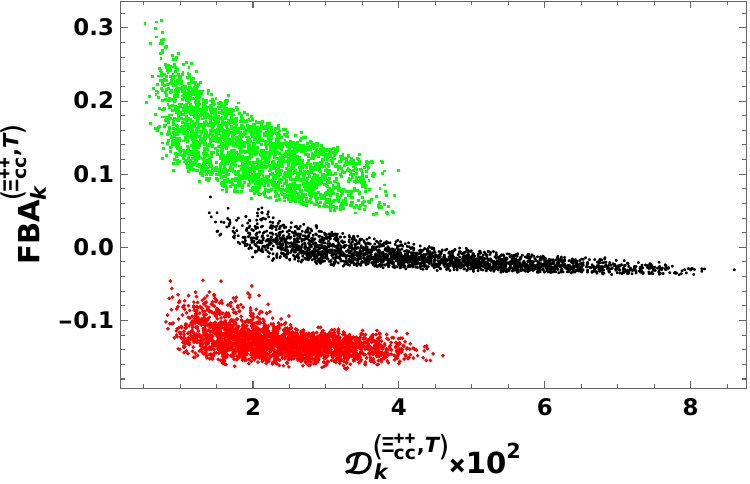}
\includegraphics[width=1.8in,height=1.25in]{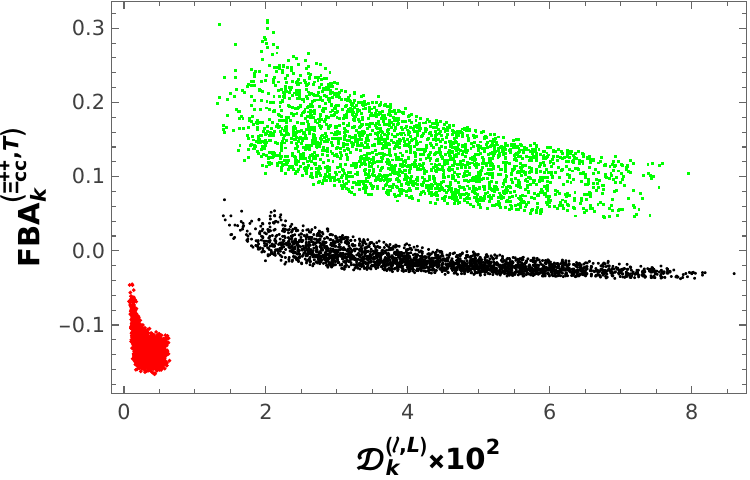}
\includegraphics[width=1.8in,height=1.25in]{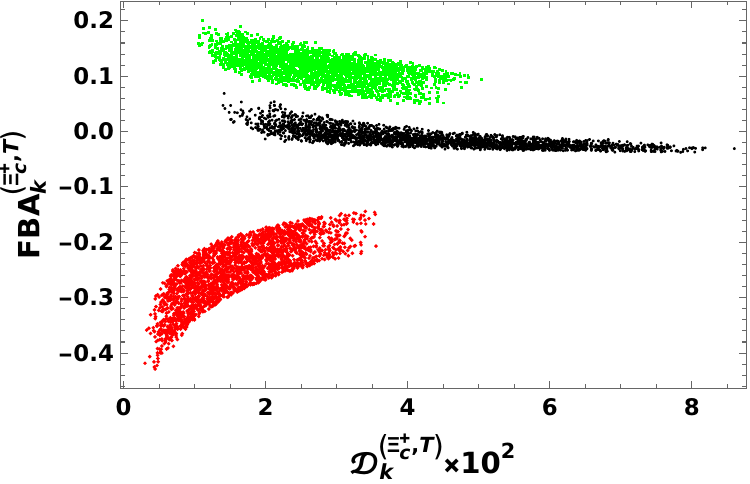}
\includegraphics[width=1.8in,height=1.25in]{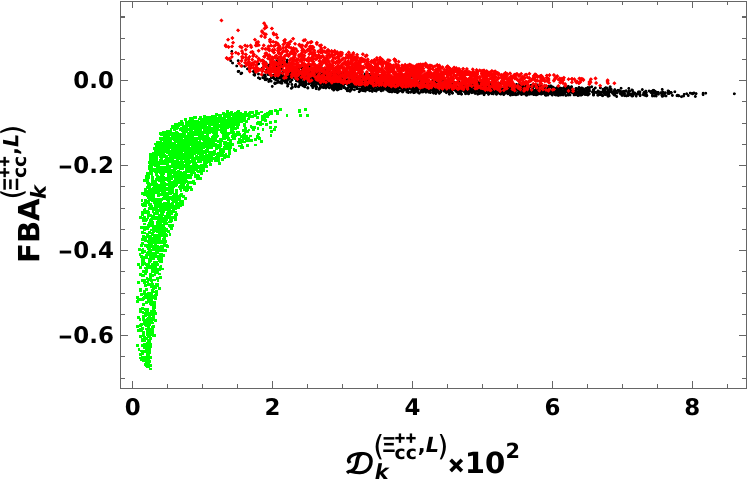}
\includegraphics[width=1.8in,height=1.25in]{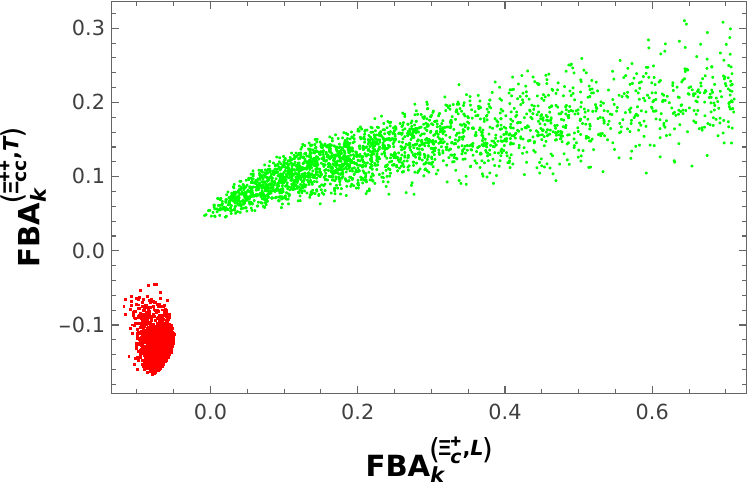}
\includegraphics[width=1.8in,height=1.25in]{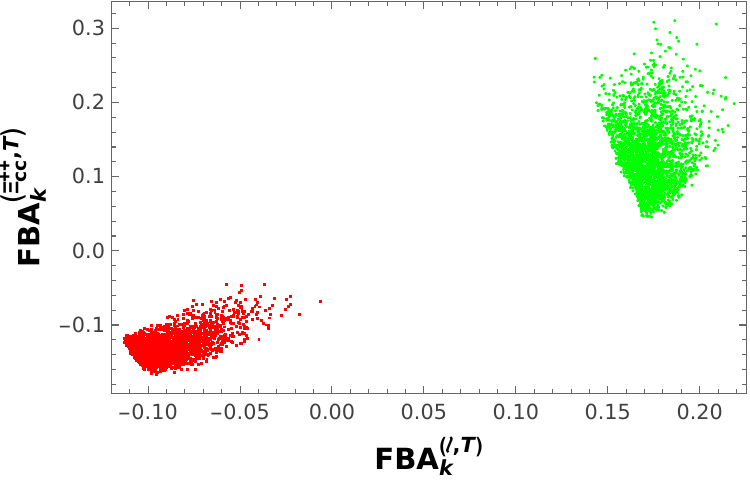}
\includegraphics[width=1.8in,height=1.25in]{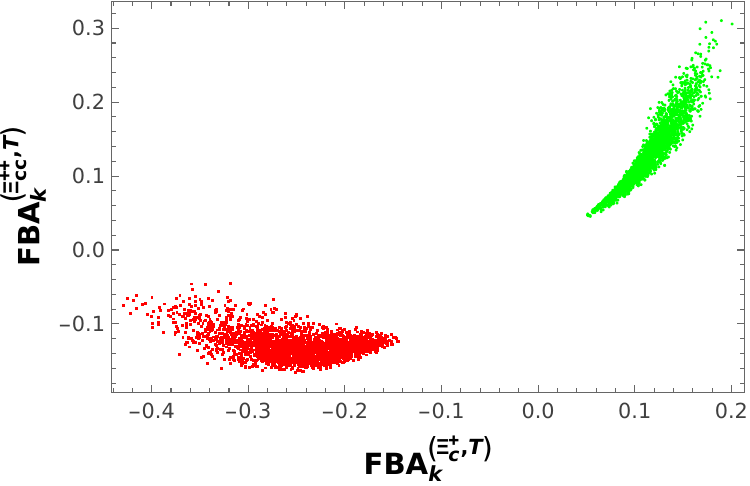}
\includegraphics[width=1.8in,height=1.25in]{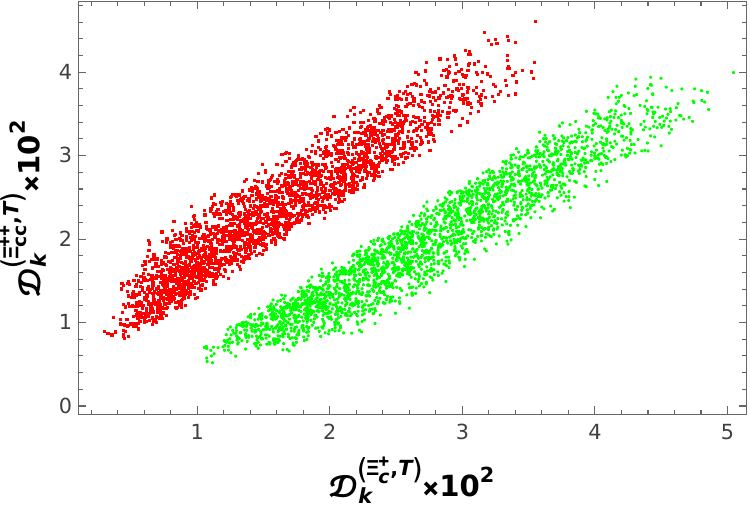}
\includegraphics[width=1.8in,height=1.25in]{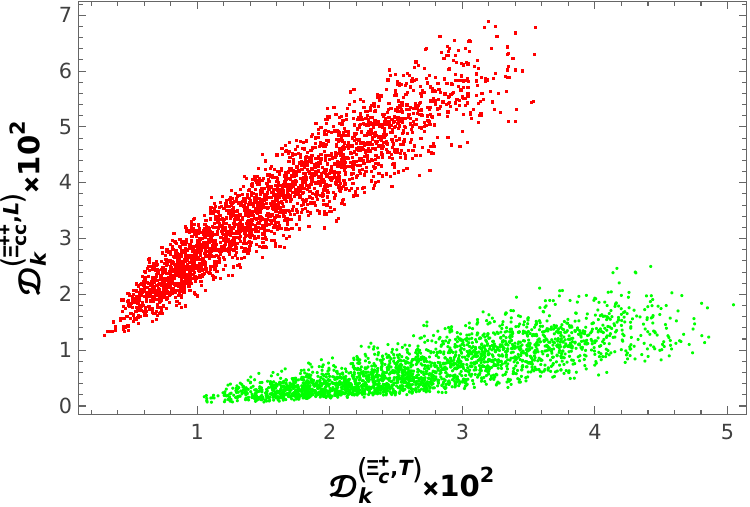}
\includegraphics[width=1.8in,height=1.25in]{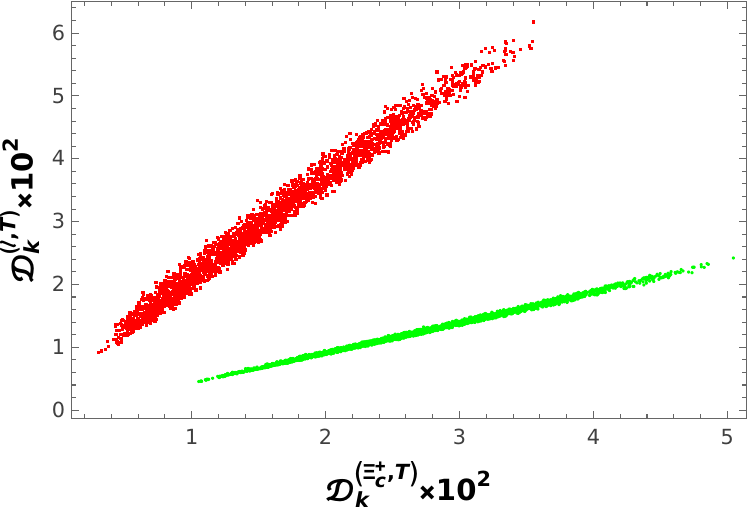}
\includegraphics[width=1.8in,height=1.25in]{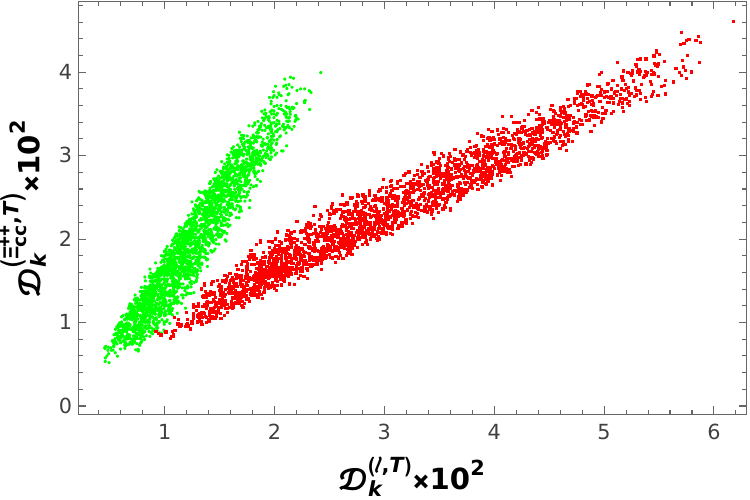}
\includegraphics[width=1.8in,height=1.25in]{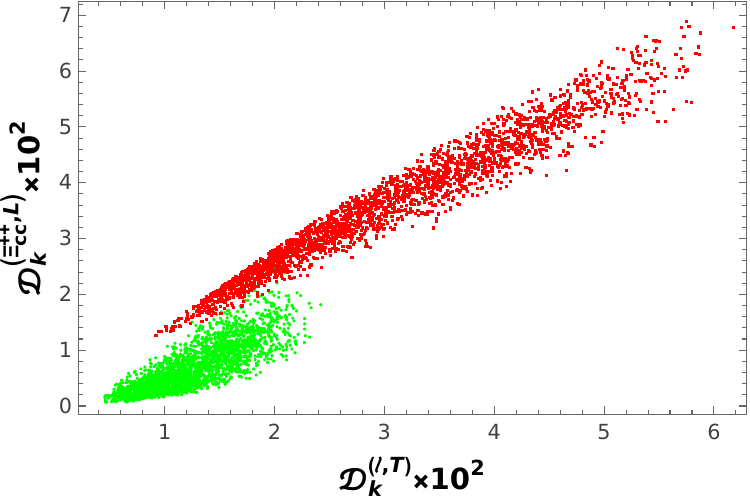}
\includegraphics[width=1.8in,height=1.25in]{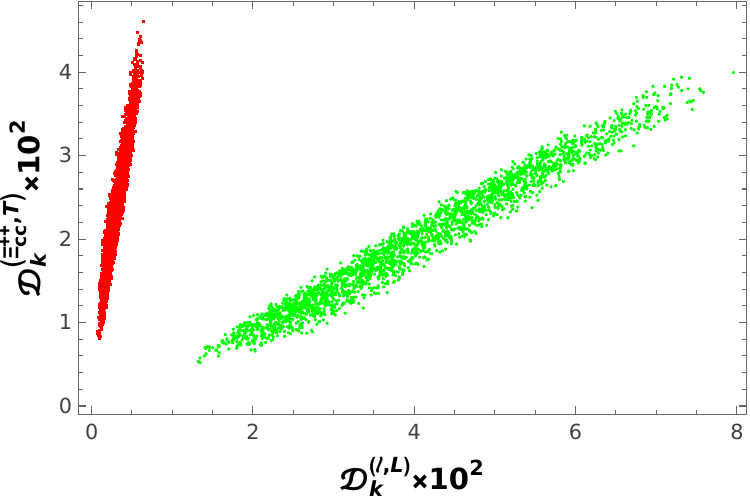}
\caption{Correlations between FBA$^{(m,n)}_{k}$ and $\mathcal{D}^{(m',n')}_{k}$, between FBA$^{(m,n)}_{k}$ and FBA$^{(m',n')}_{k}$, and between $\mathcal{D}^{(m,n)}_{k}$ and $\mathcal{D}^{(m',n')}_{k}$ for particles $m,m'$ ($m,m'\in\{\ell,\Xi_{cc}^{++},\Xi_{c}^{+}\}$) in their respective polarization direction $n,n'$ ($n,n'\in\{L,T\}$). The black, green, and red points indicate unpolarized, positive polarization, and negative polarization cases, respectively.}
\label{correlation}
\end{figure}

Finally, we have computed the correlations between different observables of the $\Xi_{cc}^{++} \to \Xi_c^+ \bar\mu \nu_\mu$ decay, and have presented some representative plots in Fig.~\ref{correlation}. The correlation plots between the differential branching ratio and the forward-backward asymmetry for the positively polarized particles show that the magnitude of the latter decreases as the former increases. The differences between the longitudinal and transverse polarization configurations show the sensitivities to the helicity structure of the weak interaction. The correlations among different differential branching ratios are nearly linear across the physical parameter space. Such linear trends indicate that the underlying helicity amplitudes contributing to different polarization channels are strongly correlated. The separation between the $k=+$ and $k=-$ polarization states in these plots further highlights the sensitivity of polarized observables. These correlations provide valuable complementary information for testing the SM and probing possible NP effects.

\section{Summary and Conclusion}\label{conclude}

In this work, we have studied the semileptonic decay $\Xi_{cc}^{++} \rightarrow \Xi_c^{+} \bar{\ell}\nu_{\ell}$, which proceeds through the quark-level transition $c \to s$. This decay channel provides an important framework for probing the dynamics of heavy baryons and for testing the structure of weak interactions in the baryon sector. In particular, we investigated the role of polarization of particles associated with this decay on various physical observables.

Using the form factors obtained from QCD sum rules, we calculated the $q^{2}$-dependent observables, which include the differential branching ratio $\mathcal{D}(q^2)$ and the forward-backward asymmetry $\text{FBA}(q^2)$. We analyzed these observables for different particle polarization configurations, with both longitudinal and transverse polarization states being considered. Our analysis shows that polarization effects can significantly modify the behavior of branching ratios and forward-backward asymmetries across the full $q^2$ range. These polarized observables exhibit distinct patterns and hierarchies for different spin configurations. We also evaluated the spin polarization asymmetries $P^{(m,n)}$ of the muon and baryons, which further illustrate the sensitivity of the decay observables to the polarization states of the participating particles.
Furthermore, we defined and computed polarization asymmetry ratios for different combinations of longitudinal and transverse polarization states. These ratios exhibit noticeable variations depending on the polarization configuration, indicating that polarization observables, especially polarization asymmetry ratios, may provide additional sensitive probes in future experimental studies.
In addition, we computed the LFU ratio and by integrating over the full $q^2$ range, we obtained $\mathcal{R}_{\Xi_c^+}(\mu/e) = 1.00229_{-0.01551}^{+0.00374}$. 

Finally, we also examined correlations among different observables of the decay $\Xi_{cc}^{++} \to \Xi_c^+ \ell^+ \nu_\ell$, including cases with different polarization states of the involved particles. Such correlations may help in identifying clean observables that can be used to test the SM predictions more precisely.

Overall, our results demonstrate that polarization effects play a significant role in the phenomenology of the decay $\Xi_{cc}^{++} \rightarrow \Xi_c^{+} \bar{\ell}\nu_{\ell}$. The study of polarized observables and ratios in this channel may serve as a useful probe for future experimental measurements and possible physics effects beyond the SM.

\section*{Appendix}\label{A}
\appendix

\section{Tables of Numerical Results}
\label{Tables}

\begin{table}[H]
    \centering
    \resizebox{1\textwidth}{!}{
    \begin{tabular}{|c|c|c|c|c|c|c|}
    	\hline
   \rowcolor{gray!20} $\tilde{\mathcal{D}}\times 10^2$& $ \tilde{\mathcal{D}}_{+}^{(\ell,L)}\times 10^2 $ &$ \tilde{\mathcal{D}}_{+}^{(\Xi_{c}^{+},L)} \times 10^2$ & $ \tilde{\mathcal{D}}_{+}^{(\Xi_{cc}^{++},L)}\times 10^2 $ &$ \tilde{\mathcal{D}}_{+}^{(\ell,T)}\times 10^2 $&$ \tilde{\mathcal{D}}_{+}^{(\Xi_{c}^{+},T)}  \times 10^2$&$ \tilde{\mathcal{D}}_{+}^{(\Xi_{cc}^{++},T)}\times 10^2 $	\\
    \hline
     ${4.14_{-2.51}^{+4.74}}$ & ${3.84_{-2.32}^{+4.36}}$ & ${0.73_{-0.54}^{+1.90}}$ & ${0.60_{-0.53}^{+1.92}}$ & ${1.21_{-0.69}^{+1.29}}$ & ${2.63_{-1.43}^{+2.47}}$ & ${1.88_{-1.25}^{+2.32}}$ \\
    	\hline\hline
       \rowcolor{gray!20} $\tilde{\mathcal{D}}\times 10^2$& $ \tilde{\mathcal{D}}_{-}^{(\ell,L)}\times 10^2 $ &$ \tilde{\mathcal{D}}_{-}^{(\Xi_{c}^{+},L)} \times 10^2$ & $ \tilde{\mathcal{D}}_{-}^{(\Xi_{cc}^{++},L)}\times 10^2 $ &$ \tilde{\mathcal{D}}_{-}^{(\ell,T)}\times 10^2 $&$ \tilde{\mathcal{D}}_{-}^{(\Xi_{c}^{+},T)}  \times 10^2$&$ \tilde{\mathcal{D}}_{-}^{(\Xi_{cc}^{++},T)}\times 10^2 $	\\
    \hline
      ${4.14_{-2.51}^{+4.74}}$ & ${0.29_{-0.20}^{+0.37}}$ & ${3.40_{-1.98}^{+2.84}}$ & ${3.54_{-1.98}^{+2.81}}$ & ${2.93_{-1.82}^{+3.44}}$ & ${1.51_{-1.08}^{+2.27}}$ & ${2.26_{-1.26}^{+2.42}}$ \\
    	\hline
        \hline
   \rowcolor{gray!20} $\text{FBA}$& $\text{FBA}_+^{(\ell,L)}$ &$\text{FBA}_+^{(\Xi_{c}^{+},L)}$ & $\text{FBA}_+^{(\Xi_{cc}^{++},L)}$ &$\text{FBA}_+^{(\ell,T)}$&$\text{FBA}_+^{(\Xi_{c}^{+},T)}$&$\text{FBA}_+^{(\Xi_{cc}^{++},T)}$	\\
    \hline
     ${-0.015_{-0.017}^{+0.038}}$ & ${0.026_{-0.016}^{+0.036}}$ & 
     ${0.223_{-0.165}^{+0.409}}$ & ${-0.196_{-0.215}^{+0.074}}$ & 
     ${0.173_{-0.017}^{+0.043}}$ & ${0.118_{-0.018}^{+0.015}}$ & 
     ${0.131_{-0.035}^{-0.017}}$ \\
    	\hline \hline
    \rowcolor{gray!20} $\text{FBA}$& $\text{FBA}_-^{(\ell,L)}$ &$\text{FBA}_-^{(\Xi_{c}^{+},L)}$ & $\text{FBA}_-^{(\Xi_{cc}^{++},L)}$ &$\text{FBA}_-^{(\ell,T)}$&$\text{FBA}_-^{(\Xi_{c}^{+},T)}$&$\text{FBA}_-^{(\Xi_{cc}^{++},T)}$	\\
    \hline
      ${-0.015_{-0.017}^{+0.038}}$ & ${-0.552_{-0.050}^{+0.004}}$ & ${-0.066_{-0.003}^{+0.006}}$ & ${0.016_{-0.011}^{+0.026}}$ & ${-0.092_{-0.013}^{+0.026}}$ & ${-0.246_{-0.033}^{+0.036}}$ & ${-0.136_{-0.026}^{+0.114}}$ \\
    	\hline
    \end{tabular} }
    \caption{Ranges of branching ratios and forward-backward asymmetry for $q^2=[0.0112-1.346]$ GeV$^2$.}
    \label{BRbarplotsdatasetfull}
\end{table} 

\begin{table}[H]
    \centering
    \resizebox{1\textwidth}{!}{
    \begin{tabular}{|c|c|c|c|c|c|c|}
    	\hline
   \rowcolor{gray!20} $\tilde{\mathcal{D}}\times 10^2$& $ \tilde{\mathcal{D}}_{+}^{(\ell,L)}\times 10^2 $ &$ \tilde{\mathcal{D}}_{+}^{(\Xi_{c}^{+},L)} \times 10^2$ & $ \tilde{\mathcal{D}}_{+}^{(\Xi_{cc}^{++},L)}\times 10^2 $ &$ \tilde{\mathcal{D}}_{+}^{(\ell,T)}\times 10^2 $&$ \tilde{\mathcal{D}}_{+}^{(\Xi_{c}^{+},T)}  \times 10^2$&$ \tilde{\mathcal{D}}_{+}^{(\Xi_{cc}^{++},T)}\times 10^2 $	\\
    \hline
     $2.78_{-1.58}^{+3.27}$ & $2.49_{-1.41}^{+2.92}$ & $0.39_{-0.33}^{+1.55}$ & $0.36_{-0.33}^{+1.55}$ & $0.68_{-0.38}^{+0.79}$ & $1.59_{-0.84}^{+1.64}$ & $1.37_{-0.83}^{+1.66}$ \\
    	\hline\hline
       \rowcolor{gray!20} $\tilde{\mathcal{D}}\times 10^2$& $ \tilde{\mathcal{D}}_{-}^{(\ell,L)}\times 10^2 $ &$ \tilde{\mathcal{D}}_{-}^{(\Xi_{c}^{+},L)} \times 10^2$ & $ \tilde{\mathcal{D}}_{-}^{(\Xi_{cc}^{++},L)}\times 10^2 $ &$ \tilde{\mathcal{D}}_{-}^{(\ell,T)}\times 10^2 $&$ \tilde{\mathcal{D}}_{-}^{(\Xi_{c}^{+},T)}  \times 10^2$&$ \tilde{\mathcal{D}}_{-}^{(\Xi_{cc}^{++},T)}\times 10^2 $	\\
    \hline
      $2.78_{-1.58}^{+3.27}$ & $0.29_{-0.17}^{+0.36}$ & $2.39_{-1.26}^{+1.73}$ & $2.42_{-1.26}^{+1.72}$ & $2.09_{-1.21}^{+2.49}$ & $1.18_{-0.74}^{+1.63}$ & $1.41_{-0.76}^{+1.61}$ \\
    	\hline
        \hline
   \rowcolor{gray!20} $\text{FBA}$& $\text{FBA}_+^{(\ell,L)}$ &$\text{FBA}_+^{(\Xi_{c}^{+},L)}$ & $\text{FBA}_+^{(\Xi_{cc}^{++},L)}$ &$\text{FBA}_+^{(\ell,T)}$&$\text{FBA}_+^{(\Xi_{c}^{+},T)}$&$\text{FBA}_+^{(\Xi_{cc}^{++},T)}$	\\
    \hline
     ${-0.066_{-0.006}^{+0.012}}$ & ${ 0.009_{-0.005}^{+0.010}}$ & 
     ${-0.0004_{-0.058}^{+0.304}}$ & ${-0.105_{+0.015}^{+0.021}}$ &
     ${ 0.250_{-0.006}^{+0.013}}$ & ${ 0.015_{-0.015}^{+0.014}}$ &
     ${-0.004_{-0.049}^{-0.001}}$ \\
    	\hline \hline
    \rowcolor{gray!20} $\text{FBA}$& $\text{FBA}_-^{(\ell,L)}$ &$\text{FBA}_-^{(\Xi_{c}^{+},L)}$ & $\text{FBA}_-^{(\Xi_{cc}^{++},L)}$ &$\text{FBA}_-^{(\ell,T)}$&$\text{FBA}_-^{(\Xi_{c}^{+},T)}$&$\text{FBA}_-^{(\Xi_{cc}^{++},T)}$	\\
    \hline
      ${-0.066_{-0.006}^{+0.012}}$ & ${-0.704_{-0.017}^{+0.003}}$ & ${-0.077_{-0.001}^{+0.002}}$ & ${-0.060_{-0.003}^{+0.007}}$ & ${-0.170_{-0.004}^{+0.006}}$ & ${-0.175_{-0.019}^{+0.021}}$ & ${-0.127_{-0.013}^{+0.071}}$ \\
    	\hline
    \end{tabular} }
    \caption{Ranges of branching ratios and forward-backward asymmetry for $q^2=[0.0112-0.4]$ GeV$^2$.}
    \label{BRbarplotsdataset1}
\end{table} 

\begin{table}[H]
    \centering
    \resizebox{1\textwidth}{!}{
    \begin{tabular}{|c|c|c|c|c|c|c|}
    	\hline
   \rowcolor{gray!20} $\tilde{\mathcal{D}}\times 10^2$& $ \tilde{\mathcal{D}}_{+}^{(\ell,L)}\times 10^2 $ &$ \tilde{\mathcal{D}}_{+}^{(\Xi_{c}^{+},L)} \times 10^2$ & $ \tilde{\mathcal{D}}_{+}^{(\Xi_{cc}^{++},L)}\times 10^2 $ &$ \tilde{\mathcal{D}}_{+}^{(\ell,T)}\times 10^2 $&$ \tilde{\mathcal{D}}_{+}^{(\Xi_{c}^{+},T)}  \times 10^2$&$ \tilde{\mathcal{D}}_{+}^{(\Xi_{cc}^{++},T)}\times 10^2 $	\\
    \hline
     $5.89_{-3.58}^{+6.90}$ & $5.55_{-3.35}^{+6.46}$ & $0.99_{-0.79}^{+2.93}$ & $0.83_{-0.78}^{+2.96}$ & $1.77_{-1.02}^{+1.98}$ & $3.65_{-1.99}^{+3.50}$ & $2.80_{-1.85}^{+3.47}$ \\
    	\hline\hline
       \rowcolor{gray!20} $\tilde{\mathcal{D}}\times 10^2$& $ \tilde{\mathcal{D}}_{-}^{(\ell,L)}\times 10^2 $ &$ \tilde{\mathcal{D}}_{-}^{(\Xi_{c}^{+},L)} \times 10^2$ & $ \tilde{\mathcal{D}}_{-}^{(\Xi_{cc}^{++},L)}\times 10^2 $ &$ \tilde{\mathcal{D}}_{-}^{(\ell,T)}\times 10^2 $&$ \tilde{\mathcal{D}}_{-}^{(\Xi_{c}^{+},T)}  \times 10^2$&$ \tilde{\mathcal{D}}_{-}^{(\Xi_{cc}^{++},T)}\times 10^2 $	\\
    \hline
      $5.89_{-3.58}^{+6.90}$ & $0.34_{-0.23}^{+0.44}$ & $4.90_{-2.80}^{+3.97}$ & $5.06_{-2.80}^{+3.94}$ & $4.13_{-2.56}^{+4.92}$ & $2.25_{-1.60}^{+3.40}$ & $3.10_{-1.73}^{+3.43}$ \\
    	\hline
        \hline
   \rowcolor{gray!20} $\text{FBA}$& $\text{FBA}_+^{(\ell,L)}$ &$\text{FBA}_+^{(\Xi_{c}^{+},L)}$ & $\text{FBA}_+^{(\Xi_{cc}^{++},L)}$ &$\text{FBA}_+^{(\ell,T)}$&$\text{FBA}_+^{(\Xi_{c}^{+},T)}$&$\text{FBA}_+^{(\Xi_{cc}^{++},T)}$	\\
    \hline
     ${-0.015_{-0.014}^{+0.033}}$ & ${ 0.021_{-0.013}^{+0.030}}$ & 
     ${ 0.145_{-0.123}^{+0.473}}$ & ${-0.121_{-0.126}^{+0.040}}$ &  
     ${ 0.173_{-0.015}^{+0.034}}$ & ${ 0.109_{-0.017}^{+0.017}}$ &   
     ${ 0.109_{-0.045}^{-0.012}}$ \\
    	\hline \hline
    \rowcolor{gray!20} $\text{FBA}$& $\text{FBA}_-^{(\ell,L)}$ &$\text{FBA}_-^{(\Xi_{c}^{+},L)}$ & $\text{FBA}_-^{(\Xi_{cc}^{++},L)}$ &$\text{FBA}_-^{(\ell,T)}$&$\text{FBA}_-^{(\Xi_{c}^{+},T)}$&$\text{FBA}_-^{(\Xi_{cc}^{++},T)}$	\\
    \hline
      ${-0.015_{-0.014}^{+0.033}}$ & ${-0.596_{-0.036}^{+0.004}}$ & 
      ${-0.047_{-0.004}^{+0.006}}$ & ${ 0.003_{-0.009}^{+0.021}}$ & 
      ${-0.095_{-0.011}^{+0.023}}$ & ${-0.216_{-0.045}^{+0.035}}$ & 
      ${-0.127_{-0.022}^{+0.112}}$ \\
    	\hline
    \end{tabular} }
    \caption{Ranges of branching ratios and forward-backward asymmetry for $q^2=[0.4-0.8]$ GeV$^2$.}
    \label{BRbarplotsdataset2}
\end{table} 

\begin{table}[H]
    \centering
    \resizebox{1\textwidth}{!}{
    \begin{tabular}{|c|c|c|c|c|c|c|}
    	\hline
   \rowcolor{gray!20} $\tilde{\mathcal{D}}\times 10^2$& $ \tilde{\mathcal{D}}_{+}^{(\ell,L)}\times 10^2 $ &$ \tilde{\mathcal{D}}_{+}^{(\Xi_{c}^{+},L)} \times 10^2$ & $ \tilde{\mathcal{D}}_{+}^{(\Xi_{cc}^{++},L)}\times 10^2 $ &$ \tilde{\mathcal{D}}_{+}^{(\ell,T)}\times 10^2 $&$ \tilde{\mathcal{D}}_{+}^{(\Xi_{c}^{+},T)}  \times 10^2$&$ \tilde{\mathcal{D}}_{+}^{(\Xi_{cc}^{++},T)}\times 10^2 $	\\
    \hline
     $3.82_{-2.39}^{+4.19}$ & $3.56_{-2.20}^{+3.86}$ & $0.79_{-0.50}^{+1.40}$ & $0.59_{-0.49}^{+1.43}$ & $1.17_{-0.67}^{+1.15}$ & $2.61_{-1.45}^{+2.31}$ & $1.58_{-1.12}^{+1.95}$ \\
    	\hline\hline
       \rowcolor{gray!20} $\tilde{\mathcal{D}}\times 10^2$& $ \tilde{\mathcal{D}}_{-}^{(\ell,L)}\times 10^2 $ &$ \tilde{\mathcal{D}}_{-}^{(\Xi_{c}^{+},L)} \times 10^2$ & $ \tilde{\mathcal{D}}_{-}^{(\Xi_{cc}^{++},L)}\times 10^2 $ &$ \tilde{\mathcal{D}}_{-}^{(\ell,T)}\times 10^2 $&$ \tilde{\mathcal{D}}_{-}^{(\Xi_{c}^{+},T)}  \times 10^2$&$ \tilde{\mathcal{D}}_{-}^{(\Xi_{cc}^{++},T)}\times 10^2 $	\\
    \hline
      $3.82_{-2.39}^{+4.19}$ & $0.26_{-0.19}^{+0.33}$ & $3.03_{-1.89}^{+2.79}$ & $3.23_{-1.90}^{+2.76}$ & $2.65_{-1.72}^{+3.04}$ & $1.20_{-0.94}^{+1.89}$ & $2.24_{-1.28}^{+2.25}$ \\
    	\hline
        \hline
   \rowcolor{gray!20} $\text{FBA}$& $\text{FBA}_+^{(\ell,L)}$ &$\text{FBA}_+^{(\Xi_{c}^{+},L)}$ & $\text{FBA}_+^{(\Xi_{cc}^{++},L)}$ &$\text{FBA}_+^{(\ell,T)}$&$\text{FBA}_+^{(\Xi_{c}^{+},T)}$&$\text{FBA}_+^{(\Xi_{cc}^{++},T)}$	\\
    \hline
     ${ 0.012_{-0.025}^{+0.065}}$ & ${ 0.041_{-0.024}^{+0.061}}$ &   
     ${ 0.375_{-0.197}^{+0.319}}$ & ${-0.313_{-0.255}^{+0.113}}$ &  
     ${ 0.140_{-0.027}^{+0.064}}$ & ${ 0.173_{-0.015}^{+0.015}}$ &  
     ${ 0.241_{-0.034}^{+0.024}}$ \\
    	\hline \hline
    \rowcolor{gray!20} $\text{FBA}$& $\text{FBA}_-^{(\ell,L)}$ &$\text{FBA}_-^{(\Xi_{c}^{+},L)}$ & $\text{FBA}_-^{(\Xi_{cc}^{++},L)}$ &$\text{FBA}_-^{(\ell,T)}$&$\text{FBA}_-^{(\Xi_{c}^{+},T)}$&$\text{FBA}_-^{(\Xi_{cc}^{++},T)}$	\\
    \hline
      ${ 0.012_{-0.025}^{+0.065}}$ & ${-0.385_{-0.031}^{+0.001}}$ & 
      ${-0.082_{-0.003}^{+0.006}}$ & ${ 0.072_{-0.022}^{+0.052}}$ & 
      ${-0.044_{-0.020}^{+0.053}}$ & ${-0.336_{-0.078}^{+0.051}}$ & 
      ${-0.149_{-0.037}^{+0.135}}$ \\
    	\hline
    \end{tabular} }
    \caption{Ranges of branching ratios and forward-backward asymmetry for $q^2=[0.8-1.346]$ GeV$^2$.}
    \label{BRbarplotsdataset3}
\end{table}

\begin{table}[H]
    \centering
    \resizebox{1\textwidth}{!}{
    \begin{tabular}{|>{\columncolor{gray!20}}c|c|c|c|c|}
    	\hline
  Observables&$[0.0112-0.4]$  GeV$^2$ &$[0.4-0.8]$ GeV$^2$& $[0.8-1.346]$ GeV$^2$& $[0.0112-1.346]$ GeV$^2$ \\
    \hline\hline
    $P^{(\ell,L)}$  & {(0.786,0.802)} & {(0.878,0.903)} & {(0.852,0.905)} & {(0.850,0.882)} \\\hline
    $P^{(\Xi_c^+,L)}$  & {(-0.890,-0.359)} & {(-0.822,-0.387)} & {(-0.604,-0.455)} & {(-0.758,-0.406)} \\\hline
    $P^{(\Xi_{cc}^{++},L)}$  & {(-0.936,-0.367)} & {(-0.953,-0.408)} & {(-0.863,-0.496)} & {(-0.917,-0.432)} \\\hline
    $P^{(\ell,T)}$  & {(-0.514,-0.487)} & {(-0.415,-0.358)} & {(-0.422,-0.307)} & {(-0.437,-0.367)} \\\hline
    $P^{(\Xi_c^+,T)}$  & {(0.069,0.255)} & {(0.118,0.438)} & {(0.228,0.631)} & {(0.149,0.468)} \\\hline
    $P^{(\Xi_{cc}^{++},T)}$  & { (-0.099,0.002)} & {(-0.183,-0.021)} & {(-0.347,-0.119)} & {(-0.224,-0.053)} \\\hline
    \end{tabular} }
    \caption{Ranges of polarization asymmetries in different $q^{2}$ bins.}
    \label{Table_asymmetries}
\end{table}

\begin{table}[H]
    \centering
    \resizebox{1\textwidth}{!}{
    \begin{tabular}{|c|c|c|c|c|c|c|c|c|}
    	\hline
   \rowcolor{gray!20} $\mathcal{R}_{TL}^{(\ell,\ell)}[+]$  &$\mathcal{R}_{TL}^{(\ell,\ell)}[-]$ & $\mathcal{R}_{TL}^{(\Xi_c^+,\Xi_c^+)}[+]$ &$\mathcal{R}_{TL}^{(\Xi_c^+,\Xi_c^+)}[-]$ & $\mathcal{R}_{TL}^{(\ell,\Xi_c^+)}[+]$ &$\mathcal{R}_{TL}^{(\ell,\Xi_c^+)}[-]$\\
    \hline
     (0.304,0.336) & (9.58,11.6) & (1.94,6.08) & (0.302,0.605) & (0.948,2.62) & (0.778,1.02) \\
    	\hline\hline
       \rowcolor{gray!20} $\mathcal{R}_{TL}^{(\Xi_{cc}^{++},\Xi_c^+)}[+]$ &$\mathcal{R}_{TL}^{(\Xi_{cc}^{++},\Xi_c^+)}[-]$ & $\mathcal{R}_{LL}^{(\ell,\Xi_c^+)}[+]$&$\mathcal{R}_{LL}^{(\ell,\Xi_c^+)}[-]$&$\mathcal{R}_{TT}^{(\Xi_c^+,\ell)}[+]$&$\mathcal{R}_{TT}^{(\Xi_c^+,\ell)}[-]$\\
    \hline
      (1.60,3.21) & (0.696,0.749) & (3.12,7.79) & {(0.067,0.107)} & (2.04,2.32) & (0.389,0.592) \\
    	\hline
        \hline
   \rowcolor{gray!20} $\mathcal{R}_{LL}^{(\Xi_{cc}^{++},\ell)}[+]$&$\mathcal{R}_{LL}^{(\Xi_{cc}^{++},\ell)}[-]$ &$\mathcal{R}_{TT}^{(\Xi_{cc}^{++},\ell)}[+]$&$\mathcal{R}_{TT}^{(\Xi_{cc}^{++},\ell)}[-]$&$\mathcal{R}_{LL}^{(\Xi_{cc}^{++},\Xi_c^+)}[+]$&$\mathcal{R}_{LL}^{(\Xi_{cc}^{++},\Xi_c^+)}[-]$ \\
    \hline
     (0.044,0.307) & (9.55,16.2) & (1.23,1.68) & (0.733,0.895) & (0.343,0.956) & (1.02,1.09) \\
    \hline
    \end{tabular} }
    \caption{Ranges of the polarization ratios for different polarization combinations of the muon $\ell$, $\Xi_c^{+}$, and $\Xi_{cc}^{++}$.}
    \label{TableBRbarplotsratio}
\end{table}

\begin{table}[H]
    \centering
    \resizebox{1\textwidth}{!}{
    \begin{tabular}{|>{\columncolor{gray!20}}c|c|c|c|c|}
    	\hline
  Observables&$[0.0112-0.4]$ GeV$^2$ &$[0.4-0.8]$ GeV$^2$& $[0.8-1.346]$ GeV$^2$& $[0.0112-1.346]$ GeV$^2$ \\
    \hline\hline
    $\mathcal{R}_{P_{TL}^{(\ell,\ell)}}$ &  {(-0.927,-0.926)} &  {(-0.953,-0.947)} & {(-0.948,-0.937)} &  (-0.944,-0.938) \\\hline
    $\mathcal{R}_{P_{TL}^{(\Xi_c^+,\Xi_c^+)}}$  &{  (0.417,0.933)} &{  (0.483,0.927)} & { (0.619,0.894)} &  (0.524,0.905) \\\hline
    $\mathcal{R}_{P_{TL}^{(\Xi_{cc}^{++},\Xi_{cc}^{++})}}$  &{  (0.368,0.923)} &{  (0.390,0.932)} &{  (0.400,0.737)} &  (0.388,0.872) \\\hline
    $\mathcal{R}_{P_{TL}^{(\ell,\Xi_c^+)}}$  & { (-0.190,0.711)} & { (-0.033,0.658)} &{  (0.041,0.364)} &  (-0.037,0.542) \\\hline
    $\mathcal{R}_{P_{LT}^{(\ell,\Xi_{cc}^{++})}}$  &{  (0.785,0.834)} &{  (0.882,0.932)} & { (0.882,0.953)} &  (0.864,0.923) \\\hline
    $\mathcal{R}_{P_{TL}^{(\Xi_{cc}^{++},\Xi_{c}^{+})}}$  &  {(0.360,0.867) }&  {(0.369,0.752) }&  {(0.325,0.355) }&  (0.361,0.644) \\\hline
    $\mathcal{R}_{P_{LL}^{(\ell,\Xi_c^+)}}$  &{  (0.893,0.987)} &{  (0.944,0.990)} &{  (0.942,0.976)} & (0.934,0.983) \\\hline
    $\mathcal{R}_{P_{LL}^{(\Xi_{cc}^{++},\ell)}}$  & {(-0.993,-0.895)} &  {(-0.998,-0.947)} &  {(-0.993,-0.948)} & (-0.995,-0.938) \\\hline
    $\mathcal{R}_{P_{TT}^{(\Xi_{cc}^{++},\ell)}}$  &{  (0.408,0.515)} &{  (0.187,0.397)} & { (-0.045,0.319)} & (0.156,0.393) \\\hline
    \end{tabular} }
    \caption{Ranges of the polarization asymmetry ratios in different $q^{2}$ bins for different polarization combinations of the muon $\ell$, $\Xi_c^{+}$, and $\Xi_{cc}^{++}$.}
    \label{Table_asymmetries_ratios}
\end{table}

\section{Decay Amplitudes for Polarized Particles}\label{app_ampl}

{\allowdisplaybreaks
\begin{align}
    |\mathcal{A}_k^{(\ell,L)}|^2 & = \frac{4}{3q^4}(q^2-m_\mu^2) \Bigg\{ 2\bigg\{12q^4(k+1)M_{\Xi_c^+}M_{\Xi_{cc}^{++}}[-\mathcal{F}_1^2+\mathcal{G}_1^2] + (k-1)m_{\mu}^2  \nonumber \\
    & \times \bigg[2(\mathcal{F}_1^2 + \mathcal{G}_1^2) + \frac{1}{M_{\Xi_c^+}M_{\Xi_{cc}^{++}}} \big[Q_+  \mathcal{F}_2\mathcal{F}_3 +Q_- \mathcal{G}_2\mathcal{G}_3 \big] \bigg]\Big[q^4 + q^2(M_{\Xi_c^+}^2+M_{\Xi_{cc}^{++}}^2) \nonumber \\
    & -2(M_{\Xi_c^+}^2- M_{\Xi_{cc}^{++}}^2)^2\Big] -2q^2(k+1)(\mathcal{F}_1^2 + \mathcal{G}_1^2)  \big[2q^4-q^2(M_{\Xi_c^+}^2+M_{\Xi_{cc}^{++}}^2) \nonumber \\
    & - (M_{\Xi_c^+}^2 - M_{\Xi_{cc}^{++}}^2)^2\big] \bigg\} + \bigg\{ Q_+ Q_- q^2(k+1) \bigg[ \frac{Q_+ \mathcal{F}_2}{M_{\Xi_{cc}^{++}}}\left(\frac{\mathcal{F}_2}{M_{\Xi_{cc}^{++}}} + \frac{\mathcal{F}_3}{M_{\Xi_c^+}} \right) \nonumber \\ 
    & +\frac{Q_-\mathcal{G}_2}{M_{\Xi_{cc}^{++}}}\left(\frac{\mathcal{G}_2}{M_{\Xi_{cc}^{++}}} + \frac{\mathcal{G}_3}{M_{\Xi_c^+}} \right) \bigg] \bigg\} + \bigg\{ -2(k-1)m_\mu^2\Omega \bigg[Q_+\left(\frac{\mathcal{F}_2^2}{M_{\Xi_{cc}^{++}}^2} + \frac{\mathcal{F}_3^2}{M_{\Xi_c^+}^2} \right) \nonumber \\
    & + Q_- \left(\frac{\mathcal{G}_2^2}{M_{\Xi_{cc}^{++}}^2} + \frac{\mathcal{G}_3^2}{M_{\Xi_c^+}^2} \right) \bigg] + \frac{2}{M_{\Xi_{cc}^{++}}^2}q^2(k-1) m_\mu^2\left[Q_+ \mathcal{F}_2^2 + Q_- \mathcal{G}_2^2 \right] \nonumber \\
    & \times (2M_{\Xi_c^+}^2-M_{\Xi_{cc}^{++}}^2)+ \frac{1}{M_{\Xi_c^+}^2}\big[Q_+ \mathcal{F}_3^2 + Q_- \mathcal{G}_3^2 \big][2(k-1)(M_{\Xi_c^+}^2+2M_{\Xi_{cc}^{++}}^2) \nonumber \\
    & -q^2(k+1)][(M_{\Xi_c^+}^2-M_{\Xi_{cc}^{++}}^2) + q^2] \bigg\} + 4 \bigg\{ -2 (k-1)m_\mu^2  \Bigg[ \left[Q_+ \mathcal{F}_1\left( \frac{\mathcal{F}_2}{M_{\Xi_{cc}^{++}}} + \frac{\mathcal{F}_3}{M_{\Xi_c^+}}\right)\right] \nonumber \\
    & + \bigg[Q_- \mathcal{G}_1\bigg( \frac{\mathcal{G}_2}{M_{\Xi_{cc}^{++}}} + \frac{\mathcal{G}_3}{M_{\Xi_c^+}}\bigg)  \bigg] \Bigg](M_{\Xi_c^+}+M_{\Xi_{cc}^{++}}) (M_{\Xi_c^+}-M_{\Xi_{cc}^{++}})^2 + Q_+Q_- q^2(k+1) \nonumber \\
    & \times (M_{\Xi_c^+}+M_{\Xi_{cc}^{++}}) \bigg[\bigg[\mathcal{F}_1\left( \frac{\mathcal{F}_2}{M_{\Xi_{cc}^{++}}}+ \frac{\mathcal{F}_3}{M_{\Xi_c^+}}\right)\bigg] + \bigg[\mathcal{G}_1\left( \frac{\mathcal{G}_2}{M_{\Xi_{cc}^{++}}} + \frac{\mathcal{G}_3}{M_{\Xi_c^+}}\right)\bigg] \bigg] \nonumber \\
    & + q^2(k-1) m_\mu^2 \bigg[Q_+ \mathcal{F}_1\bigg(\frac{\mathcal{F}_2}{M_{\Xi_{cc}^{++}}}(2M_{\Xi_c^+} -M_{\Xi_{cc}^{++}}) -\frac{\mathcal{F}_3}{M_{\Xi_c^+}}(M_{\Xi_c^+}-2M_{\Xi_{cc}^{++}}) \bigg) \nonumber \\
    & + Q_- \mathcal{G}_1\bigg(\frac{\mathcal{G}_2}{M_{\Xi_{cc}^{++}}} (2M_{\Xi_c^+}+ M_{\Xi_{cc}^{++}}) -\frac{\mathcal{G}_3}{M_{\Xi_c^+}}(M_{\Xi_c^+}+2M_{\Xi_{cc}^{++}}) \bigg)\bigg] \bigg\} \Bigg\}, 
\end{align}
}
where $\Omega \equiv [(M_{\Xi_c^+}^2-M_{\Xi_{cc}^{++}}^2)^2-q^4]$.

{\allowdisplaybreaks
\begin{align}
    |\mathcal{A}_k^{(\Xi_c^+,L)}|^2 & = \frac{-2}{3q^4}(q^2-m_\mu^2)\Bigg\{ 4\Omega[\mathcal{F}_1^2 + \mathcal{G}_1^2]\big[[3m_\mu^2(k-1)-q^2(k+1)] \big] + \lambda\big[m_\mu^2(k-1) \nonumber \\
    & + q^2(k+1)\big] + 48q^4M_{\Xi_c^+}M_{\Xi_{cc}^{++}}(k+1)[\mathcal{F}_1^2-\mathcal{G}_1^2] + 4(M_{\Xi_c^+}+M_{\Xi_{cc}^{++}}) \nonumber \\
    & \times \Big[m_\mu^2\lambda (k-1)- Q_+  q^2(k+1)-\lambda\Big]\mathcal{F}_1\bigg[\frac{\mathcal{F}_2}{M_{\Xi_{cc}^{++}}} +\frac{\mathcal{F}_3}{M_{\Xi_c^+}}\bigg] + 4 Q_+  m_\mu^2(k-1) \nonumber \\
    & \times (M_{\Xi_c^+}-M_{\Xi_{cc}^{++}})\mathcal{F}_1 \bigg[\frac{\mathcal{F}_2}{M_{\Xi_{cc}^{++}}}(M_{\Xi_c^+}^2-M_{\Xi_{cc}^{++}}^2 -q^2) + \frac{\mathcal{F}_3}{M_{\Xi_c^+}}(M_{\Xi_c^+}^2-M_{\Xi_{cc}^{++}}^2+q^2) \bigg] \nonumber \\
    & + m_\mu^2(k-1) \bigg[ \frac{1}{M_{\Xi_{cc}^{++}}^2} \big[ Q_+  \mathcal{F}_2^2 + Q_- \mathcal{G}_2^2\big]\big[\lambda + 3(M_{\Xi_c^+}M_{\Xi_{cc}^{++}} -q^2)^2\big] \nonumber \\
    & + \frac{2}{M_{\Xi_c^+}M_{\Xi_{cc}^{++}}} [ Q_+  \mathcal{F}_2\mathcal{F}_3 + Q_- \mathcal{G}_2\mathcal{G}_3] \big[3\Omega +\lambda\big] +\frac{1}{M_{\Xi_c^+}^2}[ Q_+  \mathcal{F}_3^2 + Q_- \mathcal{G}_3^2] \nonumber \\
    & \times \big[3\tilde{\Omega}+\lambda\big] \bigg] - q^2(k+1)(3 Q_+  Q_- -\lambda) \bigg[ Q_+ \bigg(\frac{\mathcal{F}_2^2}{M_{\Xi_{cc}^{++}}^2} +\frac{\mathcal{F}_3^2}{M_{\Xi_c^+}^2}\bigg)^2 \nonumber \\
    & +  Q_- \bigg(\frac{\mathcal{G}_2^2}{M_{\Xi_{cc}^{++}}^2}+\frac{\mathcal{G}_3^2}{M_{\Xi_c^+}^2}\bigg)^2 - 8q^2(M_{\Xi_c^+} -M_{\Xi_{cc}^{++}}) \mathcal{G}_1 \bigg(\frac{\mathcal{G}_2^2}{M_{\Xi_{cc}^{++}}^2}+\frac{\mathcal{G}_3^2}{M_{\Xi_c^+}^2}\bigg) \bigg] \nonumber \\
    & +  4m_\mu^2\mathcal{G}_1\bigg[\big[m_\mu^2(k-1)(M_{\Xi_c^+}-M_{\Xi_{cc}^{++}})\big[3 Q_-  (M_{\Xi_c^+}+M_{\Xi_{cc}^{++}})^2 + \lambda\big]\big] \nonumber \\
    & - 2q^2 \big[3k Q_- (M_{\Xi_c^+}+M_{\Xi_{cc}^{++}})^2 - \lambda\big]\left[\frac{\mathcal{G}_2}{M_{\Xi_{cc}^{++}}}+\frac{\mathcal{G}_3}{M_{\Xi_c^+}}\right] - 3m_\mu^2(k -1)q^2 Q_- \nonumber \\
    & \times (M_{\Xi_c^+}+M_{\Xi_{cc}^{++}})\bigg[\frac{\mathcal{G}_2}{M_{\Xi_{cc}^{++}}} -\frac{\mathcal{G}_3}{M_{\Xi_c^+}}\bigg] \bigg] + 2q^4\bigg[ \bigg( 1-k\frac{M_{\Xi_c^+}}{M_{\Xi_{cc}^{++}}} \bigg) \mathcal{G}_2 \nonumber \\
    & - \bigg(1 - k \frac{M_{\Xi_{cc}^{++}}}{M_{\Xi_c^+}}\bigg)\mathcal{G}_3\bigg]\Bigg\},
\end{align}
}
where $\lambda \equiv M_{\Xi_{cc}^{++}}^4 - 2(M_{\Xi_c^+}^2 + s)M_{\Xi_{cc}^{++}}^2 + (M_{\Xi_{cc}^{++}}^2 - s)^2$ and $\tilde{\Omega} \equiv [(M_{\Xi_c^+}^2-M_{\Xi_{cc}^{++}}^2)^2+q^4]$.

{\allowdisplaybreaks
\begin{align}
    |\mathcal{A}_k^{(\Xi_{cc}^{++},L)}|^2 & = \frac{1}{3q^4} (q^2-m_\mu^2) \Bigg\{ \bigg\{ \Big[ \mathcal{F}_1^2+\mathcal{G}_1^2 \Big] \Big[ (2m_\mu^2+q^2)(M_{\Xi_c^+}^2-M_{\Xi_{cc}^{++}}^2)^2 + q^2(q^2-m_\mu^2) \nonumber \\
    & \times (M_{\Xi_c^+}^2+M_{\Xi_{cc}^{++}}^2) - q^4(2q^2+m_\mu^2) \Big] - 6M_{\Xi_c^+}M_{\Xi_{cc}^{++}} q^4 \Big[ \mathcal{F}_1^2-\mathcal{G}_1^2 \Big] \bigg\} + \frac{4}{M_{\Xi_{cc}^{++}}^2} \nonumber \\
    & \times \bigg\{ \Big[ Q_+\mathcal{F}_2^2 + Q_-\mathcal{G}_2^2 \Big] \Big[ (2m_\mu^2+q^2)\lambda + 6m_\mu^2M_{\Xi_{cc}^{++}}^2 q^2 \Big] \bigg\} + \frac{4}{M_{\Xi_c^+}^2}\bigg\{ \Big[ Q_+\mathcal{F}_3^2 + Q_-\mathcal{G}_3^2 \Big] \nonumber \\
    & \times \Big[ (2m_\mu^2+q^2)\lambda + 6m_\mu^2M_{\Xi_c^+}^2 q^2 \Big] \bigg\} + 16 \bigg\{ -m_\mu^2 q^2 \bigg[ Q_+ \bigg[ \frac{\mathcal{F}_1\mathcal{F}_2}{M_{\Xi_{cc}^{++}}}(2M_{\Xi_c^+}-M_{\Xi_{cc}^{++}}) \nonumber \\
    & - \frac{\mathcal{F}_1\mathcal{F}_3}{M_{\Xi_c^+}}(M_{\Xi_c^+}-2M_{\Xi_{cc}^{++}}) \bigg] + Q_- \bigg[ \frac{\mathcal{G}_1\mathcal{G}_2}{M_{\Xi_{cc}^{++}}}(2M_{\Xi_c^+}+M_{\Xi_{cc}^{++}}) - \frac{\mathcal{G}_1\mathcal{G}_3}{M_{\Xi_c^+}}(M_{\Xi_c^+} \nonumber \\
    & +2M_{\Xi_{cc}^{++}}) \bigg] \bigg] + Q_+(M_{\Xi_c^+}+M_{\Xi_{cc}^{++}}) \Big[ (2m_\mu^2+q^2)(M_{\Xi_c^+}-M_{\Xi_{cc}^{++}})^2-q^4 \Big] \bigg(\frac{\mathcal{F}_1\mathcal{F}_2}{M_{\Xi_{cc}^{++}}} \nonumber \\
    & + \frac{\mathcal{F}_1\mathcal{F}_3}{M_{\Xi_c^+}}\bigg) + Q_-(M_{\Xi_c^+}-M_{\Xi_{cc}^{++}}) \Big[ (2m_\mu^2+q^2)(M_{\Xi_c^+}+M_{\Xi_{cc}^{++}})^2-q^4 \Big] \bigg(\frac{\mathcal{G}_1\mathcal{G}_2}{M_{\Xi_{cc}^{++}}} \nonumber \\
    & + \frac{\mathcal{G}_1\mathcal{G}_3}{M_{\Xi_c^+}}\bigg) \bigg\} + \frac{8}{M_{\Xi_c^+}M_{\Xi_{cc}^{++}}} \bigg\{ \Big[ Q_+\mathcal{F}_2\mathcal{F}_3 + Q_-\mathcal{G}_2\mathcal{G}_3 \Big] \Big[ q^4(q^2-m_\mu^2) - q^2(2q^2+m_\mu^2) \nonumber \\
    & \times (M_{\Xi_c^+}^2+M_{\Xi_{cc}^{++}}^2) + (2m_\mu^2+q^2)(M_{\Xi_c^+}^2-M_{\Xi_{cc}^{++}}^2)^2 \Big] \bigg\} + 32k\Big[(2m_\mu^2+q^2)(M_{\Xi_c^+}^2 \nonumber \\
    & \times -M_{\Xi_{cc}^{++}}^2+q^2) + q^2(q^2-m_\mu^2)\Big]\sqrt{\lambda}\mathcal{F}_1\mathcal{G}_1 + \frac{8k\sqrt{\lambda}}{M_{\Xi_{cc}^{++}}} \bigg\{ \Big[ (M_{\Xi_c^+}-M_{\Xi_{cc}^{++}})^2 \big[ q^4 \nonumber \\
    & + 2M_{\Xi_{cc}^{++}}(2m_\mu^2+q^2) (M_{\Xi_c^+}+M_{\Xi_{cc}^{++}}) \big] - q^4\big[ q^2 - m_\mu^2 + 2M_{\Xi_{cc}^{++}}(M_{\Xi_c^+}+M_{\Xi_{cc}^{++}}) \big] \Big] \nonumber \\
    & \times \bigg[ \frac{\mathcal{F}_1\mathcal{G}_2}{M_{\Xi_{cc}^{++}}} + \frac{\mathcal{F}_1\mathcal{G}_3}{M_{\Xi_c^+}} \bigg] - \Big[ (M_{\Xi_c^+}-M_{\Xi_{cc}^{++}})^2 \big[ q^4 - 2M_{\Xi_{cc}^{++}} (2m_\mu^2+q^2)(M_{\Xi_c^+}-M_{\Xi_{cc}^{++}}) \big] \nonumber \\
    & - q^4\big[ q^2 - m_\mu^2 - 2M_{\Xi_{cc}^{++}}(M_{\Xi_c^+}-M_{\Xi_{cc}^{++}}) \big] \Big] \bigg[ \frac{\mathcal{F}_2\mathcal{G}_1}{M_{\Xi_{cc}^{++}}} + \frac{\mathcal{F}_3\mathcal{G}_1}{M_{\Xi_c^+}} \bigg] - m_\mu^2q^2 \bigg[ \frac{\mathcal{F}_1\mathcal{G}_2}{M_{\Xi_{cc}^{++}}} \nonumber \\
    & \times\Big[ (M_{\Xi_c^+}+M_{\Xi_{cc}^{++}})^2 -2M_{\Xi_{cc}^{++}}^2 \Big] - \frac{\mathcal{F}_2\mathcal{G}_1}{M_{\Xi_{cc}^{++}}} \Big[ (M_{\Xi_c^+}-M_{\Xi_{cc}^{++}})^2 -2M_{\Xi_{cc}^{++}}^2 \Big] + \frac{\mathcal{F}_1\mathcal{G}_3}{M_{\Xi_c^+}} \nonumber \\
    & \times \Big[ (M_{\Xi_c^+}-M_{\Xi_{cc}^{++}})^2 -2M_{\Xi_{cc}^{++}}(M_{\Xi_c^+}-2M_{\Xi_{cc}^{++}}) \Big] - \frac{\mathcal{F}_3\mathcal{G}_1}{M_{\Xi_c^+}} \Big[ (M_{\Xi_c^+}+M_{\Xi_{cc}^{++}})^2  \nonumber \\
    & +2M_{\Xi_{cc}^{++}}(M_{\Xi_c^+}+2M_{\Xi_{cc}^{++}}) \Big] \bigg] + \frac{q^2(q^2-m_\mu^2)}{M_{\Xi_c^+}M_{\Xi_{cc}^{++}}} \Big[ Q_+ (M_{\Xi_c^+} \mathcal{F}_2 \mathcal{G}_1 + M_{\Xi_{cc}^{++}}\mathcal{F}_3 \mathcal{G}_1) \nonumber \\
    & - Q_- (M_{\Xi_c^+} \mathcal{F}_1 \mathcal{G}_2 + M_{\Xi_{cc}^{++}}\mathcal{F}_1 \mathcal{G}_3) \Big] \bigg\} + 8k\sqrt{\lambda} \bigg\{ (2m_\mu^2 + q^2)\lambda \bigg( \frac{\mathcal{F}_2\mathcal{G}_2}{M_{\Xi_{cc}^{++}}^2} + \frac{\mathcal{F}_3\mathcal{G}_3}{M_{\Xi_c^+}^2} \bigg) \nonumber \\
    & + 6m_\mu^2q^2(\mathcal{F}_2\mathcal{G}_2 + \mathcal{F}_3\mathcal{G}_3) \bigg\} + \frac{8k\sqrt{\lambda}}{M_{\Xi_c^+}M_{\Xi_{cc}^{++}}} \bigg\{ \Big[ \mathcal{F}_2\mathcal{G}_3+\mathcal{F}_3\mathcal{G}_2 \Big] \Big[ (2m_\mu^2+q^2)(M_{\Xi_c^+}^2-M_{\Xi_{cc}^{++}}^2)^2 \nonumber \\
    & - q^2(M_{\Xi_c^+}+M_{\Xi_{cc}^{++}})(m_\mu^2+2q^2) + q^4(q^2-m_\mu^2) \Big] \bigg\} \Bigg\}.
\end{align}
}

{\allowdisplaybreaks
\begin{align}
    |\mathcal{A}_k^{(\ell,T)}|^2 & = \frac{1}{3q^4} (q^2-m_\mu^2) \Bigg\{ 4\bigg\{ \Big[ \mathcal{F}_1^2+\mathcal{G}_1^2 \Big] \Big[ (M_{\Xi_c^+}^2-M_{\Xi_{cc}^{++}}^2)\big[ 3\pi kqm_\mu \sqrt{\lambda} + 4(2m_\mu^2+q^2) \nonumber \\
    & \times (M_{\Xi_c^+}^2-M_{\Xi_{cc}^{++}}^2) \big] + 4q^2(q^2-m_\mu^2)(M_{\Xi_c^+}^2+M_{\Xi_{cc}^{++}}^2) - 4q^4(2q^2+m_\mu^2) \Big] \nonumber \\
    & - 24M_{\Xi_c^+}M_{\Xi_{cc}^{++}} q^2\Big[\mathcal{F}_1^2-\mathcal{G}_1^2\Big] \bigg\} + \frac{1}{M_{\Xi_{cc}^{++}}^2} \bigg\{ \Big[ Q_+\mathcal{F}_2^2 + Q_-\mathcal{G}_2^2 \Big] \Big[ 3\pi kqm_\mu \sqrt{\lambda} \nonumber \\
    & \times \big( M_{\Xi_c^+}^2-M_{\Xi_{cc}^{++}}^2-q^2 \big) + 4\lambda(2m_\mu^2+q^2) + 24m_\mu^2M_{\Xi_{cc}^{++}}^2q^2 \Big] \bigg\} + \frac{1}{M_{\Xi_c^+}^2} \nonumber \\
    & \times \bigg\{ \Big[ Q_+\mathcal{F}_3^2 + Q_-\mathcal{G}_3^2 \Big] \Big[ 3\pi kqm_\mu \sqrt{\lambda} \big( M_{\Xi_c^+}^2-M_{\Xi_{cc}^{++}}^2+q^2 \big) + 4\lambda(2m_\mu^2+q^2) \nonumber \\
    & + 24m_\mu^2M_{\Xi_c^+}^2q^2 \Big] \bigg\} + 4 \bigg\{ 3\pi kqm_\mu\lambda \bigg[ (M_{\Xi_c^+}^2-M_{\Xi_{cc}^{++}}^2) \bigg[ (M_{\Xi_c^+}+M_{\Xi_{cc}^{++}})\bigg( \frac{\mathcal{F}_1\mathcal{F}_2}{M_{\Xi_{cc}^{++}}} \nonumber \\
    & + \frac{\mathcal{F}_1\mathcal{F}_3}{M_{\Xi_c^+}} \bigg) + (M_{\Xi_c^+}-M_{\Xi_{cc}^{++}})\bigg( \frac{\mathcal{G}_1\mathcal{G}_2}{M_{\Xi_{cc}^{++}}} + \frac{\mathcal{G}_1\mathcal{G}_3}{M_{\Xi_c^+}} \bigg) \bigg] - q^2 \bigg( \frac{M_{\Xi_c^+}}{M_{\Xi_{cc}^{++}}}(\mathcal{F}_1\mathcal{F}_2+\mathcal{G}_1\mathcal{G}_2) \nonumber \\
    & - \frac{M_{\Xi_{cc}^{++}}}{M_{\Xi_c^+}}(\mathcal{F}_1\mathcal{F}_3+\mathcal{G}_1\mathcal{G}_3) \bigg) \bigg] - 4\bigg[ Q_+ \bigg[m_\mu^2q^2\bigg( \frac{\mathcal{F}_1\mathcal{F}_2}{M_{\Xi_{cc}^{++}}}(2M_{\Xi_c^+}-M_{\Xi_{cc}^{++}}) -\frac{\mathcal{F}_1\mathcal{F}_3}{M_{\Xi_c^+}} \nonumber \\
    & \times (M_{\Xi_c^+}-2M_{\Xi_{cc}^{++}}) \bigg) - (M_{\Xi_c^+}+M_{\Xi_{cc}^{++}}) \Big[ (M_{\Xi_c^+}-M_{\Xi_{cc}^{++}})^2(2m_\mu^2+q^2)-q^4 \Big] \nonumber \\
    & \times \bigg(\frac{\mathcal{F}_1\mathcal{F}_2}{M_{\Xi_{cc}^{++}}} + \frac{\mathcal{F}_1\mathcal{F}_3}{M_{\Xi_c^+}}\bigg) \bigg] + Q_- \bigg[m_\mu^2q^2\bigg( \frac{\mathcal{G}_1\mathcal{G}_2}{M_{\Xi_{cc}^{++}}}(2M_{\Xi_c^+}+M_{\Xi_{cc}^{++}}) -\frac{\mathcal{G}_1\mathcal{G}_3}{M_{\Xi_c^+}} \nonumber \\
    & \times (M_{\Xi_c^+}+2M_{\Xi_{cc}^{++}}) \bigg) - (M_{\Xi_c^+}-M_{\Xi_{cc}^{++}}) \Big[ (M_{\Xi_c^+}+M_{\Xi_{cc}^{++}})^2(2m_\mu^2+q^2)-q^4 \Big] \nonumber \\
    & \times \bigg(\frac{\mathcal{G}_1\mathcal{G}_2}{M_{\Xi_{cc}^{++}}} + \frac{\mathcal{G}_1\mathcal{G}_3}{M_{\Xi_c^+}}\bigg) \bigg] \bigg] \bigg\} + \frac{2}{M_{\Xi_c^+}M_{\Xi_{cc}^{++}}} \bigg\{ \Big[ Q_+\mathcal{F}_2\mathcal{F}_3 + Q_-\mathcal{G}_2\mathcal{G}_3 \Big] \Big[ 3\pi kqm_\mu\sqrt{\lambda} \nonumber \\
    & \times (M_{\Xi_c^+}^2-M_{\Xi_{cc}^{++}}^2) - 4 \big[ q^4(q^2+m_\mu^2) + q^2(M_{\Xi_c^+}^2+M_{\Xi_{cc}^{++}}^2)(2q^2-m_\mu^2) \nonumber \\
    & + (M_{\Xi_c^+}^2-M_{\Xi_{cc}^{++}}^2)(2m_\mu^2+q^2) \big]  \Big] \bigg\} + 24\pi kq^3m_\mu\sqrt{\lambda}\mathcal{F}_1\mathcal{G}_1  \Bigg\}.
\end{align}
}

{\allowdisplaybreaks
\begin{align}
    |\mathcal{A}_k^{(\Xi_{c}^{+},T)}|^2 & = \frac{1}{3q^4} (q^2-m_\mu^2) \Bigg\{ -4\bigg\{3\pi kq^3\Big[ Q_-(M_{\Xi_c^+}+M_{\Xi_{cc}^{++}})\mathcal{F}_1^2+Q_+(M_{\Xi_c^+}-M_{\Xi_{cc}^{++}})\mathcal{G}_1^2 \Big] \nonumber \\
    & - 4q^4 \Big[ \big[ (M_{\Xi_c^+}-M_{\Xi_{cc}^{++}})^2 - 4M_{\Xi_c^+}M_{\Xi_{cc}^{++}} \big] \mathcal{F}_1^2 + \big[ (M_{\Xi_c^+}+M_{\Xi_{cc}^{++}})^2 + 4M_{\Xi_c^+}M_{\Xi_{cc}^{++}} \big] \nonumber \\
    & \times \mathcal{G}_1^2  \Big] + 2\Big[ 2q^4(m_\mu^2 + 2q^2) + 2m_\mu^2q^2(M_{\Xi_c^+}^2+M_{\Xi_{cc}^{++}}^2) - (2m_\mu^2+q^2)(M_{\Xi_c^+}^2-M_{\Xi_{cc}^{++}}^2) \Big] \nonumber \\
    & \times \Big[ \mathcal{F}_1^2 + \mathcal{G}_1^2\Big] \bigg\} + \frac{4}{M_{\Xi_{cc}^{++}}^2}\bigg\{ \Big[ Q_+\mathcal{F}_2^2+Q_-\mathcal{G}_2^2 \Big] \Big[ (2m_\mu^2+q^2)\lambda + 6m_\mu^2M_{\Xi_{cc}^{++}}^2q^2 \Big] \bigg\} \nonumber \\
    & + \frac{4}{M_{\Xi_c^+}^2}\bigg\{ \Big[ Q_+\mathcal{F}_3^2+Q_-\mathcal{G}_3^2 \Big] \Big[ (2m_\mu^2+q^2)\lambda + 6m_\mu^2M_{\Xi_c^+}^2q^2 \Big] \bigg\} + 2 Q_+\mathcal{F}_1 \nonumber \\
    & \times \bigg\{ 8m_\mu^2q^2 \bigg[ \frac{\mathcal{F}_2}{M_{\Xi_{cc}^{++}}}(-2M_{\Xi_c^+}+M_{\Xi_{cc}^{++}}) + \frac{\mathcal{F}_3}{M_{\Xi_c^+}}(M_{\Xi_c^+}-2M_{\Xi_{cc}^{++}}) \bigg] - \Big[ 3\pi kq^3Q_- \nonumber \\
    & -8(M_{\Xi_c^+}+M_{\Xi_{cc}^{++}})\big[ (2m_\mu^2+q^2)(M_{\Xi_c^+}-M_{\Xi_{cc}^{++}})^2 - q^4 \big] \Big] \bigg[ \frac{\mathcal{F}_2}{M_{\Xi_{cc}^{++}}} + \frac{\mathcal{F}_3}{M_{\Xi_c^+}} \bigg] \bigg\} \nonumber \\
    & - 2 Q_-\mathcal{G}_1 \bigg\{ 8m_\mu^2q^2 \bigg[ \frac{\mathcal{G}_2}{M_{\Xi_{cc}^{++}}}(2M_{\Xi_c^+}+M_{\Xi_{cc}^{++}}) - \frac{\mathcal{G}_3}{M_{\Xi_c^+}}(M_{\Xi_c^+}+2M_{\Xi_{cc}^{++}}) \bigg] + \Big[3\pi kq^3Q_+ \nonumber \\
    & - 8(M_{\Xi_c^+}-M_{\Xi_{cc}^{++}})\big[ (2m_\mu^2+q^2)(M_{\Xi_c^+}+M_{\Xi_{cc}^{++}})^2 - q^2 \big]\Big]\bigg[ \frac{\mathcal{G}_2}{M_{\Xi_{cc}^{++}}} +\frac{\mathcal{G}_3}{M_{\Xi_c^+}} \bigg] \bigg\} \nonumber \\
    & - 24\pi kqM_{\Xi_c^+} m_\mu^2(M_{\Xi_c^+}^2-M_{\Xi_{cc}^{++}}^2-q^2)\mathcal{F}_1\mathcal{G}_1 + 6\pi kqm_\mu^2\mathcal{F}_1^2\bigg\{ 2q^2(M_{\Xi_c^+}-M_{\Xi_{cc}^{++}}) \nonumber \\
    & \times \bigg[\frac{M_{\Xi_c^+}}{M_{\Xi_{cc}^{++}}}\mathcal{G}_2 + \frac{M_{\Xi_{cc}^{++}}}{M_{\Xi_c^+}}\mathcal{G}_3\bigg] - (M_{\Xi_c^+}^2-M_{\Xi_{cc}^{++}}^2)(M_{\Xi_c^+}-M_{\Xi_{cc}^{++}})^2 \bigg[ \frac{\mathcal{G}_2}{M_{\Xi_{cc}^{++}}} + \frac{\mathcal{G}_3}{M_{\Xi_c^+}} \bigg] \nonumber \\
    & - q^4\bigg[ \frac{\mathcal{G}_2}{M_{\Xi_{cc}^{++}}} - \frac{\mathcal{G}_3}{M_{\Xi_c^+}} \bigg] \bigg\} + 6\pi kqm_\mu^2\mathcal{G}_1^2\bigg\{ 2q^2(M_{\Xi_c^+}+M_{\Xi_{cc}^{++}}) \bigg[\frac{M_{\Xi_c^+}}{M_{\Xi_{cc}^{++}}}\mathcal{F}_2 - \frac{M_{\Xi_{cc}^{++}}}{M_{\Xi_c^+}}\mathcal{F}_3\bigg] \nonumber \\
    & - (M_{\Xi_c^+}^2-M_{\Xi_{cc}^{++}}^2)(M_{\Xi_c^+}+M_{\Xi_{cc}^{++}})^2 \bigg[ \frac{\mathcal{F}_2}{M_{\Xi_{cc}^{++}}} + \frac{\mathcal{F}_3}{M_{\Xi_c^+}} \bigg] - q^4\bigg[ \frac{\mathcal{F}_2}{M_{\Xi_{cc}^{++}}} - \frac{\mathcal{F}_3}{M_{\Xi_c^+}} \bigg] \bigg\} \Bigg\}.
\end{align}
}

{\allowdisplaybreaks
\begin{align}
    |\mathcal{A}_k^{(\Xi_{cc}^{++},T)}|^2 & = \frac{1}{3q^4} (q^2-m_\mu^2) \Bigg\{ 4\bigg\{ 3\pi kq^3 \Big[ -Q_-(M_{\Xi_c^+}+M_{\Xi_{cc}^{++}})\mathcal{F}_1^2 + Q_+(M_{\Xi_c^+}-M_{\Xi_{cc}^{++}})\mathcal{G}_1^2 \Big] \nonumber \\
    & + 4\Big[ \mathcal{F}_1^2+\mathcal{G}_1^2 \Big] \Big[ (2m_\mu^2 + q^2)(M_{\Xi_c^+}^2-M_{\Xi_{cc}^{++}}^2)^2 + q^2 (q^2-m_\mu^2)(M_{\Xi_c^+}^2+M_{\Xi_{cc}^{++}}^2) \nonumber \\
    & - q^4(2q^2+m_\mu^2)\Big] -24M_{\Xi_c^+}M_{\Xi_{cc}^{++}} q^4\Big[\mathcal{F}_1^2-\mathcal{G}_1^2\Big] \bigg\} + \frac{4}{M_{\Xi_{cc}^{++}}^2} \bigg\{ \Big[ Q_+\mathcal{F}_2^2 + Q_-\mathcal{G}_2^2 \Big] \nonumber \\
    & \times \Big[ (2m_\mu^2 + q^2)\lambda + 6m_\mu^2M_{\Xi_{cc}^{++}}^2q^2 \Big] \bigg\} + \frac{4}{M_{\Xi_c^+}^2} \bigg\{ \Big[ Q_+\mathcal{F}_3^2 + Q_-\mathcal{G}_3^2 \Big]\Big[ (2m_\mu^2 + q^2)\lambda \nonumber \\
    & + 6m_\mu^2M_{\Xi_c^+}^2q^2 \Big] \bigg\} - 2 \bigg\{ 3\pi kq^3\lambda \bigg[ \frac{1}{M_{\Xi_{cc}^{++}}} \Big[ \mathcal{F}_1\mathcal{F}_2-\mathcal{G}_1\mathcal{G}_2 \Big] + \frac{1}{M_{\Xi_c^+}} \Big[ \mathcal{F}_1\mathcal{F}_3-\mathcal{G}_1\mathcal{G}_3 \Big] \bigg] \nonumber \\
    & + 8 m_\mu^2 q^2 \bigg[ Q_+ \bigg[ \frac{\mathcal{F}_1\mathcal{F}_2}{M_{\Xi_{cc}^{++}}}(2M_{\Xi_c^+}-M_{\Xi_{cc}^{++}}) - \frac{\mathcal{F}_1\mathcal{F}_3}{M_{\Xi_c^+}}(M_{\Xi_c^+}-2M_{\Xi_{cc}^{++}}) \bigg] + Q_- \bigg[ \frac{\mathcal{G}_1\mathcal{G}_2}{M_{\Xi_{cc}^{++}}} \nonumber \\
    & \times (2M_{\Xi_c^+}+M_{\Xi_{cc}^{++}}) - \frac{\mathcal{G}_1\mathcal{G}_3}{M_{\Xi_c^+}}(M_{\Xi_c^+}+2M_{\Xi_{cc}^{++}}) \bigg] \bigg] - 8Q_+(M_{\Xi_c^+}+M_{\Xi_{cc}^{++}})\Big[ (2m_\mu^2+q^2) \nonumber \\
    & \times (M_{\Xi_c^+}-M_{\Xi_{cc}^{++}})^2-q^4 \Big] \bigg( \frac{\mathcal{F}_1\mathcal{F}_2}{M_{\Xi_{cc}^{++}}} + \frac{\mathcal{F}_1\mathcal{F}_3}{M_{\Xi_c^+}} \bigg) - 8Q_-(M_{\Xi_c^+}-M_{\Xi_{cc}^{++}})\Big[ (2m_\mu^2+q^2) \nonumber \\
    & \times (M_{\Xi_c^+}+M_{\Xi_{cc}^{++}})^2-q^4 \Big] \bigg( \frac{\mathcal{G}_1\mathcal{G}_2}{M_{\Xi_{cc}^{++}}} + \frac{\mathcal{G}_1\mathcal{G}_3}{M_{\Xi_c^+}} \bigg) \bigg\} + \frac{8}{M_{\Xi_c^+}M_{\Xi_{cc}^{++}}} \bigg\{ \Big[ Q_+\mathcal{F}_2\mathcal{F}_3 \nonumber \\
    & + Q_-\mathcal{G}_2\mathcal{G}_3 \Big] \Big[ (2m_\mu^2+q^2)(M_{\Xi_c^+}^2-M_{\Xi_{cc}^{++}}^2)^2 - q^2(2q^2+m_\mu^2)(M_{\Xi_c^+}^2+M_{\Xi_{cc}^{++}}^2) \nonumber \\
    & + q^4(q^2-m_\mu^2) \Big] \bigg\} -24\pi kq M_{\Xi_{cc}^{++}}m_\mu^2(M_{\Xi_c^+}^2-M_{\Xi_{cc}^{++}}^2+q^2)\mathcal{F}_1\mathcal{G}_1 + 6\pi kqm_\mu^2 \nonumber \\
    & \times \bigg\{ q^4 \bigg[ \frac{1}{M_{\Xi_{cc}^{++}}}\Big[ \mathcal{F}_1\mathcal{G}_2-\mathcal{F}_2\mathcal{G}_1 \Big] - \frac{1}{M_{\Xi_c^+}}\Big[ \mathcal{F}_1\mathcal{G}_3-\mathcal{F}_3\mathcal{G}_1 \Big] \bigg] - 2q^2 \bigg[ (M_{\Xi_c^+}-M_{\Xi_{cc}^{++}}) \nonumber \\
    & \times \bigg( \frac{M_{\Xi_c^+}}{M_{\Xi_{cc}^{++}}}\mathcal{F}_1\mathcal{G}_2 + \frac{M_{\Xi_{cc}^{++}}}{M_{\Xi_c^+}}\mathcal{F}_1\mathcal{G}_3 \bigg) + (M_{\Xi_c^+}+M_{\Xi_{cc}^{++}})\bigg( -\frac{M_{\Xi_c^+}}{M_{\Xi_{cc}^{++}}}\mathcal{F}_2\mathcal{G}_1 \nonumber \\
    & + \frac{M_{\Xi_{cc}^{++}}}{M_{\Xi_c^+}}\mathcal{F}_3\mathcal{G}_1 \bigg) \bigg] + (M_{\Xi_c^+}^2-M_{\Xi_{cc}^{++}}^2) \bigg[ (M_{\Xi_c^+}-M_{\Xi_{cc}^{++}})^2 \bigg(\frac{\mathcal{F}_1\mathcal{G}_2}{M_{\Xi_{cc}^{++}}} + \frac{\mathcal{F}_1\mathcal{G}_3}{M_{\Xi_c^+}}\bigg) \nonumber \\
    & - (M_{\Xi_c^+}+M_{\Xi_{cc}^{++}})^2 \bigg(\frac{\mathcal{F}_2\mathcal{G}_1}{M_{\Xi_{cc}^{++}}} + \frac{\mathcal{F}_3\mathcal{G}_1}{M_{\Xi_c^+}}\bigg) \bigg]  \bigg\}  \Bigg\}.
\end{align}
}

\providecommand{\url}[1]{#1}\providecommand{\href}[2]{#2}\begingroup\raggedright\endgroup

\end{document}